\documentclass[showpacs,pre]{revtex4}
\usepackage{graphicx}
\usepackage{amsmath}
\usepackage{amssymb}

\begin{document}

\def\vec#1{{\bf #1}}

\def\btt#1{{\tt$\backslash$#1}}

\newcommand{\R}{{\mathbb{R}}}
\def\Z{\mathbb Z}
\def\C{\mathbb C}
\def\Q{\mathbb Q}
\def\N{\mathbb N}
\def\mm{}  

\def\be#1{\begin{equation}  \label{#1}}               %% begin{equation} label
\def\beq{\begin{equation}}                           %% begin{equation} ohne label
\def\ee{\end{equation}} 
\def\bd{\begin{displaymath}}                           %% begin{eq. ohne Nr.}
\def\ed{\end{displaymath}}                             %% end{eq. ohne Nr.}
\def\ba#1{\left(  \begin{array}{#1}}                   % array def
\def\ea{\end{array}  \right)}
\def\eps{\varepsilon}
\def\al{\alpha }
\def\la{\lambda}
\def\norm#1#2{\| #1 \|_{#2}}                          %% L(#2)-Norm
\def\nor#1{\vert #1 \vert}                            %% Betrag
\def\An{A_{\nu}}
\def\Am{A_{\mu}}
\def\Zn{Z_{\nu}}
\def\Zm{Z_{\mu}}                               %% Betrag

\def\intdOdO{\int\!\!\!\!\int\limits_{\!\!\!0}^{4\pi}\!\! d\Omega d\Omega'}
\def\intdOdOdO{\int\!\!\!\!\int\!\!\!\!\int\limits_{\!\!\!\!\!\!\!\!0}^{\!\!\!
4\pi}\!\! 
d\Omega d\Omega'd\Omega''} 

\def\intdrr{\int\!\!\!\!\int\limits_{\!\!\!\Omega}\!\!}
\def\intdrrr{\int\!\!\!\!\int\!\!\!\!\int\limits_{\!\!\!\!\Omega}\!\!} 

\def\rmean{\rho} 

\def\ueber#1#2{{\setbox0=\hbox{$#1$}%
  \setbox1=\hbox to\wd0{\hss$\scriptscriptstyle #2$\hss}%
  \offinterlineskip
  \vbox{\box1\kern0.4mm\box0}}{}}

\title{Local orientations of fluctuating fluid interfaces}

\author{Klaus  Mecke$^{1}$ and Siegfried Dietrich$^{2,3}$}

\affiliation{
(1) Institut f\"ur Theoretische Physik, 
Universit\"at Erlangen-N\"urnberg, 
Staudtstrasse 7, D-91058 Erlangen,  Germany \\ 
(2) Max-Planck-Institut f\"ur Metallforschung, 
Heisenbergstr. 3, 
D-70569
Stuttgart, 
Germany  \\
(3) ITAP,       
Universit\"at Stuttgart, 
Pfaffenwaldring 57, 
D-70569 Stuttgart, 
Germany}

\date{\today}

\begin{abstract} 
Thermal fluctuations cause the local normal vectors of fluid interfaces to deviate from the 
vertical direction defined by the flat mean interface position. This leads to a nonzero mean 
value of the corresponding polar tilt angle which renders a characterization of the thermal 
state of an interface. Based on the concept of an effective interface Hamiltonian we determine 
the variances of the local interface position and of its lateral derivatives. This leads to the 
probability distribution functions for the metric of the interface and for the tilt angle which 
allows us to calculate its mean value and its mean square deviation. We compare the temperature 
dependences of these quantities as predicted by the simple capillary wave model, by an improved 
phenomenological model, and by the microscopic effective interface Hamiltonian derived from 
density functional theory. The  mean tilt angle discriminates  clearly between these theoretical 
approaches and emphasizes the importance of the variation of the surface tension at small 
wave lengths. Also the tilt angle two-point correlation function is determined which renders 
an additional structural characterization of interfacial fluctuations.  
Various experimental accesses to measure the local orientational fluctuations 
are discussed. 

% \noindent 
% PACS: \  68.03.-g, 68.37.-d, 68.05.Cf, 68.15.+e, 68.47.Pe 

\end{abstract}


\maketitle

\section{Introduction}
\label{introduction}

If two phases of condensed matter, which coexist thermodynamically at first-order phase transitions, are brought into spatial contact under suitable boundary conditions, an interface is formed through which the system interpolates smoothly between the structural properties of the adjacent bulk phases. The formation of such an interface is accompanied by a cost in free energy given by the product of the mean interfacial area and the surface tension. Since the surface tension is always positive the system acts as to minimize the interfacial area under given boundary conditions which gives rise to a plethora of capillary phenomena (for reviews see, e.g., Refs. \cite{mandelstam,rowlinson,nelson,desai,chavrolin}). If the first-order phase transitions end at a critical point where the differences between the coexisting phases disappear, the interface broadens and ceases to exist; accordingly the surface tension vanishes there with a universal power law as function of temperature. There are numerous experimental and theoretical techniques to measure and to calculate surface tensions the knowledge of which allows one to tune capillary phenomena in a specific way \cite{mandelstam,rowlinson,nelson,desai,chavrolin}. 

With the advent of sophisticated spectroscopic techniques such as X-ray and neutron scattering under grazing incidence it has become possible to resolve the structural properties of interfaces at an atomic scale, paving the way for a microscopic understanding of interfacial properties in terms of mean density profiles, concentration profiles, and - in the case of nonspherical particles - orientational profiles (see, e.g., Ref. \cite{haase} and references therein). 

These mean profiles, i.e., one-point correlation functions, can be inferred from deviations of reflectivity coefficients from the corresponding Fresnel formulae holding for steplike variations of the structural properties. The analysis of the diffuse scattering around the specular beam provides an even deeper insight into the structural properties of interfaces by probing the structure factor and thus the two-point correlation function in the lateral directions and close to the interface \cite{haase}. 

This interfacial structure factor is of particular interest because in addition to the bulk-like 
density fluctuations, which occur on length scales up to the bulk correlation length $\xi$ and 
reach the surface of each of the two semi-infinite phases, there are capillary wavelike fluctuations 
occuring at length scales larger than $\xi$ but tied to the interfacial region only. Without gravity or lateral confinement these capillary wavelike fluctuations lead to a unlimited thickening of the interface 
and to a unlimited increase in the spatial dependence of the two-point correlation function on 
the lateral distance, known as the roughening of the interface. If one of the two adjacent phases is a solid, this broadening occurs only above the so-called roughening transition whereas for interfaces between fluid phases it occurs at all temperatures. Thus, in the following we focus on fluid interfaces, in particular the liquid-vapor interface of a simple fluid. 

The physical picture of the capillary wavelike fluctuations is that of an intrinsic profile - generated by bulklike fluctuations as if the interface would be flat - exhibiting local, laterally varying height positions. This picture reconciles van der Waals' original idea of an interface \cite{vanderwaals} and the purely capillary wavelike picture of a fluctuating steplike interface put forward by Buff, Lovett, and Stillinger \cite{buff}. 

Based on microscopic density functional theory one can determine the effective interface Hamiltonian, and thus the statistical weight for such a given configuration, which can be expressed in terms of a wavelength dependent surface tension; it equals the macroscopic surface tension at large wavelengths, attains a deep minimum at medium lengths, and increases for short wavelengths \cite{meckedietrich}. The ensuing theoretical predictions for the lateral interfacial structure factor have been confirmed quantitatively for water \cite{nature}, OMCTS, CCl$_4$, squalane,  glycol \cite{mora}, and  Ga \cite{li} by using X-ray scattering under grazing incidence. 

Given this successful agreement between theory and experiment one can turn to the question whether this wavelength dependent surface tension represents primarily a transparent parametrization of the interfacial structure factor or whether the corresponding effective interface Hamiltonian has the quality of transferability, i.e., predicts also other key features of fluid interfaces. 

In this context we consider a snapshot of an interfacial configuration. The points in space, at which the local density equals, say, the mean value of the two bulk densities, form a two-dimensional isodensity surface which can be identified as the local interface position. This surface is characterized by giving its local and laterally varying height with respect to the mean flat interface position. 

Alternatively, at each point of the surface one can attach its normal vector which forms the polar angle $\theta$ with the vertical direction given, e.g., by the direction of gravity and which is the normal for the flat {\it mean} interface. The thermal fluctuations of the interface cause corresponding fluctuations of the surface normals and thus of the polar tilt angle $\theta$. Since $\theta$ is always positive its mean value $<\theta>$ is nonzero; $<\theta>  \; =  0$ would imply the complete absence of capillary wavelike fluctuations. Therefore, $<\theta>$ and the mean squared deviation $<\theta^2>-<\theta>^2$ characterize the thermal state of the interface with a particular emphasis on the capillary wavelike fluctuations. Based on the aforementioned effective interface Hamiltonian one can derive the probability distribution function for $\theta$ and thus calculate its mean value and its moments. Their increase with temperature $T$ signals the unfreezing of capillary waves. 
In this sense $<\theta>(T)$ represents an additional characteristic quantity for defining the thermal state of interfaces, on equal footing with the surface tension. Moreover, the function $<\theta>(T)$ probes directly the concept of the wavelength dependent surface energy. 

Unrelated to this latter aspect, in a series of papers Simpson and Rowlen studied the surface 
roughness contributions 
to measurements, based on linear 
dichroism, of the molecular orientations  of optically active molecules adsorbed on rough solid 
surfaces \cite{simpson1,simpson2,simpson3}. 
In order to infer from such data the local molecular orientation they must be corrected by the 
effects due to the frozen orientational fluctuations of the rough supporting substrate. Thus in 
this context the tilt angle distribution of the interface plays an important role. Other experiments 
consider orientational fluctuations in monolayers or thin multilayer films of non-spherical particles 
adsorbed on fluid interfaces \cite{iwamoto91,iwamoto,iwamoto99,schalke,shen}. 
These orientational fluctuations are determined by fixed 
relations between the orientations of the adsorbed molecules and the fluctuating local orientations 
of the supporting fluid interface. Here we focus on the latter contribution. There is a wide range 
of experimental techniques which provide access to the molecular orientations 
at interfaces \cite{simpson2} such as fluorescence, second-harmonic generation (SHG), sum-frequency generation (SFG), linear dichroism (LD), and angle-resolved photoacoustic spectroscopy (ARPAS). For instance, in Ref. \cite{firestone} the spectroscopically determined orientation of complex molecules such as proteins attached to rough solid substrates is studied. If adsorbed with low concentrations at fluid interfaces such tracer molecules probe the orientational fluctuations of the interface normals.  Promising further perspectives are provided by studying fluid interfaces of colloidal suspensions by laser scanning confocal microscopy which allows one to observe, down to the scale of the colloidal particles, thermally induced capillary waves in real space and to analyze the distribution of the local interface orientations \cite{aarts}.

In Sec. \ref{orientation} we introduce the geometry of the local orientations of fluid interfaces followed in Sec. \ref{capillary} by a discussion of the various theoretical approaches describing capillary wavelike fluctuations as they are used for calculating the Gaussian interfacial fluctuations in Sec. \ref{gaussian}. The dependence of the mean tilt angle and its mean squared deviation on physical parameters is analyzed in Sec. \ref{results}. The relation of these results with experiments is discussed in Sec. \ref{experiments}. We summarize our results in Sec. \ref{conclusion}.

\section{Local orientations of fluid interfaces }
\label{orientation}

The capillary wave approach to study liquid interfaces, which was 
put forward by Buff, Lovett, and Stillinger (BLS)
\cite{buff}, describes the actual 
smooth density profile as the thermal average of a locally fluctuating steplike
interface between the two coexisting 
phases. The local position of the liquid-vapor interface is  described by 
a single-valued function $f(\vec{R})$ depending on the lateral coordinates  $\vec{R}=(x,y)$ 
with $R=\nor{\vec{R}}$  in the
$x$-$y$-plane parallel to the mean interface at $z=0$ (see
Fig. 1). This so-called Mong\'e parametrization neglects overhangs  of the interface as well as bubbles of one phase
inside the other, i.e., domains topologically separated from the
interface; apart from the close neighborhood of the critical point such configurations carry a sufficiently small statistical weight to be negligible. 
 The orientation of the mean interface described  by the unit-vector 
$\vec{e}_z$ is chosen to be parallel to  the  z-axis, i.e.,  perpendicular to the
$x$-$y$-plane. 
This direction is determined either by corresponding boundary conditions or the 
direction of the gravitational force. 
Although the {\it mean} orientation $\vec{e}_z$ is fixed one does observe 
local variations of the interfacial orientation due to thermal fluctuation of $f(\vec{R})$. 
These deviation may be described by a local tilt angle 
\be{tiltangle} 
\theta(\vec{R}) = \arccos \left( \vec{e}_z\cdot \vec{n}(\vec{R}) \right) 
\ee
where 
\be{normal} 
\vec{n}(\vec{R})  = {1 \over \sqrt{1 + f_x^2+f_y^2} }\left(\begin{matrix}-f_x\cr -f_y\cr 1 \end{matrix} \right)
\ee
denotes the normal vector of the interface at the lateral position $\vec{R}$ which is given by the partial derivatives 
$f_i := {\partial f (\vec{R})  \over \partial i}$, $i=x,y$, of the interface position $f(\vec{R})$.  In addition to the polar angle $0\leq \theta(\vec{R})\leq \pi/2$ one may define 
\be{phi} 
\phi(\vec{R}) = \arccos \left( {\vec{n}(\vec{R})  \cdot \vec{e}_x \over \sin\theta(\vec{R})} \right)  \;, 
\ee
i.e., the azimuthal angle between the projection of the normal vector $\vec{n}$ onto the $x$-$y$-plane 
and the x-direction in the $x$-$y$-plane. 
The local tilt angle $\theta(\vec{R})$ and $\phi(\vec{R})$ describe uniquely the interface orientation 
because 
\be{normal2} 
\vec{n}(\vec{R}) = \left( \begin{matrix} \cos\phi(\vec{R}) \sin \theta(\vec{R}) \cr 
\sin\phi(\vec{R}) \sin \theta(\vec{R}) \cr 
\cos \theta(\vec{R})  \end{matrix} \right) 
\ee 
can be reconstructed from the knowledge of  the angles $\theta$ and $\phi$.

\begin{figure}[htbp]
\begin{center}
\includegraphics[width=0.62\linewidth]{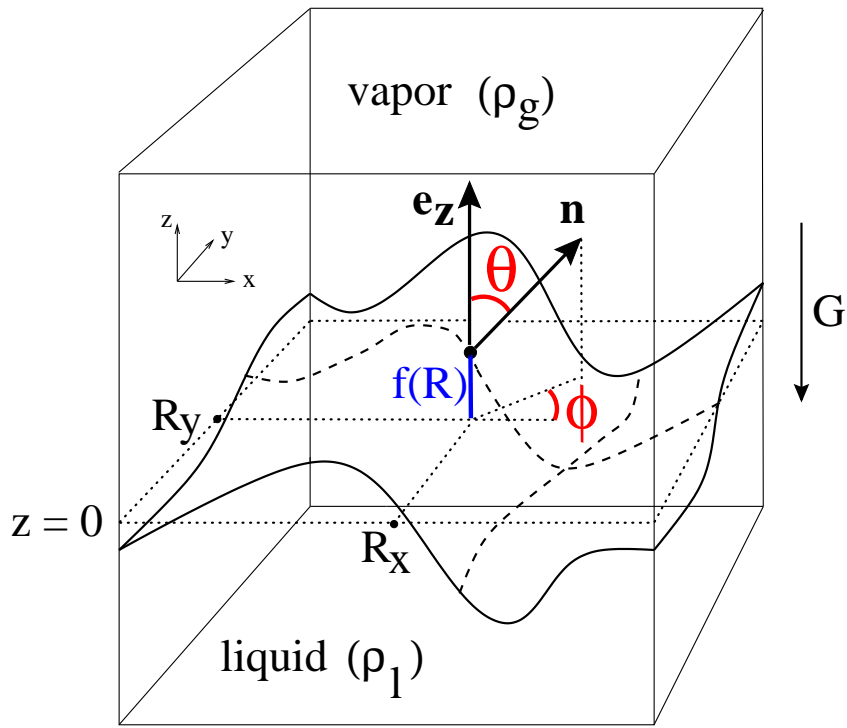}
\end{center}
\caption{
Schematic picture of an interface configuration between the coexisting liquid and
vapor bulk phases with number densities $\rho_l$ and $\rho_g$,
respectively. The interface does not contain overhangs or bubbles.  
Thus the local position of the liquid-vapor interface can be described
by a single-valued function $f(\vec{R})$, where $\vec{R}=(R_x,R_y)$ denotes the
lateral coordinates. The dashed curves indicate the intersections 
between the two-dimensional mani\-fold $f(\vec{R})$ and  planes
$x=R_x=const$ and $y=R_y=const$, respectively. Gravity $\vec{G}$
leads to a mean interface position at $z=0$ (dotted lines). The volume of the sample
is $V=AL$ where $A$ is the lateral area and $L$ is the vertical height. The local 
orientation of the interface is described by the normal vector $\vec{n}(\vec{R})$ 
characterized by the angles $\Theta(\vec{R})$ and $\phi(\vec{R})$ 
(see Eq. (\protect\ref{tiltangle})-(\protect\ref{normal2})). 
}
\label{interfacefig}
\end{figure}

An important motivation to study the  local tilt angle $0\leq \theta(\vec{R}) \leq \pi/2$  is the non-vanishing 
thermal average 
\be{thermaltilt} 
<\theta>(T) \;\;\; >  \;\;\; 0 
\ee
of the tilt angle which is strictly positive at finite temperatures $T$. Thus the interface is always locally 
tilted, so 
that one may observe relevant thermal fluctuation features already in the mean value $<\theta>(T)$, 
in contrast to 
the mean interface position $<f>(\vec{R})=0$ which always vanishes. 
A vanishing mean value of $\theta$ would imply the complete absence of capillary wavelike 
fluctuations.  
Thus, the 'structure' of the interface reveals itself via  the interface position $f(\vec{R})$ 
only through  its variance, i.e., the mean squared deviation 
\be{variance} 
\sigma^2(T) \;\; = \;\; <(f(\vec{R}))^2> \;\; = \;\; \int_{\R^2}{d^2q \over (2\pi)^2 } \; \tilde{\sigma}^2(q)
\ee
from its vanishing mean value, 
where $\tilde{\sigma}^2(q)$ denotes the fluctuation contributions stemming from interfacial modes 
with lateral wavevector $\vec{q}$ with $q  = |\vec{q}|$ (see, c.f., Eq. (\ref{prob3})).  

Recent progress  in grazing incidence X-ray scattering experiments  \cite{nature,mora,li} 
enables  an almost complete 
determination of $\tilde{\sigma}^2(\vec{q})$ and therefore of 
the structure and the fluctuations of the interfacial position $f(\vec{R})$. 
In particular, as predicted theoretically \cite{meckedietrich}, it has been demonstrated 
by  X-ray scattering experiments 
that  for wavevectors above  a few $nm^{-1}$ there is 
a decrease of up to $75 \%$  of the surface energy due 
to the attractive dispersion forces. 
But for some fluids the fluctuation spectrum at large wavevectors $\vec{q}$ could not be resolved 
close to the physical limit $q_{max} = {2\pi \over a}$ of a few \AA$^{-1}$ 
where $a$  denotes the microscopic 
size of the molecules.   
As we shall show in the following the fluctuations of the local tilt angle $\theta(\vec{R})$, i.e., 
its mean value $<\theta>$ 
 and its mean squared deviation  
$<(\delta \theta)^2>$  with $\delta \theta = \theta - <\theta>$  are  much more sensitive
 to the structure of the interface at large wavevectors, so that the analysis of them  promises 
a  significant improvement of the understanding of fluid interfaces.

\section{Capillary wave theory}
\label{capillary}

The cost 
in free energy to generate an undulation of a flat interface, i.e., to generate a non-flat 
interface configuration 
depends on  the amplitude $f(\vec{R})$ of the undulation but also on physical parameters 
such as temperature $T$, the gravitational constant $G$, the  surface tension $\gamma_0$, and 
the bending rigidity $\kappa_0$ or ultimately the interaction potential between the fluid particles.  
The usual approach to describe this cost in free energy assumes that the amplitudes 
are small so that the Hamiltonian may be expanded into powers of $f(\vec{R})$ and its 
derivatives $\nabla f(\vec{R})$ and $\Delta f(\vec{R})$ yielding  ($m$ denotes the mass of the particles and $\Delta\rho=\rho_l-\rho_g$ the difference in number densities of the coexisting liquid and vapor phase) 
\be{hamiltonian} 
\begin{array}{c} 
\displaystyle  {\cal H}^{(H)}[f(\vec{R})] =  
   {1 \over 2} \int_{\R^2} d^2R  \left[  mG\Delta\rho \left(f(\vec{R})\right)^2  \right. \cr
  \left. \displaystyle
+ \gamma_0 \left(\nabla f(\vec{R})\right)^2 + \kappa_0 \left(\Delta f(\vec{R})\right)^2  \right] \;. 
\end{array} 
\ee
This gradient expansion is known also as the 
Helfrich Hamiltonian for fluctuating membranes \cite{gompper,helfrich,david}. 

It is transparent to study ${\cal H}^{(H)}$ in Fourier space 
in which  the bending modes decouple.  To this end 
we introduce the Fourier transformed functions
\be{fourier} 
\begin{array}{ll} 
\displaystyle
\tilde{f}(\vec{q}) & =\displaystyle \int_{\R^2}d^2R\; e^{-i\vec{q}\cdot\vec{R}}f(\vec{R})\ \ , \cr 
  & \cr 
\displaystyle
f(\vec{R}) & =\displaystyle\int_{\R^2}{d^2q \over (2\pi)^2 }
e^{i\vec{q}\cdot\vec{R}}\tilde{f}(\vec{q})      \cr 
\end{array} 
\ee
yielding the Fourier transformed Hamiltonian  
\be{helfham} 
{\cal H}^{(H)}[\tilde{f}(\vec{q})] =   {1 \over 2} \int_{\R^2} {d^2q \over (2\pi)^2}\;
 |\tilde{f}(\vec{q})|^2 \left(  mG\Delta\rho +
\gamma^{(H)}(q) q^2 \right)  
\ee
with the momentum dependent  surface tension 
\be{sigma} 
\gamma^{(H)}(q) =\gamma_0 + \kappa_0 q^2 \;\;. 
\ee
We note  that the use of an upper cut-off $q< q_{\max}=2\pi/a$ for the wavevectors is 
indispensible because the concept of capillary wave fluctuations becomes meaningless 
on length scales  smaller 
than the microscopic size $a$ of the molecules.

Instead of starting from the {\em ansatz} in Eq. (\ref{hamiltonian}) an alternative strategy 
(see Ref. \cite{meckedietrich}) to {\em derive} ${\cal H}[f(\vec{R})]$ is based on  a microscopic
density functional theory  for inhomogeneous simple 
fluids, which is a  successful approach for the description of
nonuniform fluids \cite{evans,evans2}. The different kinds of occuring density
fluctuations - bulk 
bubbles and interface undulations - are separated by determining the
intrinsic density profile $\rho_f(\vec{r})$ via minimizing the  grand canonical density 
functional 
$\Omega[\rho(\vec{r})]$ under the
constraint of a locally prescribed  interface 
position $f(\vec{R})$, i.e., the location of the isodensity contour of the mean
density is fixed as function of the lateral coordinates $\vec{R}$. 
Due to the long-ranged character of the underlying van der Waals interparticle
potential  $W(\vec{r})$ between the fluid particles nonanalytic contributions occur 
and therefore a gradient expansion
of the Hamiltonian breaks down.  
Instead, in Fourier space one obtains in terms of the amplitudes 
$\tilde{f}(\vec{q})$ describing the interface position  the Hamiltonian  
\be{gaussham} 
{\cal H}^{(G)}[\tilde{f}(\vec{q})] =  \int_{\R^2} {d^2q \over (2\pi)^2}\;
 {1 \over 2}  E(q) \; |\tilde{f}(\vec{q})|^2 
\ee
where the superscript  $(G)$ indicates that this expression refers  to a Gaussian
approximation bilinear  in the amplitudes $f(\vec{q})$ of the bending modes.  
The momentum dependent energy  (per (lenght)$^4$) 
\be{gaussenergy} 
E(q) = E_0 +\gamma(q) q^2  
\ee  
with $E_0=mG\Delta\rho$  can be written in terms of a 
momentum dependent  surface tension $\gamma(q)$, which is not simply given by the Helfrich 
expression (\ref{sigma}) but by an explicit functional of the molecular interaction potential 
$W(\vec{r})$. Based on the simple  functional \cite{evans2}
$$ \Omega\left[\rho(\vec{r})\right]=  \int_{V} d^3r \left(
f_h(\rho(\vec{r})) 
+ \mu \rho(\vec{r}) + \rho(\vec{r}) V(\vec{r}) \right)  $$
\be{functional}
+{1 \over 2}\int_{V} d^3r\int_{V} d^3r'
w(|\vec{r}-\vec{r}'|)\rho(\vec{r})\rho (\vec{r}') \;\;,  
\ee 
where $V$ is the volume of the sample, $\rho(\vec{r})$ is the number 
density of the fluid particles at  $\vec{r}=(x,y,z)$, $r=|\vec{r}|$, 
$f_h(\rho)$ is the reference bulk free energy  of a system determined by
the short-ranged, repulsive   contribution  $w_s(r)$ to the 
interaction potential $W(\vec{r})$, and $w(\vec{r})$ is the 
attractive part of $W(\vec{r})$, one finds  
within the reliable 
so-called product approximation   \cite{meckedietrich}
\be{sigmass}
\begin{array}{ll} 
\displaystyle
\gamma(q) \simeq  & \displaystyle \left(\gamma_0 + \tilde{\kappa}_0^{(H)}q^2 \right)
{\tilde{w}(q)-\tilde{w}(0) \over {1\over 2} 
\tilde{w}''(0) q^2}  \cr 
  &  \displaystyle  +
\left(\kappa_0 
-\tilde{\kappa}_0^{(HH)} {\tilde{w}(q)\over \tilde{w}(0)} \right) q^2 + {\cal
O}(q^4)  
\end{array} 
\ee 
with the surface tension and  bending rigidities 
\be{kappa0}
\begin{array}{lll} 
\displaystyle
\gamma_0 & \displaystyle = {1 \over 12}  \tilde{w}''(0) {(\Delta\rho)^2 \over \xi}   &  \;\;>\;0\quad,   \cr 
  & \cr 
\displaystyle
\tilde{\kappa}_0^{(H)} & \displaystyle  =  {1 \over 24} \tilde{w}''(0)
(\Delta\rho)^2 \xi  C_H  & \;\; >\;0\quad, \cr  
  & \cr 
\displaystyle
\tilde{\kappa}_0^{(HH)}  & \displaystyle = -{4 \over 65} \tilde{w}(0)  (\Delta\rho)^2 \xi^3
C_H^2   & \;\;>\;0\quad , \cr    
  & \cr 
\displaystyle
\kappa_0  & \displaystyle = {4\over 65} (\Delta\rho)^2 \xi^3
C_H^2   {\partial^2 f_h(\rho) \over 
\partial\rho^2}  & \;\;>\;0\quad . 
\end{array} 
\ee
For the bulk free energy density $f_h(\rho)$ one may use   the Carnahan-Starling expression  
\cite{evans2}
\be{carnahan}
f_h(\rho) = k_B T \rho \left\{ \ln(\rho\lambda^3) -1 + {4 \eta -
3\eta^2 \over (1-\eta)^2} \right\} 
\ee
where $\lambda$ is the thermal de Broglie wavelength and $\eta= {\pi
\over 6} \rho r^3_0$ the packing fraction. (For the definition of $r_0$ see, c.f., Eq. (\ref{potential})). 
Within this product approximation the momentum dependence of
$\gamma(q)$ in Eq. (\ref{sigmass}) is determined by the three-dimensional Fourier
transform 
\be{fourierpot} 
\tilde{w}(\vec{Q})=\int_{\R^3}d^3\vec{r}\;
e^{-i\vec{Q}\cdot\vec{r}}w(|\vec{r}|)  \;, 
\ee
of the interparticle potential $w(r)$
evaluated for the absolute value $q$ of the lateral momentum
$\vec{q}$ with 
$\tilde{w}''(0) > 0$ and  $\tilde{w}(0) < 0$. 

Considering particles which interact via van der
Waals forces  one may adopt for the attractive part of the interaction 
potential $w(r)$ the form 
\be{potential}
w(r)= - {w_0 r_0^6 \over (r_0^2+r^2)^3} \quad
\longrightarrow_{\!\!\!\!_{\!\!\!\!\!\!\!\!\!\! 
r \rightarrow \infty}}  \quad -A r^{-(d+\tau)} 
\ee
which exhibits  the correct large distance behavior $w(r\rightarrow \infty) 
\rightarrow  -Ar^{-6}$  with $A=w_0r_0^6$  for $d=3$, $\tau=3$ 
\cite{barker}. The length
$r_0$ corresponds to the diameter of the particles  and
thus   serves as a lower limit for the 
length scale of the  density fluctuations and of the 
capillary waves considered below.   Then, Eq. (\ref{sigmass}) reduces to  the explicit expression 
\be{finalsigma} 
\begin{array}{ll} 
\displaystyle 
{\gamma(q)\over \gamma_0} & \displaystyle  = 
(2-C_H\left(\xi q \right)^2){1-(1+qr_0)e^{-qr_0}\over (qr_0)^2} + 0.74\, C_H^2 \left(\xi q\right)^2 \left({1 \over 2}
 + \left({\xi \over r_0}\right)^2 \left(1-(1+qr_0)e^{-qr_0}\right) \right)  \cr 
   &  \cr 
   &\displaystyle 
\longrightarrow_{\!\!\!\!_{\!\!\!\!\!\!\!\!\!\! 
q\rightarrow 0}}  \quad  1 - {2\over 3} qr_0 + {3\over 4} \left[1-{2\over 3} C_H(1-0.74C_H)\left({\xi\over r_0}\right)^2\right](qr_0)^2 + {\cal O}((qr_0)^3)    \cr 
   &  \cr 
   &\displaystyle 
\longrightarrow_{\!\!\!\!_{\!\!\!\!\!\!\!\!\!\! 
q\rightarrow \infty}}  \quad   0.74 C_H^2\left(\left({\xi \over r_0}\right)^4+ {1\over 2} \left({\xi \over r_0}\right)^2\right)(qr_0)^2 -C_H\left({\xi\over r_0}\right)^2 +{2\over (qr_0)^2} + {\cal O}\left(q^3 e^{-qr_0} \right)   \cr 
\end{array} 
\ee
depending solely on the bulk correlation length $\xi$, the diameter $r_0$ of the molecules, 
and the dimensionless parameter $C_H\approx 0.25$ describing the distortion of the intrinsic 
density profile due to the bending of the interface. Upon increasing $q$, $\gamma(q)/\gamma_0$ 
decreases below $1$, reaches a minimum and increases again  (for more details see Ref. \cite{meckedietrich}).
This decrease of $\gamma(q)$ is in accordance with recent simulation 
data \cite{stecki,milchev} which, however, have not yet confirmed the 
predicted and experimentally observed reincrease of $\gamma(q)$ at large $q$. 
  
For the density functional theory given in Eq. (\ref{functional}) and the 
interaction potential in Eq. (\ref{potential}) one has  
$\tilde{w}(q)=\tilde{w}(0)(1+qr_0)e^{-qr_0}$ with $\tilde{w}(0)
=-\pi^2r_0^3w_0/4$,  so that one finally finds the temperature dependences  
\be{surfacetens}  
\begin{array}{ll} 
  \displaystyle
\gamma_0 & \displaystyle = {\pi^2 \over 48} w_0 {r_0^5 \over \xi} (\Delta\rho)^2 \;,  \cr 
 &\cr  
\displaystyle
\tilde{\kappa}_0^{(H)} & \displaystyle = {\xi^2 \over 2} C_H \gamma_0 \;, \cr
 &\cr  
\displaystyle
\tilde{\kappa}_0^{(HH)}  & \displaystyle =  { 48\over 65} {\xi^4\over r_0^2} C_H^2 \gamma_0 \;, \cr
 &\cr  
\displaystyle
\kappa_0  & \displaystyle = \tilde{\kappa}_0^{(HH)} \left( 1+ {1\over 2} {r_0^2 \over \xi^2}\right) \;, \cr 
\end{array} 
\ee
and 
close to the critical point at $T_c$: 
\be{finalsigma0} 
\begin{array}{ll}   
\displaystyle
\Delta \rho & \displaystyle = {1.23 \over r_0^3} \left(1- {T \over T_c}\right)^{1\over 2} \;,  \cr 
  & \cr 
\displaystyle
\xi  & \displaystyle = {r_0 \over 2} {T \over T_c} \left(1- {T \over T_c}\right)^{-{1\over 2}} \;.    \cr  
\end{array} 
\ee 
Although it is straightforward to determine $\xi$ and
$\Delta\rho$ numerically we found that these approximate but analytic 
expressions are quite accurate  for temperatures between the triple point 
$T_{tr}\simeq (2/3)T_c$  and the critical point (apart from the critical region in 
which deviations from the above mean-field behavior occur), so that they can be  
used in the following. 
For temperatures below $T_{tr}$ - where no fluid interface exists - the
model defined by Eq. (\ref{functional}) ceases to be applicable because it does
not capture the freezing transition, which would require a more
sophisticated version of the density functional.

In order to study the dependences on the numerous physical parameters we 
define two length scales: the  capillary length $L_c$ and the 
stiffness length $L_\kappa$, respectively,   
\be{caplength} 
L_c = \sqrt{\gamma_0 \over E_0} \quad,  \quad\; L_\kappa = \sqrt{\kappa_0 \over \gamma_0} \;\;, 
\ee
which in addition to the microscopic cut-off lengths $r_0$ (range of the hard-core repulsion potential) 
and $a={2\pi \over q_{max}}$ 
(size of the molecules), 
and the bulk correlation length $\xi$  
 are relevant for  the cost in
free energy for bending 
an interface with
wave vector $\bf q$ 
(see Eqs.  (\ref{gaussham}) and (\ref{gaussenergy})).  
For simple liquids  
the values for $r_0$, $a$, and $L_\kappa$ are of the same order of magnitude  
and may be 
identified in a first approximation. But since the distributions of the local orientations 
of the interface turn out to depend sensitively 
on these microscopic length scales we prefer to keep them distinguished until their  influence 
has been elucidated. Moreover, colloidal suspensions (see, e.g., 
Ref. \cite{aarts}) might offer the possibility to keep these length scales 
well separated and to vary them to large extent independently.

\section{Gaussian interfacial fluctuations}
\label{gaussian} 

Before we study in detail the dependence of the interfacial structure on physical 
parameters such as the surface tension $\gamma_0$ 
or the interparticle potential $w(\vec{r})$  in this section we discuss 
those properties of an interface which follow  from the Gaussian approximation 
given by Eq. (\ref{gaussham}). 

Since within the  Gaussian approximation $P[f]\sim {\rm exp}(-\beta {\cal H}^{(G)})$, 
see Eq. (\ref{gaussham}), the 
bending modes are decoupled  one finds for 
the probability of a surface wave $\tilde{f}(q)$ of wavevector $q$  the explicit expression 
(see Eq. (\ref{gaussham})) 
\be{prob1} 
P[\tilde{f}(q)] = {E(q) \over \pi k_BT} e^{- {E(q) \over k_BT} |\tilde{f}(q)|^2 } 
\ee
so that mean values $<A>$ for functionals $A[f(\vec{R})] $ of the interface position can 
straightforwardly be calculated by 
\be{averages} 
<A>[\beta E(q)] = \int{\cal D}f \;\; A[f(\vec{R})] 
\ee
where 
\be{integration}
\int{\cal D}f \ldots = \prod_{\vec{q}\in\R^2/2}  \int_{-\infty}^\infty  d\left({\cal R}e \tilde{f}(\vec{q})\right) 
 \int_{-\infty}^\infty d\left({\cal I}m\tilde{f}(\vec{q})\right)  \ldots {E(q) \over \pi k_BT} e^{- {E(q) \over k_BT}\left[\left({\cal R}e \tilde{f}(\vec{q})\right)^2+ \left({\cal I}m\tilde{f}(\vec{q})\right)^2 \right] } 
\ee
denotes the integration measure for the interfacial degrees of freedom  
which solely depends  on $\beta E(q)$ 
with $\beta=1/(k_BT)$ as the inverse thermal energy.  Because of the 
relation $\tilde{f}(-\vec{q}) = \tilde{f}^*(\vec{q})$, the product in Eq. (\ref{integration}) runs only 
over suitably discretized  vectors $\vec{q}$ in the half space $\R^2/2$. 
Details on the Wiener measure in terms of Fourier 
coefficients can be found in Ref. \cite{chaichian}. 

In a first step we consider local second order moments $\sigma^2_{2n,2m}$ (see, c.f.,  Eq. (\ref{prob7})) 
characterizing the distribution functions of lateral derivatives 
of $f(\vec{R})$  at one and the same lateral position $\vec{R}$, whereas 
Subsec. \ref{nonlocal} focuses on correlations of $f(\vec{R})$ at 
points $\vec{R} = \vec{R}_1$,  $\vec{R}_2$ 
with non-vanishing distance vector  $\vec{a}=\vec{R}_2-\vec{R}_1$.

\subsection{Local  variances}
\label{local} 

The probability $P[f_0]$ to find  the interface at the lateral coordinate 
$\vec{R}$ at a specified position $f_0 =
f(\vec{R}=\vec{0})$  is given by  (see Eq. (\ref{fourier}))  
\be{prob2} 
\begin{array}{ll} 
\displaystyle
P[f_0;\sigma]  & \displaystyle = \int{\cal D}f\; 
\delta\left(f_0 - \int_{\R^2} {d^2\vec{q} \over (2\pi)^2} \tilde{f}(q)\right)
\\
  &  \\ 
 & \displaystyle = \int\limits_{-\infty}^\infty {dp \over 2\pi} e^{ipf_0 }\prod_{\vec{q}\in \R^2/2} 
 \int_{-\infty}^\infty  d\left({\cal R}e \tilde{f}(\vec{q})\right) 
 \int_{-\infty}^\infty d\left({\cal I}m\tilde{f}(\vec{q})\right)  \cr
 & \displaystyle \quad  \quad\times {E(q) \over \pi k_BT} e^{ -{E(q)
\over k_BT}\left[\left({\cal R}e \tilde{f}(\vec{q})\right)^2+ \left({\cal I}m\tilde{f}(\vec{q})\right)^2 \right]  
 -2 i p \left({\cal R}e \tilde{f}(\vec{q}) \right) } \\ 
  &  \\  
 & \displaystyle = {1  \over \sqrt{2\pi \sigma^2}} e^{-{1 \over 2} { f_0^2
\over \sigma^2 }} 
\end{array} 
\ee
with the fluctuation spectrum (see Eq. (\ref{variance})), 
$\tilde{\sigma}^2(q) = k_BT/E(q)$ and the variance 
\be{prob3}
\sigma^2 = \int {d^2\vec{q} \over (2\pi)^2} {k_BT \over E(q)} =
\int\limits_{q_{min}}^{q_{max}} dq \; {k_BT \over 2\pi} {q \over E(q)} \;\;. 
\ee
Integrating over all wavevectors requires  special care since  modes may be 
constrained  by physical boundary conditions. 
As stated after Eq. (\ref{sigma})  $q_{max}=2\pi/a$ denotes the 
maximum possible wavevector because fluctuations 
on length scales smaller than the microscopic size $a$ of the molecules are not compatible 
with the concept of capillary waves. 
Accordingly, one may introduce a lower cut-off $q_{min}=2\pi/L$ due to the finite lateral size $L$ 
of the interface. If $L$ is larger than the capillary length $L_c$ (see Eq. (\ref{caplength})) 
one may safely set $q_{min}=0$ so that fluctuations with long wavelengths are suppressed  only 
by gravity.

Indeed, $P[f_0;\sigma]$ is properly normalized so that $\int_{-\infty}^\infty df_0 P[f_0;\sigma] =1$. 
Accordingly, the probability $P[f^{(n,m)}_0;\sigma_{2n,2m}]$ to find a prescribed derivative 
\be{prob6} 
f^{(n,m)}_0= \left. {\partial^{n+m}f(\vec{R}) \over \partial x^n \partial y^m} 
\right|_{\vec{R}=\vec{0}}  
\ee
at $\vec{R}$ is given analogously to  Eq. (\ref{prob2}) with $f_0$ replaced by $f_0^{(n,m)}$ 
but with the variance 
\be{prob7}
\sigma_{2n,2m}^2 
= \int {d^2\vec{q} \over (2\pi)^2} \; q_x^{2n}q_y^{2m} {k_BT\over E(q)}  
\; = \; I_{n,m} \;  \sigma_{2n+2m+1}^2 
\ee
with 
\be{prob8} 
\sigma_{k}^2  = \int\limits_{q_{min}}^{q_{max}} {dq \over 2\pi} \; q^{k} { k_BT\over E(q)} 
\ee
and 
$I_{n,m}= \int_0^{2\pi} {d\phi\over 2\pi} \cos^{2n}\phi\sin^{2m}\phi = {1\over 2\pi} \left(J_{n,m}+J_{m,n}\right)$ for integers $n,m\in \N$  
with  $J_{n,m}= 2\int_0^{\pi/2} d\phi  \cos^{2n}\phi\sin^{2m}\phi = B(m+{1\over 2},n+{1\over 2})$ (3.621.5 
and 8.384.1 in Ref. \cite{gradshteyn})  so that ($n,m\in \N$)
\be{prob10} 
I_{n,m} 
= {1 \over \pi} {\Gamma\left(n+{1\over 2}\right)\Gamma\left(m+{1\over 2}\right) 
\over \Gamma(n+m+1)} \;\;. 
\ee 
In particular,  one finds $I_{0,0}=1$, $I_{1,0}=I_{0,1}={1\over 2}$, 
$I_{2,0}=I_{0,2}={3\over 8}$, $I_{1,1}={1\over 8}$. Therefore the probability 
$P[\vec{f}\,'_0;\sigma']=P[f^{(1,0)}_0;\sigma']P[f^{(0,1)}_0;\sigma']$ for the slope (orientation) 
$\vec{f}\,'_0 = \nabla f(\vec{R}=\vec{0})$ of the interface at $\vec{R} = 0$ 
has the variance ${\sigma'}^{2} = \sigma^2_{2,0} = \sigma^2_3/2$. Note 
that $\sigma'$,  $\sigma_3$, and $\vec{f}_0'$ are dimensionless. 
The probability $P[\Delta f;\sigma_{H} ]$ of the Laplacian $\Delta f(\vec{R})$, which in lowest order in $f$ is twice the mean 
curvature 
\be{meancurv} 
H_0 = {f_{xx}(1+f_y^2) - 2f_xf_yf_{xy} + f_{yy}(1+f_x^2) \over 2(1+f_x^2+f_y^2)^{3\over 2} }  \;, 
\ee  
has the variance 
$\sigma_{\Delta}^2 =  \sigma^2_5$: 
\be{prob7s}
\sigma_{\Delta}^2 
= \int {d^2\vec{q} \over (2\pi)^2} \; \left(q_x^{2}+q_y^{2}\right)^2 {k_BT\over E(q)}  
\; = \;  \int\limits_{q_{min}}^{q_{max}} {dq \over 2\pi} \; q^{5} { k_BT\over E(q)} \;\;. 
\ee

\subsection{Finite distance variances}
\label{nonlocal}

An alternative  to the local slope $\vec{f}'_0$ as a measure for the orientation of an
 interface is provided by finite differences $\delta f := f(\vec{R}+\vec{a}) - f(\vec{R})$ of 
the interface position. 
Analogous to Eq. (\ref{prob2}), the joint probability of finding the heights $f_0$ and $f_{\vec{a}}$ 
of the interface at the lateral positions $\vec{0}$ and $\vec{a}\neq \vec{0}$, respectively, reads 
\be{joint} 
\begin{array}{ll} 
\displaystyle
P[f_0,f_{\vec{a}}]  & \displaystyle  = \int{\cal D}f\; 
\delta\left(f_0 - \int {d^2\vec{q} \over (2\pi)^2} \tilde{f}(q)\right)
\delta\left(f_{\vec{a}} - \int {d^2\vec{q} \over (2\pi)^2} \tilde{f}(q)e^{i\vec{q}\cdot\vec{a}} \right)
\\
  &  \\
 & \displaystyle = \int\limits_{-\infty}^\infty {dp \over 2\pi}\int\limits_{-\infty}^\infty {dp' \over 2\pi} e^{ipf_0 +ip'f_{\vec{a}}}
\prod_{\vec{q}\in \R^2/2}  {E(q) \over \pi k_BT}
 \int_{-\infty}^\infty  d\left({\cal R}e \tilde{f}(\vec{q})\right) 
 \int_{-\infty}^\infty d\left({\cal I}m\tilde{f}(\vec{q})\right)  \cr
 & \displaystyle \quad  \quad\times e^{ -{E(q)
\over k_BT}\left[\left({\cal R}e \tilde{f}(\vec{q})\right)^2+ \left({\cal I}m\tilde{f}(\vec{q})\right)^2 \right]  
 -2 i \left({\cal R}e \tilde{f}(\vec{q}) \right) (p+p'\cos \vec{q}\vec{a}) 
+2 i \left({\cal I}m \tilde{f}(\vec{q}) \right) p'\sin \vec{q}\vec{a}  } \\ 
  &  \\ 
 & \displaystyle = \int\limits_{-\infty}^\infty {dp \over 2\pi} \int\limits_{-\infty}^\infty {dp' \over 2\pi} 
 {\rm exp}\left\{ipf_0 + ip'f_{\vec{a}}  -{1\over 2} (p,p') {\bf \Sigma^2}  \left(\begin{matrix} p \cr p' \end{matrix} \right)  \right\}   \\
  &  \\
 &  \displaystyle = {1\over 2\pi \sqrt{\lambda_+\lambda_-}} {\rm exp}\left\{-{1\over 2} 
\left(f_0, f_{\vec{a}}\right) {\bf \Sigma^{-2}} 
\left(\begin{matrix}f_0\cr f_{\vec{a}}\end{matrix}\mm 
\right) \right\} 
\end{array} 
\ee
using the formula 
\be{gaussintegral} 
\int_{-\infty}^\infty \prod_{i=1}^N dx_i \;{\rm exp}\left\{-{1\over 4} x_i A_{ij}x_j + s_ix_i\right\} \;\;= \;\; {2^N \pi^{N\over 2} \over \sqrt{{\rm det} A}} \; {\rm exp}\left\{s_iA_{ij}^{-1}s_j\right\} 
\ee
for any symmetric and positive definite matrix $(A_{ij})$ 
with the matrices  of the variances (${\bf \Sigma^2}{\bf \Sigma^{-2}}={\bf I}$) 
\be{joint2} 
{\bf \Sigma^2} = \left(\begin{matrix}
\sigma^2 & \sigma^2(\vec{a}) \cr \sigma^2(\vec{a}) & \sigma^2 
\end{matrix}\mm  \right)   \quad, \quad \quad 
{\bf \Sigma^{-2}} = {1\over \lambda_+\lambda_-} \left(\begin{matrix}
\sigma^2 & -\sigma^2(\vec{a}) \cr -\sigma^2(\vec{a}) & \sigma^2 
\end{matrix}\mm  \right)   
\ee
where $\sigma^2$ is given by Eq. (\ref{prob3}). 
The eigenvalues 
\be{joint3} 
\lambda_{\pm}(\vec{a}) = \sigma^2 \pm \sigma^2(\vec{a}) \;\;\geq \;\; 0  
\ee
of  the matrices  ${\bf \Sigma^{2}}$ and $\lambda_+\lambda_-{\bf \Sigma^{-2}}$ are determined 
by Eq. (\ref{prob3}) and by the correlation function   
\be{joint4} 
\begin{array}{ll} 
\displaystyle
\sigma^2(\vec{a}) & \displaystyle  = 
\int
{d^2q \over (2\pi)^2} {k_BT \over
E(q)} \cos (\vec{q}\vec{a}) \cr
  & \cr 
   & \displaystyle  = 
\int\limits_{q_{min}}^{q_{max}} {dq \over 2\pi} {k_BT \; q \over
E(q)} J_0(qa) \;\;,  \cr 
 \end{array} 
\ee
with $\sigma^2(a=0)=\sigma^2$ and 
where $J_0(x)$ denotes the Bessel function with $J_0(x\rightarrow 0) = 1- {1\over 4} x^2 
+ {\cal O}(x^4)$ and $J_0(x\rightarrow\infty) = (\sin x + \cos x) {1\over \sqrt{\pi x} } 
+ {\cal O}(x^{-1})$. 
Since $E(q)$ increases at least as $\sim q^2$ for large $q$ the latter property of $J_0(x)$ 
allows one to set $q_{max} = \infty$ for any finite $a$.  
For $q_{min}=0$ and $q_{max}=\infty$ and $E(q)$ given by Eq. (\ref{sigma}) one finds 
for the correlation function defined in Eq. (\ref{joint4}) 
\be{joint5} 
\begin{array}{ll} 
\displaystyle
\sigma^2(\vec{a})  & \displaystyle= {k_BT \over 2\pi \gamma_0} K_0\left( {a\over L_c} \right) \;,\quad 
\kappa_0=0\;,  \cr
    & \displaystyle \rightarrow {k_BT \over 2\pi \gamma_0} \ln (L_c/a) \;,\quad L_c>>a \;, 
\end{array} 
\ee
where $K_0(x)$ denotes the modified Bessel function. 
One recovers the distribution 
of the interface position (Eq. (\ref{prob2})) 
\be{test1} 
P[f_0;\sigma]  = \int\limits_{-\infty}^\infty  df_{\vec{a}} \;\; P[f_0,f_{\vec{a}};{\bf \Sigma}]  
\ee
by integrating over the positions at $\vec{a}$.

Note that the conditional probability 
\be{test5} 
{P[f_0,f_{\vec{a}}]  \over P[f_0] } = {
{\rm exp}\left\{-{1\over 2} \left( {f_0^2/\sigma^2 \over \sigma^4/\sigma^4(\vec{a}) -1 } 
+ {f^2_{\vec{a}}/\sigma^2  - 2f_0f_{\vec{a}}\sigma^2(\vec{a})/\sigma^4 
\over 1- \sigma^4(\vec{a})/\sigma^4}  \right)\right\} 
\over \sqrt{ 2\pi \sigma^2 (1- \sigma^4(\vec{a})/\sigma^4)}} 
\ee
is not a function of $\delta f = f_0-f_{\vec{a}}$ alone, but depends  on $f_0$ 
and $f_{\vec{a}}$ separately.  For the distribution of the difference 
$\delta f(\vec{a}) := f(\vec{R}+\vec{a}) - f(\vec{R})$, i.e., the probability that at given 
points $\vec{R}$ and $\vec{R} + \vec{a}$ the height difference has a certain 
value $\delta f(\vec{a})$, one finds not the conditional probability, Eq. (\ref{test5}), but the expression  
\be{test2} 
\begin{array}{ll} 
\displaystyle
P[\delta f(\vec{a})]  & \displaystyle = \int\limits_{-\infty}^\infty  df \; P[f_0+f,f_{\vec{a}}+f]  \cr
 & \displaystyle =  \int\limits_{-\infty}^\infty  { df\over 2\pi \sqrt{\lambda_+\lambda_-}}   
{\rm exp}\left\{-{\left(f+{f_0+f_{\vec{a}}\over 2}\right)^2 \over \lambda_+} 
+ {(f_0-f_{\vec{a}})^2  \over 4\lambda_+} - {\sigma^2 (f_0-f_{\vec{a}})^2\over 2\lambda_+\lambda_-}   \right\} \cr 
 & \displaystyle = {1 \over \sqrt{2\pi \sigma_\delta^2(\vec{a}) }} 
e^{-{1\over 2} {(\delta f(\vec{a}))^2 \over  \sigma_\delta^2(\vec{a})} } 
\end{array} 
\ee
with 
\be{test3} 
\sigma_\delta^2(\vec{a})=2\lambda_- = 2(\sigma^2 - \sigma^2(\vec{a}) )  
= {1\over \pi} \int_{q_{min}}^{q_{max}}  dq {k_BT \over E(q)} q (1-J_0(qa)) \;. 
\ee
Due to the translational invariance in lateral directions this distribution for $f(\vec{R}+\vec{a}) - f(\vec{R})$ 
is independent of $\vec{R}$ so that without loss of generality one can set $\vec{R}=0$.   
The finite difference variance $\sigma_\delta^2(a)$ is shown in Fig. \ref{figsigma_finite} for 
the phenomenological capillary wave model with $E(q)/E_0=1+L_c^2q^2(1+L_\kappa^2q^2)$
 (see Eqs. (\ref{sigma}) and (\ref{gaussenergy})) and the effective length 
scales $L_c=\sqrt{\gamma_0/E_0}$ 
and $L_\kappa=\sqrt{\kappa_0/\gamma_0}$ 
given by Eq. (\ref{caplength}). 

\begin{figure}[htbp]
\begin{center}
\includegraphics[width=0.65\linewidth]{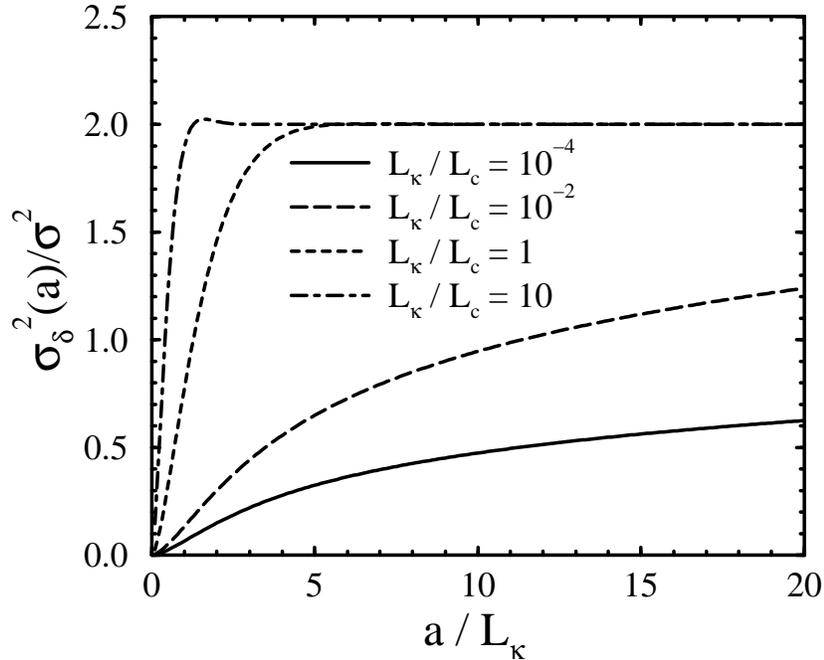}
\end{center}
\caption{Finite difference variance $\sigma^2_\delta(a)$ (Eq. (\ref{test3})) normalized by the variance $\sigma^2$ of the interface position (Eq. (\ref{prob3})) as function of $a/L_\kappa$ for 
various values of $L_\kappa/L_c$ (Eq. (\ref{caplength})) for the phenomenological 
capillary wave model 
(Eqs. (\ref{sigma}) and (\ref{gaussenergy})). The quadratic increase 
of $\sigma^2_\delta(a)$ in the limit $a\rightarrow 0$ 
(see Eq. (\protect\ref{test4})) is  visible on the scale $a/L_\kappa$ 
only for small values of $L_\kappa/L_c$.    }
\label{figsigma_finite}
\end{figure}

In the limit $|\vec{a}|\rightarrow \infty$ one recovers the variance $2\sigma^2$ corresponding to  
two independently fluctuating interfaces. 
In the limit $|\vec{a}|\rightarrow 0$ one finds 
\be{test4} 
\sigma_\delta^2(\vec{a}) \rightarrow \sigma'^2\; \vec{a}^2 + {\cal O}((q_{max}a)^4) 
\ee
in accordance with the expectation that for small 
distances $a$ the difference $f(\vec{R}+\vec{a}) - f(\vec{R})$ is determined by the  slope of $f(\vec{R})$, so that its variance is given by the variance $\sigma'=\sigma_{2,0}=\sigma_3/\sqrt{2}$ of the slope (see Eq. (\ref{prob8}))  times the distance $a$.

The comparision of Eqs. (\ref{test3}) and (\ref{prob8}) allows one to express the 
variance $\sigma'^2$ of the local slope in terms 
of the finite distance variance $\sigma_\delta^2(a)$ for the height distribution at an effective lateral 
distance $a(L_\kappa q_{max},L_cq_{max})$: 
\be{sigma_g2} 
\sigma'^2 a_0^2 = \sigma^2_\delta(a(L_\kappa q_{max},L_cq_{max})) \;\;, 
\ee
where $\sigma'^2a_0^2$ is a local estimate of the  variance $\sigma^2_\delta(a_0)$ 
at a finite distance $a_0$. 
For small values of $a_0$ one may apply the expansion in Eq. (\ref{test4}) yielding 
$a(L_\kappa q_{max},L_cq_{max})=a_0$, but 
for finite distances the effective lateral 
distance  $a(L_\kappa q_{max},L_cq_{max})$ is determined  implicitly by 
\be{implicita} 
0 \; = \; \int_0^1 dy {y\left(1-(a_0q_{max} y)^2/4-J_0(a(L_\kappa q_{max},L_cq_{max})q_{max}y)\right) \over (L_cq_{max})^{-2}+y^2+(L_\kappa q_{max})^2y^4} \;. 
\ee 
The solution of Eq. (\ref{implicita}) is plotted in Fig.   \ref{figsigmazero} for several 
values of the capillary length $L_cq_{max}$. For sufficiently small distances $a_0$ one 
finds $a(L_\kappa q_{max},L_cq_{max})\simeq a_0$. In general, the numerical equivalence 
of the finite distance approach and the local slope estimate, i.e., of $a$ and $a_0$ becomes better for 
 increasing length scales $L_c$ and $L_\kappa$.

In order to obtain a good agreement of the local slope estimate $\sigma'^2a_0^2$  
and the finite distance variance $\sigma_\delta^2(a)$ the distance $a_0$ should be smaller 
than $L_\kappa$ and also smaller than $2\pi/q_{max}$.

The finite distance approach may be used as an alternative to regularize the fluctuation 
spectrum, i.e., to suppress 
the infrared divergence caused by  the  large wavelength  mode $q\rightarrow 0$ which leads 
to an infinite variance $\sigma$ for vanishing gravity, i.e.,  $E_0=0$.  
Instead of introducing a minimum value $q_{min}$ for the allowed wavevectors - which 
is necessary for local quantities such as the variance studied in Subsect. \ref{local} 
(see Eq. (\ref{prob3}))  - one can consider correlations at finite distances $a$ for which 
one can set $q_{min} = 0$.  

Also    
the ultraviolet divergence, which is caused by  the diverging number of large 
momenta  modes and  yields  
an infinite variance $\sigma'=\sigma_3/\sqrt{2}$ of the slope (see Eq. (\ref{prob8})) - even 
for a non-vanishing bending rigidity 
$\kappa_0$, may be regularized by the finite distance approach.  
Instead of introducing a maximum value $q_{max}$ for the allowed wavevectors - which 
is necessary for local quantities such as the variances studied in Subsect. \ref{local} 
(see Eq. (\ref{prob3}))  - one can consider differences  $f(\vec{R}+\vec{a}) - f(\vec{R})$ 
at finite distances $a$ for which 
one can use  $q_{max} = \infty$ if $\kappa_0\neq 0$.  The corresponding 
variance $\sigma^2_{\delta}(a)$ of the difference $\delta f(a)$ remains finite 
for $\kappa_0> 0$ whereas even in this case the 
variance $\sigma'^2$ of the 
local slope diverges. 

But as one can see from Eq. (\ref{test4}) the finite difference variance $\sigma^2_{\delta}(a)$ is given by  
 the variance $\sigma'^2$ of the local slope for small values of $a$, so that both approaches 
are  even almost quantitatively equivalent for distances  $a \lesssim {2\pi \over q_{max}}$ or  
for large $q_{max}$. At least for variances significant  differences cannot 
be observed. Therefore in the following  we apply  the simpler and conceptually more straightforward 
approach of local orientations with wavevectors larger than $q_{max}$ being  not permitted.

\begin{figure}[htbp]
\begin{center}
\includegraphics[width=0.45\linewidth]{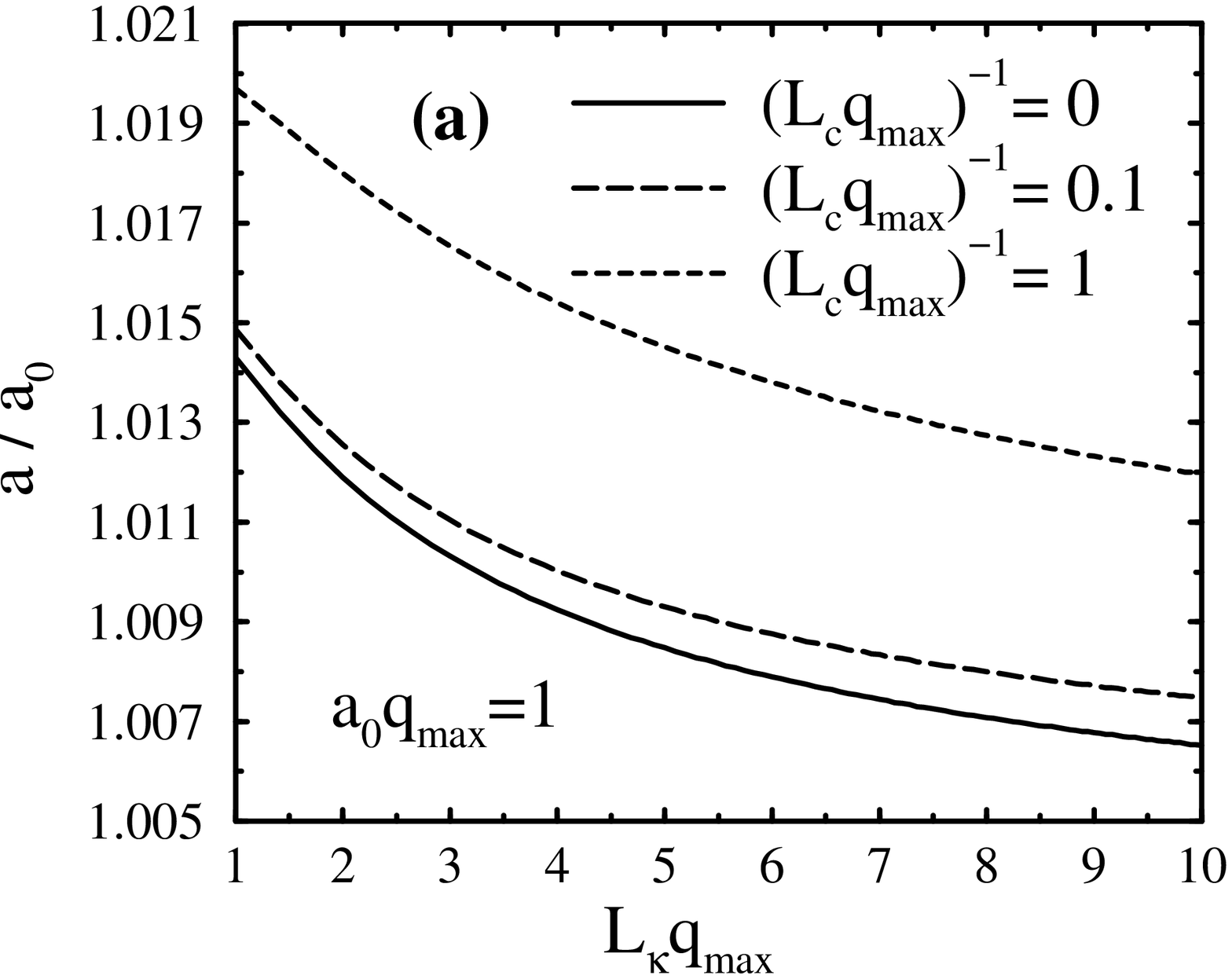}
\includegraphics[width=0.45\linewidth]{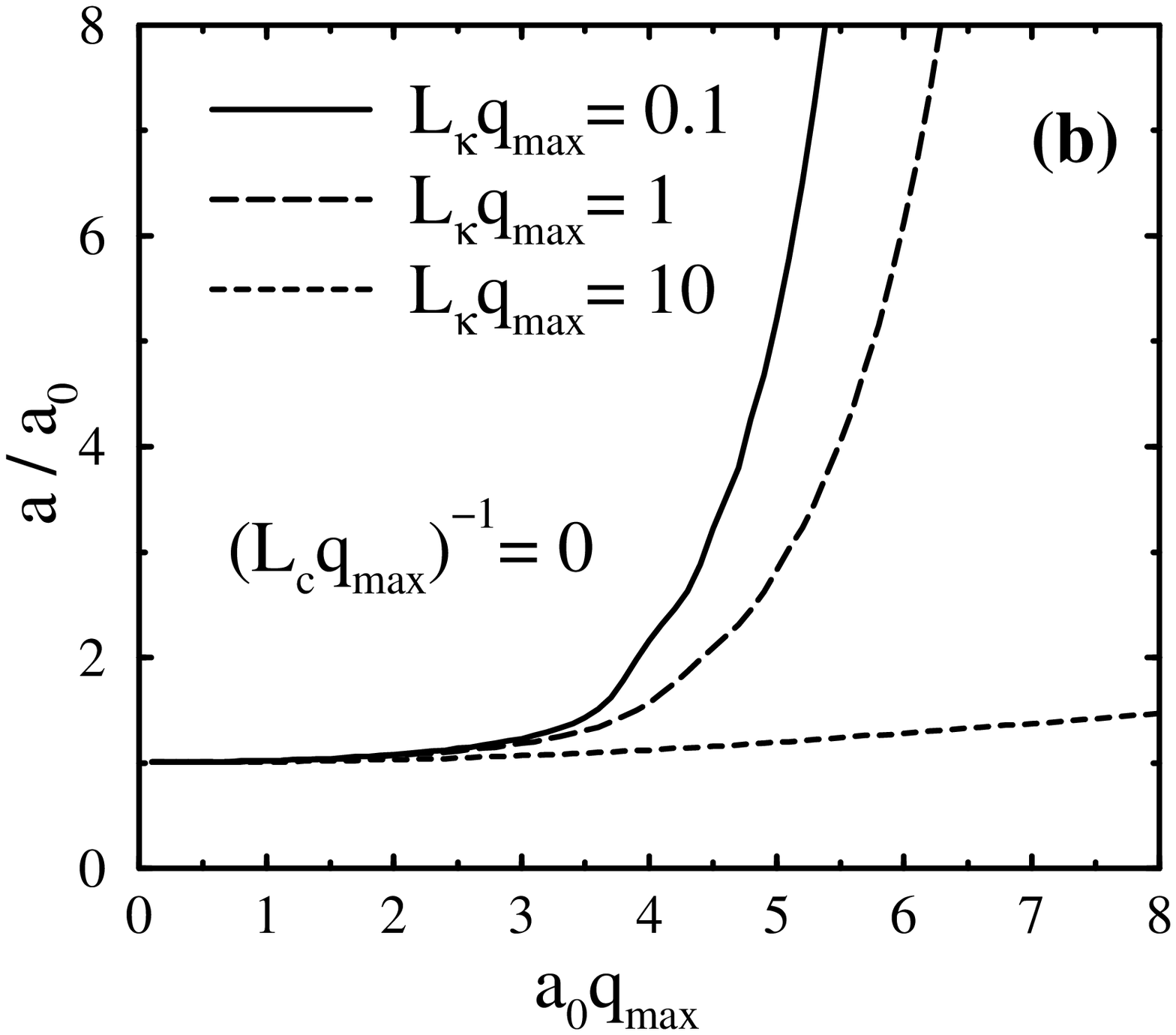}
\includegraphics[width=0.45\linewidth]{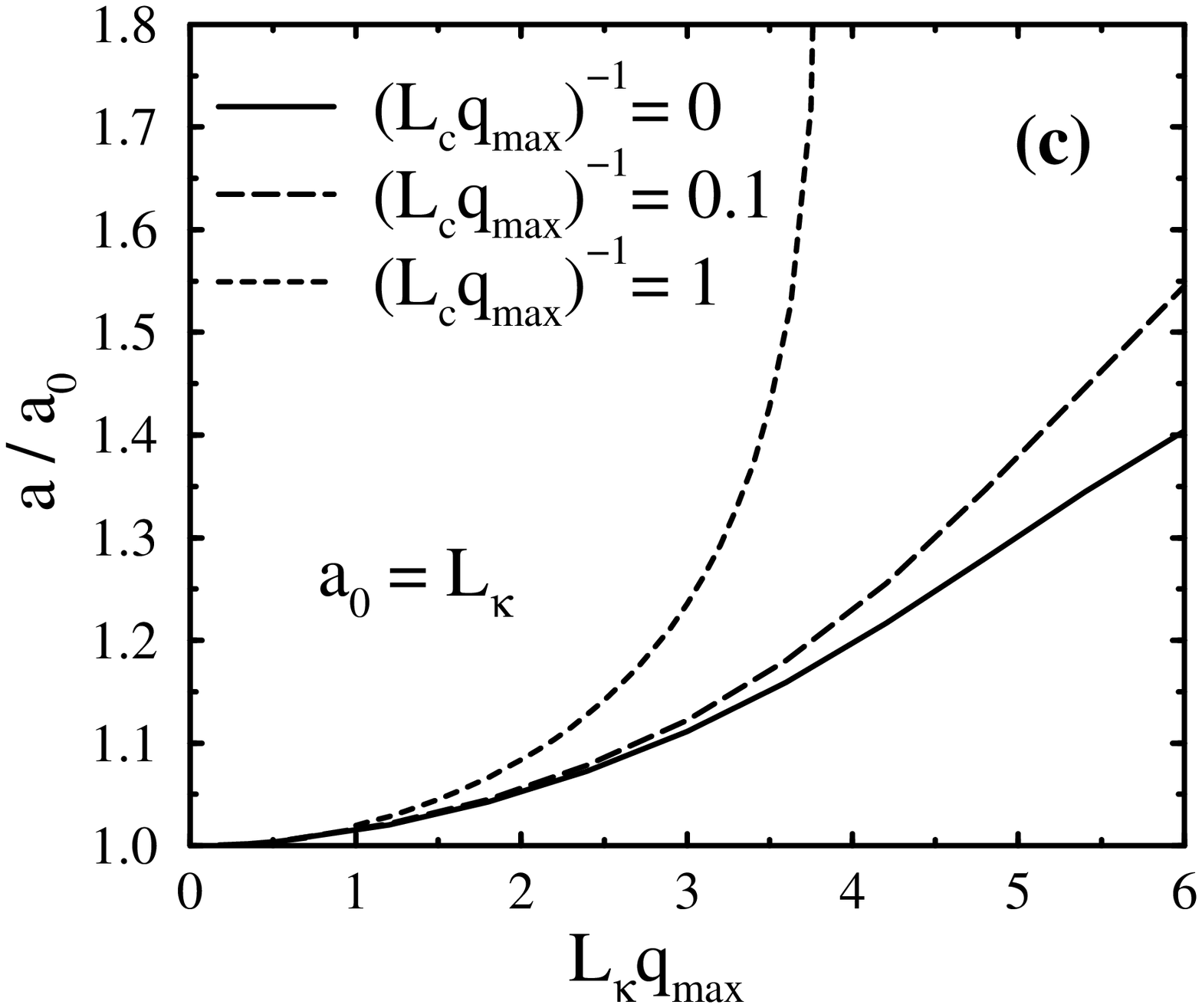}
\includegraphics[width=0.45\linewidth]{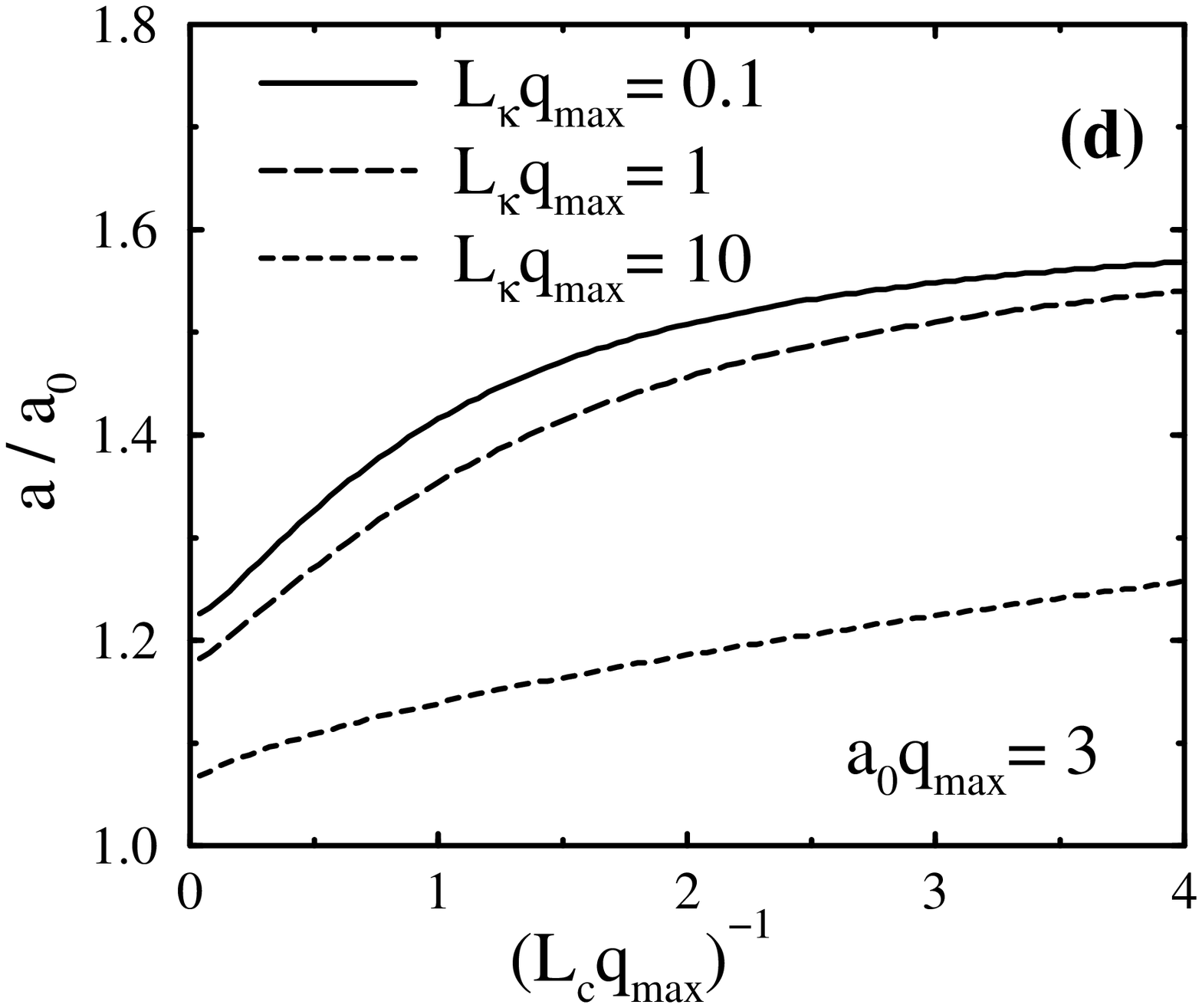}
\end{center}
\caption{Effective lateral distance $a(L_\kappa q_{max}, L_cq_{max})$ at 
which the variance $\sigma_\delta^2(a)$ (Eq. (\ref{test3})) of height differences 
equals the estimate $\sigma'^2a_0^2$ from the local slope variance (see 
Eqs. (\ref{sigma_g2}) and (\ref{prob8})): (a) for $a_0q_{max}=1$; (b) for $L_cq_{max}=\infty$;  
(c) for $a_0=L_\kappa$; and (d) for $a_0q_{max}=3$. 
The ratio $a/a_0$ decreases towards unity for increasing values of $L_\kappa$ and decreasing 
finite distances $a_0$. }
\label{figsigmazero}
\end{figure}

\subsection{Distribution of metric $g$ and  tilt angle $\theta$} 
\label{metricdist} 

The geometry of an interface can be described by the so-called 'fundamental forms' of the
 two-dimensional 
manifold $f(\vec{R})$. Thus we need to derive the probability 
distribution of the tangent vectors $\vec{t}_i$ ($i=x,y$), the normal vector $\vec{n}$ 
(see Eq. (\ref{normal})), and
the  metric tensor $g_{ij} = \vec{t}_i \cdot \vec{t}_j$.  
For the tangent vectors 
\be{tangent} 
\vec{t}_x = \left(\begin{matrix}1\cr 0\cr f_x(\vec{R}) \end{matrix} \right) \;\;\;,\;\;\; \vec{t}_y = \left(\begin{matrix}0 \cr 1 
\cr f_y(\vec{R}) \end{matrix} \right)
\ee
it is sufficient to know the distribution   $P[\partial_i f;\sigma']$ of the derivative 
given by Eq. (\ref{prob7}). Notice that we do not normalize the tangent vectors 
because we want to keep the relation $g_{ij} = \vec{t}_i \cdot \vec{t}_j$ for the metric.
Then the probability distribution for  the metric tensor reads 
\be{metric1} 
P[g_{xx}] =P[g_{yy}] = {1 \over  \sqrt{2\pi\sigma_g^2} \; \sqrt{g_{xx}-1}} 
e^{- {g_{xx}-1 \over \sigma_g^2} }  
\ee
with $g_{xx}=1+f_x^2(\vec{R}) >1$ and  $\sigma_g=\sigma_3$ given by Eq. (\ref{prob8}). 
The off-diagonal values $g_{xy} = g_{yx}= f_x(\vec{R})f_y(\vec{R})$ 
are more complicated  but can nevertheless be derived  
explicitly  ($f_x=f_0^{(1,0)}$ and $f_y=f_0^{(0,1)}$, Eq. (\ref{prob6})) 
\be{metric2}
\begin{array}{ll}  
\displaystyle
P[g_{xy}]  & \displaystyle = \int_{-\infty}^\infty  df_x \int_{-\infty}^\infty df_y P[f_x]P[f_y] \delta(g_{xy}-f_xf_y) \cr 
 & \displaystyle = \int_{-\infty}^\infty {df_x \over |f_x|} P[f_x]P[{g_{xy} \over f_x}]   \cr 
 & \displaystyle  =
{1 \over 2\pi \sigma' } \int\limits_0^\infty {dy \over y} \; e^{ 
-{1\over 2} {y \over \sigma'^2}\left(1+ {g_{xy}^2  \over y^2}\right) }  \cr 
  & \displaystyle = {1 \over \pi \sigma'^2} K_0\left(  {|g_{xy}|\over \sigma'^2}\right) 
\end{array} 
\ee
with (see Eqs.(\ref{prob7})-(\ref{prob10})) $\sigma_{2,0}^2=I_{1,0}\sigma_3^2 = {1\over 2}\sigma_3^2=\sigma'^2=\sigma_g^2/2$. 
Here, we have used the integral expressions 
\be{integral1} 
\int\limits_0^\infty dx \;  
e^{-\gamma x - \beta/(4x)} = \sqrt{\beta\over \gamma} K_1(\sqrt{\beta\gamma})  
\ee
and 
\be{integral2} 
\int\limits_0^\infty {dx \over x}  \;  
e^{-\gamma x - \beta/(4x)} = 2 K_0(\sqrt{\beta\gamma}) = -2(K'_1(\sqrt{\beta\gamma}) +
 {K_1(\sqrt{\beta\gamma}) \over \sqrt{\beta\gamma}}  )  
\ee
yielding the modified Bessel  functions $K_0(z)$ and $K_1(z)$. 
Then, 
for the  determinant  $g={\rm det} (g_{ij}) = 1+ (\nabla f(\vec{R}))^2$  
it follows straightforwardly 
the exponential distribution 
\be{metric3} 
\begin{array}{ll}  
\displaystyle
P[g]  & \displaystyle =  \int_{-\infty}^\infty df_x \int_{-\infty}^\infty df_y \; P[f_x]P[f_y] \delta(g-1-f_x^2-f_y^2)   \cr
  & \displaystyle = {1\over \pi \sigma_{2,0}^2} \int_0^{\sqrt{g-1}} {df_x \over \sqrt{g-1-f_x^2}} 
e^{-{g-1\over 2\sigma_{2,0}^2}}  \cr 
 & \displaystyle  = {1\over \sigma_g^2} e^{- {g-1\over \sigma_g^2}} 
\end{array} 
\ee
with $g>1$ and the mean value and variance  
\be{meanmetric} 
<g> = 1+ \sigma_g^2 \quad, \quad <g^2>-<g>^2 =  \sigma_g^4 \;\;,  
\ee
respectively. Since $P[g_{xy}]=P[-g_{xy}]$ the mean values of the off-diagonal elements vanish, 
so that the 
mean value of the metric reads 
\be{meanmetric2} 
<g_{ij}> = {1\over \sqrt{2}} \left( \begin{matrix}1+\sigma'^2 & 0 \cr 
0 & 1+\sigma'^2 \end{matrix}\mm  \right) \;\;. 
\ee

\begin{figure}
\begin{center}
\includegraphics[width=0.5\linewidth]{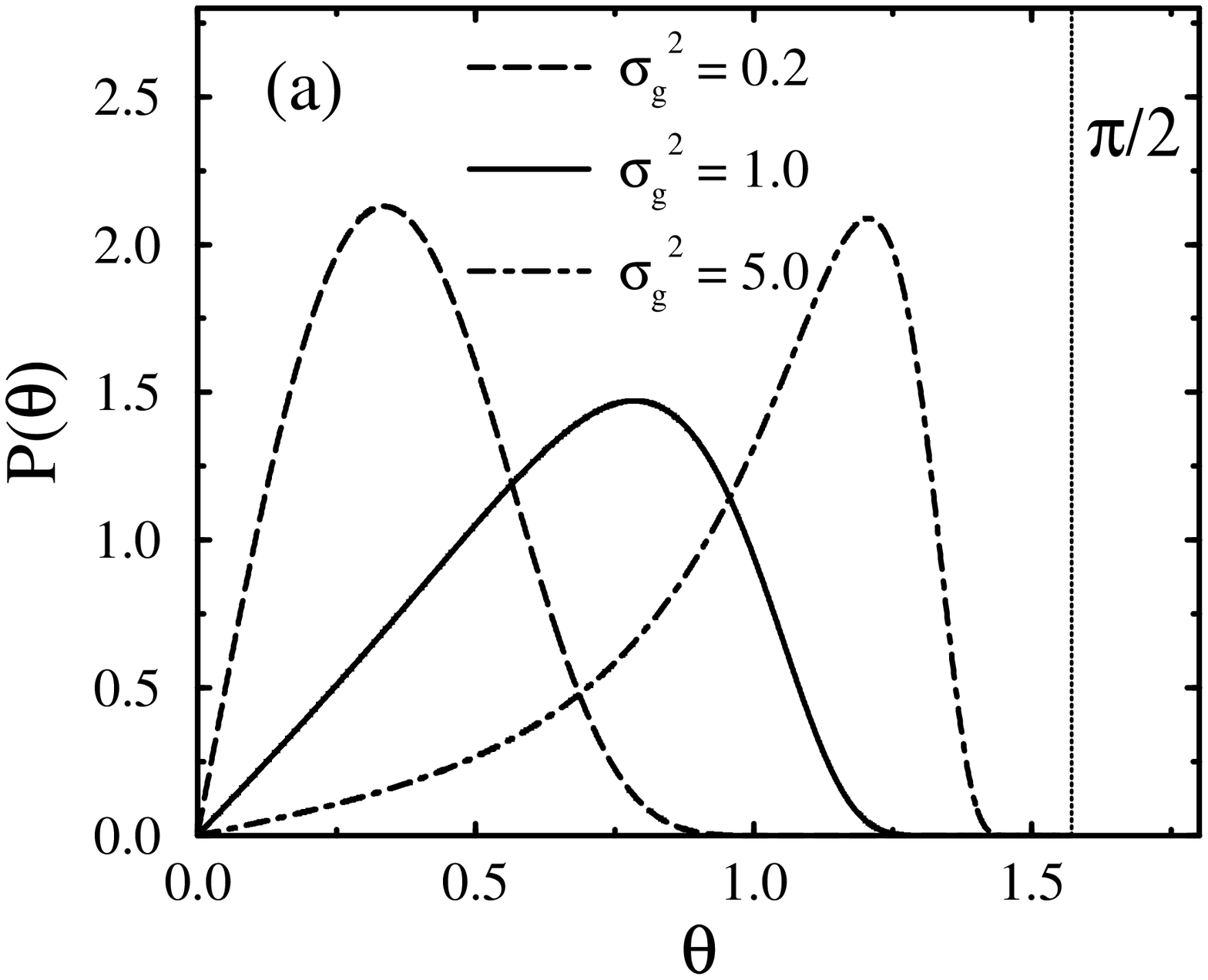}
\includegraphics[width=0.43\linewidth]{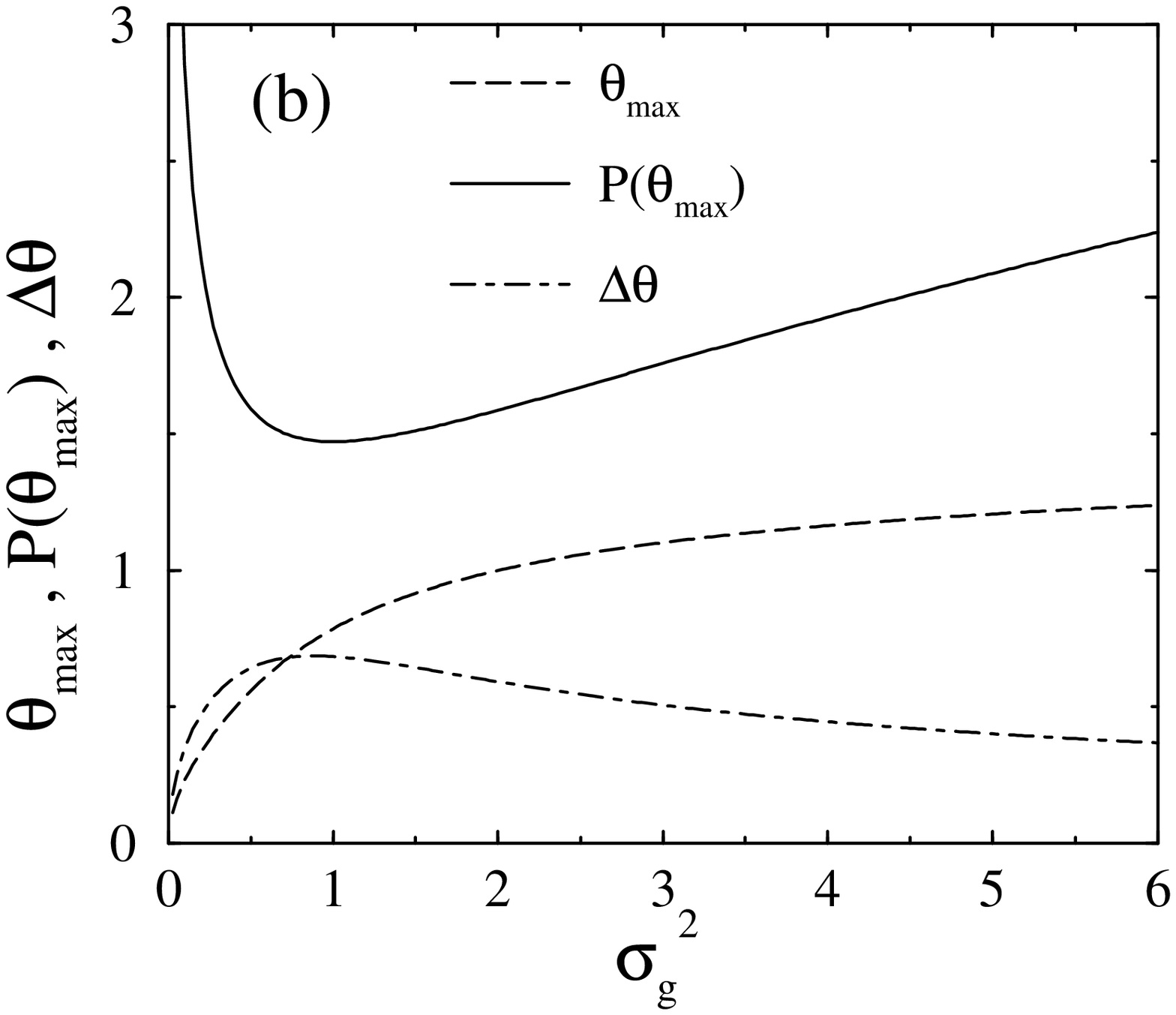}
\end{center}
\caption{(a) The probability distribution $P[\theta]$ of the tilt angle $\theta$ is given 
by Eq. (\protect\ref{angle}). (b) The position $\theta_{max}$ of the maximum  increases 
with increasing variance 
$\sigma_g^2 = 2\sigma'^2=\sigma_3^2$ (Eq. (\protect\ref{prob8})) and 
approaches the limiting value $\pi/2$ for an infinitely rough interface.  
The maximum value $P(\theta_{max})$ depends 
non-monotonically on $\sigma_g^2$.  The width $\Delta \theta$ at $P[\theta_{max}]/2$ reaches its 
maximum value $0.69$ for $\sigma_g^2=0.89$. }
\label{fig_verteilung}
\end{figure}

Using Eqs. (\ref{tiltangle}), (\ref{normal}), and (\ref{metric3})  one can  deduce  the probability 
 distribution $P[\theta]$ for the tilt angle $\theta$ between the normal $\vec{n}$ of the 
interface and the vertical axes $\vec{e}_z$. With $\theta=\arccos \left({1\over \sqrt{g}} \right)$ 
and noting that due to $g\in [1,\infty)$ one has $\theta \in [0,\pi/2)$, which excludes 
overhangs, we find ($\sigma^2_g = 2\sigma'^2=\sigma_3^2$, Eq. (\ref{prob8})) 
\be{angle} 
P[\theta] = {2\over \sigma_g^2} {\sin\theta  \over \cos^3\theta} 
e^{-{\tan^2\theta \over \sigma_g^2}}  \;\;, \quad \sigma_g^2 = \int_{q_{min}}^{q_{max}} {dq \over 2\pi} q^3 {k_BT \over E(q)} \;\;, 
\ee
which is shown in Fig. \ref{fig_verteilung} for various values of $\sigma_g^2$. Notice 
that the maximum of $P[\theta]$ at 
\be{maximum} 
\theta_{max} = {\rm arctan}\left({1\over 2} \sqrt{3\sigma_g^2-2 + \sqrt{9\sigma_g^4-4\sigma_g^2+4} }\right)
\ee
approaches the limiting values $\theta_{max} = 0$ for $\sigma_g\rightarrow 0$ and 
$\theta_{max} = \pi/2$ for $\sigma_g\rightarrow \infty$. Close to the vertical direction 
the tilt angle of the normal vector $\vec{n}$ is distributed linearly in $\theta$  
\be{angellimit1} 
P[\theta\rightarrow 0] = {2\theta \over \sigma_g^2} \left( 1 + ({4\over 3} - {1\over \sigma_g^2}) \theta^2 
+ {\cal O}(\theta^4) \right) 
\ee
whereas one finds an exponentially vanishing distribution towards the horizontal limit 
\be{angellimit2} 
P[\theta\rightarrow {\pi\over 2} ] = {2\over \sigma_g^2} {1\over ({\pi\over 2}-\theta)^3} 
e^{ - {1\over \sigma_g^2} {1\over ({\pi\over 2}-\theta)^2}}  \;\;.  
\ee

With the probability distribution given by Eq. (\ref{angle}) one finds straightforwardly  
the mean value  of the tilt angle $\theta$ (using partial integration and 3.383.10 in Ref. \cite{gradshteyn}): 
\be{meanangle} 
\begin{array}{ll} 
\displaystyle
<\theta>  & \displaystyle =   \int\limits_0^\infty dx \; e^{-x} \arctan\sqrt{x\sigma_g^2}  = {1\over 2\sigma_g} \int\limits_0^\infty dx \; x^{-{1\over 2}} {e^{-x} \over x+ \sigma_g^{-2}} \cr 
  &  \cr 
  &\displaystyle = {\pi \over 2} e^{\sigma_g^{-2}} \left(1-\Phi(\sigma_g^{-1}) \right)   \cr 
  &  \cr 
  & \displaystyle =  \left\{\begin{matrix} 
\displaystyle\sqrt{\pi}  \sum_{k=0}^\infty  (-1)^{k} 2^{-k-1}(2k-1)!!  \; \sigma_g^{2k+1}  
&, \;\; \sigma_g<1  \cr 
  \cr 
\displaystyle{\pi \over 2} e^{\sigma_g^{-2}} - \sqrt{\pi} \sum_{k=0}^\infty {2^k \over (2k+1)!!} 
\sigma_g^{-2k-1} &, \;\; \sigma_g> 1 
 \end{matrix}\mm  \right. \cr 
  &  \cr 
  & \displaystyle =  \left\{\begin{matrix} { \sqrt{\pi} \over 2} \left(\sigma_g-{1\over 2}\sigma_g^3 
+ {\cal O}(\sigma_g^5)\right) 
   \cr  
{\pi \over 2 } -  {\sqrt{\pi} \over \sigma_g} +  {\pi \over 2 \sigma_g^2}  
- {2\sqrt{\pi}\over 3} {1\over \sigma_g^3 } +    {\cal O}({1\over \sigma_g^{4}})
 \end{matrix}\mm  \right. \;\;.  \cr 
\end{array} 
\ee
Here, $\Phi(x)$ denotes the error function 
\be{error} 
\Phi(x) = {2\over \sqrt{\pi} } \int\limits_0^x e^{-t^2} dt 
\ee
with its series representations 
\be{errorseries} 
\begin{array}{ll} 
\displaystyle 
\Phi(x) & \displaystyle = {2\over \sqrt{\pi}} \sum_{k=1}^\infty (-1)^{k+1} {x^{2k-1} \over (2k-1)(k-1)!}  \cr
 & \displaystyle = {2\over \sqrt{\pi}} e^{-x^2} \sum_{k=0}^\infty {2^k x^{2k+1} \over (2k+1)!!} 
\end{array} 
\ee
and 
\be{errorseries2} 
1-\Phi(x) = {1\over \pi} e^{-x^2} \sum_{k=0}^\infty {(-1)^k \Gamma(k+{1\over 2}) \over x^{2k+1}} 
\ee
with $\Gamma(k+{1\over 2}) = 2^{-k} (2k-1)!! \sqrt{\pi}$, i.e., $\Gamma(1/2) = \sqrt{\pi}$, 
$\Gamma(3/2) = \sqrt{\pi}/2$, $\Gamma(5/2) = 3\sqrt{\pi}/4$, and $\Gamma(7/2) = 15\sqrt{\pi}/8$.  
Accordingly one finds for the mean squared deviation $<(\delta\theta)^2>=<\theta^2>-<\theta>^2$  
\be{varianceangle} 
\begin{array}{ll} 
\displaystyle
<(\delta\theta)^2>  & \displaystyle 
=   \int\limits_0^\infty dx \; e^{-x} \left(\arctan\sqrt{x\sigma_g^2}\right)^2 \; -\; <\theta>^2 \cr 
  &  \cr 
  & \displaystyle =  \left\{\begin{matrix}{4-\pi\over 4}\sigma_g^2
-{16-3\pi \over 12}\sigma_g^4 + {\cal O}(\sigma_g^6) \cr
{ 2\ln\sigma_g \over \sigma_g^2}  \;\; + \;\; {\cal O}\left({1\over \sigma_g^2}\right) 
\;\;. \hfill 
 \end{matrix}\mm  \right. \cr 
\end{array} 
\ee
These structural features $<\theta>$ and $<(\delta\theta)^2>$ of a fluctuating interface 
are shown in Fig. \ref{sigma_angle} 
as function of the variance $\sigma^2_g=2\sigma'^2$. 
Notice that even in the limit $\sigma_g\rightarrow \infty$, i.e., for a soft interface at 
high temperatures (Eq. (\ref{angle})),  
one does not obtain the mean value 
and mean squared deviation 
\be{random} 
\begin{array}{ll} 
\displaystyle
<\theta>_0 &  \displaystyle =  \int\limits_0^{\pi/2} d\theta \sin\theta \int\limits_0^{2\pi} {d\phi \over 2\pi } \;\; \theta  = 1 \cr 
 & \cr 
\displaystyle 
<(\delta \theta)^2>_0  &  \displaystyle =  \int\limits_0^{\pi/2} d\theta \sin\theta \int\limits_0^{2\pi} {d\phi \over 2\pi } \; \theta^2  \;\; - \;  <\theta>_0^2  \; = \; \pi -3 \cr 
\end{array} 
\ee
of directions $\vec{n}$ uniformly distributed on the unit semi-sphere. Thus  in the limit 
of a tensionless (i.e., $\beta E(q) = 0$)  fluctuating interface  the average 
defined in Eqs. (\ref{prob1})-(\ref{integration}) is not reduced to  
the 'random-direction'  average $<A>_0$ 
due to a geometric coherence induced by the manifold.

\begin{figure}[htbp]
\begin{center}
\includegraphics[width=0.65\linewidth]{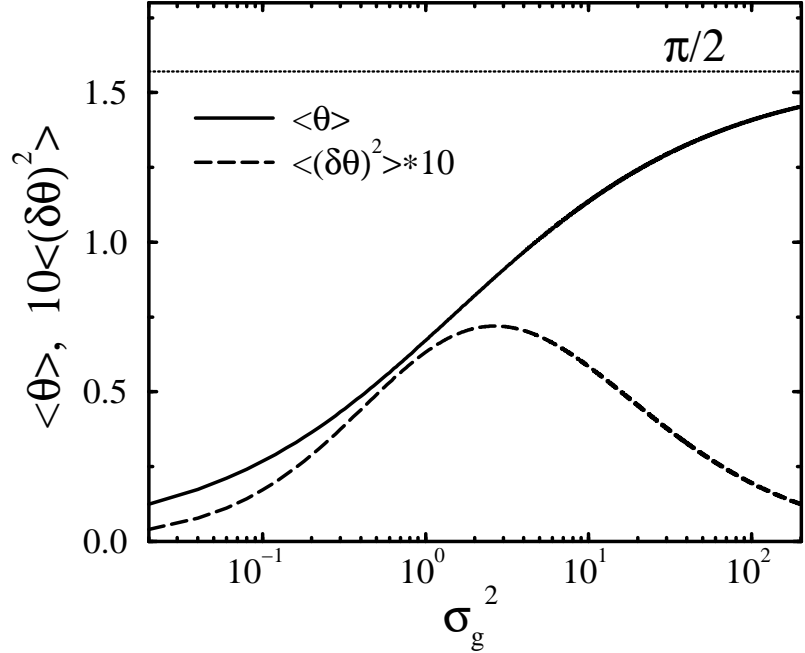}
\end{center}
\caption{Mean value $<\theta>$ and mean squared deviation $<(\delta\theta)^2>$ 
of the tilt angle between the interface normal $\vec{n}$ and the  axis $\vec{e}_z$ 
orthogonal to  the mean interface  (see Fig. \ref{interfacefig} and 
Eqs. (\protect\ref{tiltangle}), (\protect\ref{normal}), 
(\protect\ref{meanangle}), and (\protect\ref{varianceangle})) as function of the variance of the height 
fluctuations $\sigma_g^2=\sigma_3^2=
2\sigma'^2$ (see Eqs. (\protect\ref{prob8}), (\protect\ref{metric3}),  (\protect\ref{meanmetric}), and (\ref{angle})). 
The mean squared deviation obtains its maximum value $<(\delta\theta)^2>_{max}= 0.072$ for $\sigma_g^2=2.66$.}
\label{sigma_angle}
\end{figure}

\subsection{Infrared and ultraviolet limit} 
\label{infrared}

The fluctuation spectrum (see Eqs. (\ref{prob3}) and (\ref{variance})) 
\be{infrared1} 
\tilde{\sigma}^2(q)={k_BT\over E(q)} 
\ee
of the interface position $f(\vec{R})$ diverges in the limit of  
large wavelengths, $q\rightarrow 0$, and vanishing gravitation 
constant $G$, i.e., $E_0\rightarrow 0$ (see Eq. (\ref{gaussenergy})).  
This infrared divergence is a direct consequence 
of the occurrence of a so-called Goldstone mode with vanishing excitation energy (or effective mass). 
This mode exists because of  the spontaneous symmetry breaking of the 
translational 
invariance of the system due to the formation of an interface at $f(\vec{R})$ and due to the fact that the mean interface position can be shifted without any cost of free energy if $G=0$ and without pinning lateral boundary conditions. 

In contrast to $\tilde{\sigma}^2(q)$, the  fluctuation spectrum 
\be{infrared2}
\tilde{\sigma}^2_g(q) ={k_BT\over E(q)} q^2 \; 
\longrightarrow_{\!\!\!\!_{\!\!\!\!\!\!\!\!\!\! \!\!\! \!
E_0,q\rightarrow 0}}  \;  
{k_BT \over \gamma_0}  
\ee
of the metric ($\sigma_g=\sigma_3$; see Eqs. (\ref{prob8}) and (\ref{angle})), i.e., 
of the orientations and of the tilt angle, remains finite in the limit of a  
vanishing gravitational 
constant $G$ (see Eq. (\ref{gaussenergy})). It does not diverge 
because the Goldstone mode at $q=0$ is flat (corresponding to a rigid shift $f=const$) 
and does not contribute to an 
increase of the fluctuations of the local orientations, i. e., fluctuations of the  
tilt angle $\theta$. In other words, modes with wavevector $q$ contribute only proportional to 
$q^2$ to the 
distortion of a flat metric which finally suppress the spectrum in Eq. (\ref{infrared2})  
in the long-wavelength limit $q\rightarrow 0$, i.e., $\tilde{\sigma}_g^2(q \rightarrow 0)$ 
is a constant.

In contrast to the infrared case,  the ultraviolet limit $q\rightarrow \infty$ becomes more violent for 
local orientations as compared to the case of positional fluctuations 
of an interface. Heuristically one may argue 
that small-scale undulations (ripples) do not alter the position considerably but 
accentuate the more horizontal local  orientations  due 
to the many 'steep hills' or descents 
with numerous  sharply rising or falling slopes.  

As a consequence the local orientations are very sensitive to the dependence of surface energies 
at large wavevectors --  a regime which has been studied  recently by grazing 
incident X-ray scattering   \cite{nature,mora,li}.  
This kind of experiment provides only access to $\tilde{\sigma}^2(\vec{q})$ which 
makes it difficult to  resolve 
the details of $E(q)$ at large $q$. 
These features are enhanced by monitoring the fluctuations of the 
tilt angle which are governed by  
$\tilde{\sigma}^2_g(q) = q^2\tilde{\sigma}^2(q)$.

\subsection{Correlation function $<\theta(\vec{0})\theta(\vec{a})>$  of 
tilt angles}
\label{correlationfunction} 

The standard characterization of a fluid interface is given in terms of 
its surface tension, its mean position, the local height fluctuations, 
and the lateral height-height two-point correlation function. The latter 
one determines the diffuse scattering intensities of X-rays or neutrons 
from such an interface. In the present study we have put forward the 
mean tilt angle (and its variance) as an alternative characterization 
of a fluid (see Fig. \ref{sigma_angle}). 
The temperature dependence of the mean tilt angle might be considered to be 
on a par with that of the surface tension. This alternative description in 
terms of tilt angles reaches the same level of completeness as the aforementioned 
standard characterization by providing the correlation function 
$<\theta(\vec{0})\theta(\vec{a})>$ of tilt angles at different lateral positions $\vec{0}$ 
and $\vec{a}$. This goes beyond the probability distribution $P[\theta]$ 
considered so far. Based on this motivation and inspired by the eventual 
possibility to probe this correlation function also experimentally (see Ref. \cite{aarts}), 
in this subsection we determine the correlation function 
$<\theta(\vec{0})\theta(\vec{a})>$ and its dependence on the spectrum $E(q)$. 
It turns out that this correlation function probes also {\it non-local} properties 
of the interface.

To this end a family of joint probability distributions 
$P\left[\left\{f^{(n_i,m_i)}(\vec{a}_i)\right\}_{i=1}^N\right]$ has to be calculated, 
in particular  of the  first 
derivatives $\vec{f} = (f_x(\vec{0}), f_y(\vec{0}), f_x(\vec{a}), f_y(\vec{a}))$. From that 
the joint distribution $P[g_\vec{0},g_{\vec{a}}]$  of the metrics can be derived and finally    
$P[\theta(\vec{0}),\theta(\vec{a})]$ of the tilt angles at a finite distance.

The concept of orientational fluctuations is frequently used 
in the context of microemulsions, membranes, and macromolcules. There a persistence length 
is defined which characterizes  the correlation $<\vec{n}(\vec{0})\cdot\vec{n}(\vec{a})>$ of 
two normal vectors at finite but not too large distances \cite{degennes}. 
Although related the correlation  $<\theta(\vec{0})\theta(\vec{a})>
=<\arccos(\vec{n}(\vec{0})\cdot \vec{e}_z)\arccos(\vec{n}(\vec{a})\cdot\vec{e}_z)>$  is 
a more complex function. However, the analytic derivation of the 
distribution $P[\theta(\vec{0}),\theta(\vec{a})]$ allows us to determine the correlations 
exactly for arbitrary distances $\vec{a}$.

Following the same procedure as in Eqs. (\ref{prob2}) and (\ref{joint}) 
one finds the joint probability of the $N$ values $f_i:= f^{(n_i,m_i)}(\vec{a}_i)$ 
(compare Eq. (\ref{prob6})) 
at the lateral positions $\vec{a}_i$  ($i=1,\ldots,N$),  
\be{corr1} 
P\left[\left\{f^{(n_i,m_i)}(\vec{a}_i)\right\}_{i=1}^N\right] = 
{1\over \sqrt{(2\pi)^N {\rm det}({\bf \Sigma^2})}} 
{\rm exp}\left\{ - {1\over 2} \sum_{i,j=1}^N f_i \Sigma^{-2}_{ij}f_j  \right\} 
\ee
with the correlation matrix 
\be{corr2} 
\begin{array}{ll}
\Sigma^2_{ij}  & \displaystyle = 
\int {d^2q \over (2\pi)^2} {k_BT \over
E(q)} q_x^{n_i+n_j}q_y^{m_i+m_j} S_{n_i+m_i,n_j+m_j}(\vec{q}\cdot (\vec{a}_i-\vec{a}_j)) \;= \Sigma^2_{ji} \cr 
  &  \cr 
 & \displaystyle = \pm \left\{\begin{matrix} \sigma^2_{n_i+n_j,m_i+m_j}(\vec{a}_i-\vec{a}_j)\;, & n_i+n_j+m_i+m_j \; {\rm even} \;\;\;, \cr
 \tau^2_{n_i+n_j,m_i+m_j}(\vec{a}_i-\vec{a}_j) \;,   & n_i+n_j+m_i+m_j \; {\rm odd} \;\;\;,  \end{matrix}\right.  
\end{array} 
\ee
\be{corr5} 
S_{\nu,\mu}(x)   =  {(-1)^{\nu-\mu\over 2 }+(-1)^{\mu-\nu\over 2} \over 2} \cos x 
+   {(-1)^{\nu-\mu+1\over 2}-(-1)^{\mu-\nu+1 \over 2} \over 2} \sin x \;,  
\ee  
and the correlation functions   for $n+m$  even  
(compare  Eqs. (\ref{prob7}) and  (\ref{joint4})) 
\be{corr3} 
\begin{array}{ll} 
\displaystyle
\sigma_{n,m}^2(\vec{a}) & \displaystyle  = 
\int {d^2q \over (2\pi)^2} {k_BT \over
E(q)} q_x^{n}q_y^{m}\cos (\vec{q}\vec{a}) \;,\quad n+m \; {\rm even} \;, \cr
  & \cr 
   & \displaystyle  = (-1)^{n+m\over 2} {\partial^{n+m}  \over 
\partial a_x^{n}\partial a_y^{m}} \sigma^2(\vec{a})  \cr 
  &  \cr 
  & \displaystyle  =  \sum_{\underbrace{i,j=0}_{i+j \; {\rm even}}}^\infty {(-1)^{i+j\over 2} \over i!j! } I_{{n+i \over 2},{m+j\over 2}} \sigma^2_{n+m+i+j+1}  a_x^ia_y^j   
 \end{array} 
\ee    
as well as for $n+m$ odd 
\be{corr4} 
\begin{array}{ll} 
\displaystyle
\tau_{n,m}^2(\vec{a}) & \displaystyle  = 
\int {d^2q \over (2\pi)^2} {k_BT \over
E(q)} q_x^{n}q_y^{m}\sin (\vec{q}\vec{a}) \cr
  & \cr 
   & \displaystyle  = (-1)^{n+m+1\over 2} {\partial^{n+m}  
\over \partial a_x^{n} \partial a_y^{m}} \sigma^2(\vec{a})  \cr 
  &  \cr 
  & \displaystyle  =  \sum_{\underbrace{i,j=0}_{i+j \; {\rm odd}}}^\infty {(-1)^{i+j-1\over 2} \over i!j! } 
I_{{n+i \over 2},{m+j\over 2}} \sigma^2_{n+m+i+j+1}  a_x^ia_y^j \;\;. 
 \end{array} 
\ee 
For the calculation of the correlation function $<\theta(\vec{0})\theta(\vec{a})>$   
the joint probability distribution  is needed 
only for the vector $\vec{f} = (f^{(1,0)}(\vec{0}),f^{(0,1)}(\vec{0}),f^{(1,0)}(\vec{a}),f^{(0,1)}(\vec{a}))$ 
given by the  matrices 
\be{corr6} 
\Sigma^2_{ij} = \left(\begin{matrix} 
\sigma'^2 & 0  & \sigma_{2,0}^2(\vec{a})   & \sigma_{1,1}^2(\vec{a})  \cr   
0   &  \sigma'^2 & \sigma_{1,1}^2(\vec{a})   & \sigma_{0,2}^2(\vec{a})   \cr   
\sigma_{2,0}^2(\vec{a})   & \sigma_{1,1}^2(\vec{a})   & \sigma'^2  & 0  \cr   
\sigma_{1,1}^2(\vec{a})  &  \sigma_{0,2}^2(\vec{a})  & 0  & \sigma'^2  \cr   
\end{matrix} \right)  \quad {\rm and } \quad  \Sigma^{-2}_{ij} 
=  \left( \begin{matrix} A_{ij}  &  B_{ij} \cr B_{ij}  &  A_{ij} 
\end{matrix} \right)
\ee
with the matrix elements  
\begin{eqnarray} 
\label{corr8} 
A_{11} & = & {\sigma'^4\over {\rm det}(\Sigma^2)  } \left( \sigma'^4-\sigma_{0,2}^4(\vec{a})-\sigma_{1,1}^4(\vec{a}) \right)  \cr 
A_{12} & = & {\sigma'^4\over {\rm det}(\Sigma^2) } \left(\sigma_{1,1}^2(\vec{a}) (\sigma_{2,0}^2(\vec{a})+\sigma_{0,2}^2(\vec{a})) \right)   \cr 
A_{22} & = & {\sigma'^4\over {\rm det}(\Sigma^2)  } \left( \sigma'^4-\sigma_{2,0}^4(\vec{a})-\sigma_{1,1}^4(\vec{a})  \right) \cr 
B_{11} & = &  -{\sigma'^2\over {\rm det}(\Sigma^2)  } \left( \sigma_{1,1}^4(\vec{a})\sigma_{0,2}^2(\vec{a})+ \sigma_{2,0}^2(\vec{a})(\sigma'^4-\sigma_{0,2}^4(\vec{a})) \right)  \cr 
B_{12} & = &  -{\sigma'^2\over {\rm det}(\Sigma^2)  } \left( \sigma_{1,1}^2(\vec{a})\left(\sigma'^4  + \sigma_{2,0}^2(\vec{a})\sigma_{0,2}^2(\vec{a})-\sigma_{1,1}^4(\vec{a})\right) \right)  \cr 
B_{22} & = &  -{\sigma'^2\over {\rm det}(\Sigma^2)  } \left( \sigma_{1,1}^4(\vec{a})\sigma_{2,0}^2(\vec{a})+ \sigma_{0,2}^2(\vec{a})(\sigma'^4-\sigma_{2,0}^4(\vec{a}))   \right)  
\end{eqnarray}
and the determinant   
\be{corr10} 
\begin{array}{ll} 
\displaystyle 
{\rm det}(\Sigma^2) & \displaystyle  = \sigma'^8 -  \sigma'^4(\sigma_{2,0}^4(\vec{a})+\sigma_{0,2}^4(\vec{a})) 
-2\sigma_{1,1}^4(\vec{a}) (\sigma'^4+\sigma_{2,0}^2(\vec{a})\sigma_{0,2}^2(\vec{a})) \cr 
 & \displaystyle \quad \quad \;\; + \sigma_{1,1}^8(\vec{a})
+ \sigma_{2,0}^4(\vec{a})\sigma_{0,2}^4(\vec{a}) \;\;. 
\end{array} 
\ee
Repeating the derivation presented in Subsec. \ref{metricdist}  one finds the following 
explicit expression 
for the joint probability distribution 
\be{corr11} 
\begin{array}{ll} 
\displaystyle 
P[\theta_{\vec{0}},\theta_{\vec{a}}] &\displaystyle   = {4 \sin \theta_{\vec{0}} \sin \theta_{\vec{a}} \over \cos^3 \theta_{\vec{0}}  \cos^3 \theta_{\vec{a}} } {1\over  \sqrt{{\rm det}(\Sigma^2)}}
 \int\!\!\!\!\int\limits_{\!\!0}^{\!\!\pi} {d\phi  d\phi'  \over 4\pi^2}  \; 
e^{-{1 \over 2}\left( \tan\theta_{\vec{0}},\tan\theta_{\vec{a}}\right) {\bf C}(\phi,\phi') \left( \tan\theta_{\vec{0}},\tan\theta_{\vec{a}}\right)^t   }
\end{array} 
\ee 
of the tilt angles 
at different positions of the random fluctuating interface. The elements  
of the matrix ${\bf C}(\phi,\phi')$ are given by 
\be{corr12} 
\begin{array}{ll} 
\displaystyle 
C_{11}(\phi) & \displaystyle   = A_{11}\cos^2\phi +A_{22}\sin^2\phi+2A_{12}\cos\phi\sin\phi  \cr  
\displaystyle 
C_{22}(\phi') & \displaystyle   = A_{11}\cos^2\phi'+A_{22}\sin^2\phi'+2A_{12}\cos\phi'\sin\phi'  \cr 
\displaystyle 
C_{12}(\phi,\phi') & \displaystyle   = C_{21} =  B_{11}\cos\phi\cos\phi' + B_{22}\sin\phi\sin\phi' 
+B_{12}(\cos\phi\sin\phi'+\sin\phi\cos\phi' ).    
\end{array} 
\ee
\begin{figure}[htbp]
\begin{center}
\includegraphics[width=0.45\linewidth]{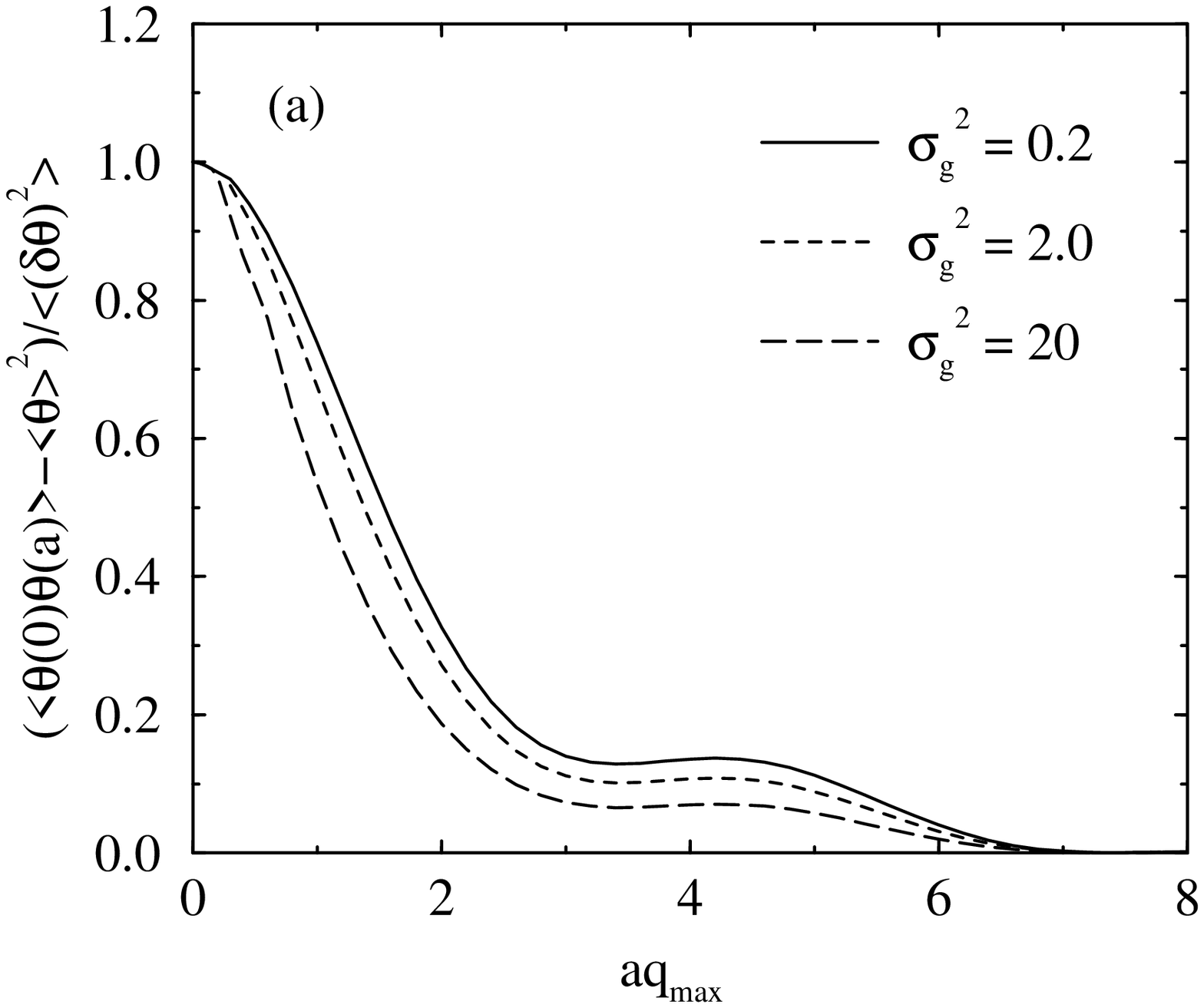}
\includegraphics[width=0.45\linewidth]{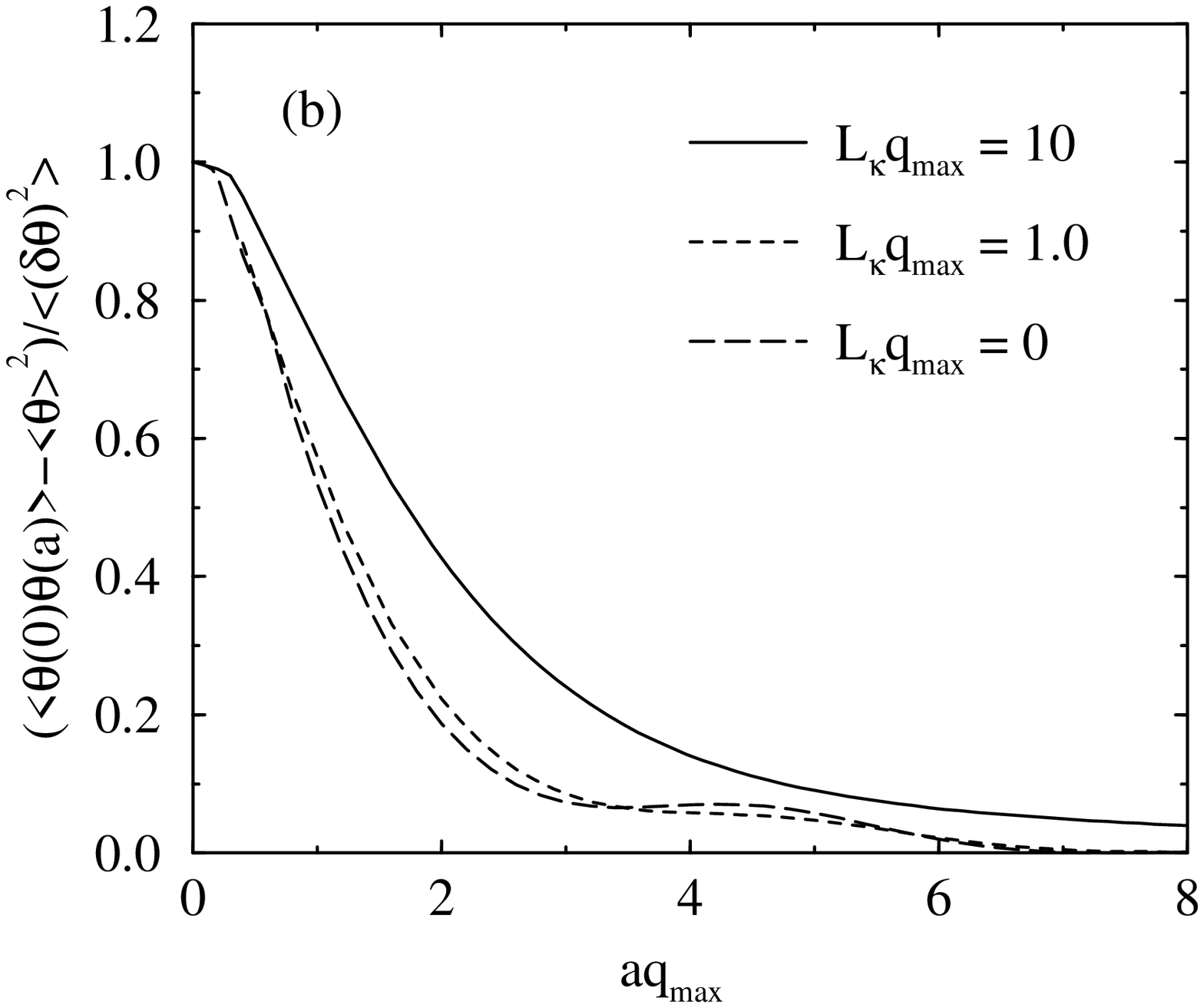}
\end{center}
\caption{The correlation function $<\theta(\vec{0})\theta(\vec{a})>$ for 
the tilt angles $\theta$ 
depends effectively only on the 
variance $\sigma_g^2$ of the metric  and on the 
functional form of the energy function 
$\bar{E}(y)=1+L_c^2q^2_{max}y^2(1+\bar{\kappa}(y)y^2)$, i.e.,  
on the wavevector-dependent  
bending rigiditiy $\bar{\kappa}(y)$. 
Here, we have chosen (a) $\bar{\kappa}(y)=0$,  for variances 
$\sigma_g^2=0.2$, $2.0$, and $20$ as well as  (b) $\sigma_g^2=20$ 
and bending rigidities $\bar{\kappa}(y)=1+ (L_{\kappa}q_{max})^2y^2$  
with $L_{\kappa}q_{max}=0$, $1$, and $10$ 
(see Eq. (\protect\ref{skg})). 
The dependence on  the ratio $L_cq_{max}$ of the 
cut-off lengths is small, so that for physical relevant  
values $L_cq_{max}>10$ no significant change is observed.  
Here, we have chosen  $L_cq_{max}=10$.  The maximum at 
$aq_{max}\approx$ does not reflect a minimum in the 
surface tension $\gamma(q)$ but seems to be  related to the 
sharp cut-off $q_{max}$ for  capillary waves. 
}
\label{figanglecorr}
\end{figure}

Finally, the  correlation function 
of the tilt angles $\theta(\vec{0})$ and $\theta(\vec{a})$ reads 
\be{corr13} 
\begin{array}{ll} 
\displaystyle 
<\theta(\vec{0})\theta(\vec{a})> & \displaystyle  = \int\!\!\!\!\int\limits_{\!\!0}^{\!\!\pi/2} 
d\theta(\vec{0})  d\theta(\vec{a}) \; \theta(\vec{0})\theta(\vec{a}) \; 
P[\theta_{\vec{0}},\theta_{\vec{a}}] \cr 
  &  \cr 
  &  \displaystyle =  \int\!\!\!\!\int\limits_{\!\!0}^{\!\!\pi} 
{d\phi  d\phi'  \over \pi^2} \int\!\!\!\!\int\limits_0^\infty 
 { dx_1 dx_2 x_1x_2{\rm arctan}(x_1) {\rm arctan}(x_2)\over \sqrt{{\rm det}(\Sigma^2)} }
  e^{-{1\over 2} \vec{x}^t{\bf C}(\phi,\phi')\vec{x} } \;,   
\end{array} 
\ee
which is shown in Fig. \ref{figanglecorr}(a) for  the phenomenological capillary wave theory 
with $\gamma(q)=\gamma_0$ and 
in Fig. \ref{figanglecorr}(b) for finite bending rigidity $\kappa_0$ 
(see Eq. (\ref{sigma})). In general, one finds that the centered and 
normalized  correlation function 
$(<\theta(\vec{0})\theta(\vec{a})-<\theta>^2)/<(\delta\theta)^2>$ 
does not depend sensitively on the particular functional form of the 
surface tension $\gamma(q)$. 
In contrast to the one-point probability distribution $P[\theta]$ given in Eq. (\ref{angle}),
there is only a double-integral representation for the two-point probability distribution 
$P[\theta_0,\theta_{\vec{a}}]$ so that  
the determination of the correlation function $<\theta(\vec{0})\theta(\vec{a})> $  involves   
a four-dimensional  numerical integration, with the integrand depending parametrically   
on the distance $\vec{a}$ and functionally on the momentum dependent energy $E(q)$. 
In order to be able to gain further insight into the expression given by Eq. (\ref{corr13}) we  
 study in the following section the dependence of the relevant variances  
$\sigma^2_{n,m}(\vec{a})$ (Eq. (\ref{corr3}))  
on the physical parameters entering into $E(q)$.

\section{Dependence on physical parameters} 
\label{results} 

Within the present Gaussian approach the observables $<\theta>$ and $<(\delta\theta)^2>$ 
are determined by the variance $\sigma_g$ (see Fig. \ref{sigma_angle}). In this section we study the 
dependence of $\sigma_g^2=\sigma^2_{k=3}$ on  the physical parameters introduced in 
Sect. \ref{capillary}. 
Since we do not consider systems with finite lateral extensions we set 
$q_{min} = 0$ and keep $G>0$. Expressing the momentum dependent surface 
tension $\gamma(\vec{q})$ in terms of a momentum dependent bending 
rigidity (see Eqs. (\ref{gaussenergy}),  (\ref{sigmass}), and  (\ref{finalsigma})), 
\be{bending} 
\gamma(q) = \gamma_0 +  \kappa(q) q^2  \;\;, 
\ee
one has in general 
\be{sigma_n} 
\sigma_{k}^2  = {k_BT \over 2\pi  \gamma_0} q_{max}^{k-1} \; S_k(L_cq_{max};[\kappa(q)]) 
\ee
with the normalized and dimensionless variances  
\be{sn} 
S_k(x;[\bar{\kappa}(q)]) =  \int\limits_0^1 dy {y^k \over x^{-2} + y^2 + y^4\bar{\kappa}(y)} 
\ee
where $\bar{\kappa}(y)=q_{max}^2 \kappa(q=q_{max}y)/\gamma_0$ is dimensionless and  
\be{sn2}
S_{k>1}(x\rightarrow \infty;[\bar{\kappa}=0]) = {1\over k-1}  \;\;. 
\ee

\subsection{Phenomenological capillary wave theory}
\label{phenomenological} 

The phenomenological capillary wave theory is defined by Eqs. (\ref{hamiltonian})-(\ref{sigma}) 
with $\kappa_0=const$. For the special case $\kappa_0=0$ the  
 normalized  variances $S_k(x,\bar{\kappa})$ (Eqs. (\ref{sigma_n}) and (\ref{sn})) for $k=1$, $2$, and $3$ 
read:   
\be{snk0} 
\begin{array}{ll} 
S_1(x;0) & = {1\over 2} \ln (1+x^2) \;, \cr 
  &  \cr 
S_2(x;0) & = 1 - {\arctan x \over x }  \;, \cr 
  &  \cr 
S_3(x;0) & = {1\over 2} \left(1- {\ln (1+x^2) \over x^2}\right)  \;. \cr 
\end{array} 
\ee
For large capillary lengths $x=q_{max}L_c>>1$ (Eq. (\ref{caplength})), which corresponds 
to the typical physical situation,   
one has (3.194.5 in Ref. \cite{gradshteyn}) 
\be{skg1} 
T_{k>1}(\bar{\kappa}_0) \equiv 
S_{k>1}(x=\infty,\bar{\kappa}_0) = {1\over k-1} \;  {}_{2}F_{1}(1,{k-1\over 2};{k+1 \over 2};-\bar{\kappa}_0)
\ee
where (Eq. (\ref{caplength}))  
\be{skg} 
\bar{\kappa}_0 = q_{max}^2 {\kappa_0 \over \gamma_0} = (q_{max}L_{\kappa})^2\;. 
\ee
For $k=2$ and $k=3$ one has 
\be{skg2} 
T_{2}(\bar{\kappa}_0) = S_2(x=\infty,\bar{\kappa}_0) = {1\over \sqrt{\bar{\kappa}_0}} \arctan \sqrt{\bar{\kappa}_0} 
\ee
and 
\be{skg3} 
T_{3}(\bar{\kappa}_0)  = S_3(x=\infty,\bar{\kappa}_0) = {1\over 2\bar{\kappa}_0} \ln (1+\bar{\kappa}_0)\;, 
\ee
respectively. For $k=1$, $S_k$ diverges in the limit $L_c\rightarrow\infty$: 
\be{sn3}
\begin{array}{ll} 
\displaystyle
S_1(x;\bar{\kappa}_0 )   & \displaystyle = {1\over 2\sqrt{\Delta}} \ln {2\bar{\kappa}_0 (1+\sqrt{\Delta}) +1 - \Delta \over 2 \bar{\kappa}_0(1-\sqrt{\Delta}) + 1-\Delta} \;\;,\quad 
 \Delta= 1-{4\bar{\kappa}_0 \over x^2}    \;, \cr 
  &  \cr 
  & \displaystyle 
\longrightarrow_{\!\!\!\!_{\!\!\!\!\!\!\!\!\!\! 
x \rightarrow \infty}}   \ln x - {1\over 2} \ln(1+\bar{\kappa}_0) + {\cal O}(x^{-2}) \cr
  &  \cr 
  & \displaystyle \quad\quad  = \ln {L_c\over L_\kappa} - {1\over 2} \ln \left(1+{\gamma_0 \over \kappa_0q^2_{max}}\right) 
+ {\cal O}\left(\left({L_{\kappa}\over L_c}\right)^2\right)
\end{array} 
\ee
with $\Delta>0$, i.e., $L_\kappa<{1\over 2} L_c$. 
The function $S_1(x,\bar{\kappa}_0)$ is shown in Fig. \ref{figs1}.

\begin{figure}[htbp]
\begin{center}
\includegraphics[width=0.65\linewidth]{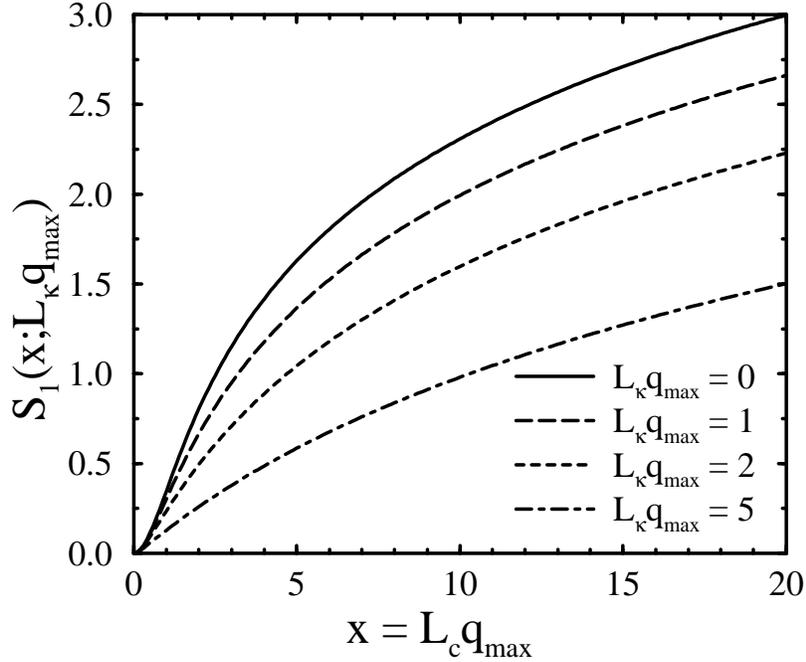}
\end{center}
\caption{Dependence of the normalized variance $\sigma^2/(k_BT/(2\pi \gamma_0)) 
= S_1(x;\bar{\kappa}_0=(L_\kappa q_{max})^2)$ 
(Eqs. (\ref{prob3}), (\ref{sn3}), and (\ref{prob3s})) on 
the capillary length $x=L_cq_{max}$ in units of the minimum wavelength  $2\pi/q_{max}$.  }
\label{figs1}
\end{figure}

The functions $S_1(x;\bar{\kappa}_0)$ and $S_3(x;\bar{\kappa}_0)$ determine the 
variances $\sigma^2=\sigma_1^2$ and $\sigma_g^2=\sigma_3^2=2\sigma'^2$, respectively: 
\be{prob3s} 
\sigma^2 = {k_BT \over 2\pi \gamma_0} S_1(L_cq_{max};\bar{\kappa}_0) 
\ee
and 
\be{probsig} 
\sigma_g^2 ={k_BT \over 2\pi \gamma_0} q^2_{max} S_3(L_cq_{max};\bar{\kappa}_0)  
\ee
with 
\be{sigma_s3} 
S_3(x;\bar{\kappa}_0) = {1\over 2\bar{\kappa}_0} \left({1\over 2} \ln \left[1+(1+\bar{\kappa}_0)x^2\right]
  -S_1(x;\bar{\kappa}_0)\right)  \;\;. 
\ee
The variance $\sigma_g^2$ characterizes the distribution function of the tilt angle $\theta$ 
(see Eq. (\ref{angle}) 
and Figs. \ref{fig_verteilung} and \ref{sigma_angle}) and is shown in Fig. \ref{figsigma}. 
\begin{figure}[htbp]
\begin{center}
\includegraphics[width=0.65\linewidth]{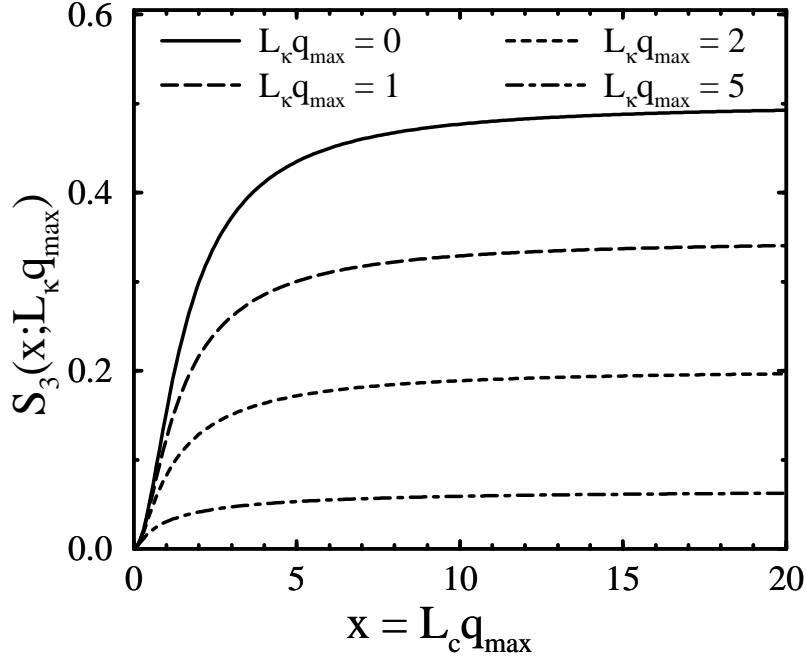}
\end{center}
\caption{The dependence of the normalized variance  
$\sigma^2_g/(k_BTq^2_{max}/(2\pi\gamma_0))= S_3(L_cq_{max};\bar{\kappa}_0)$  on 
the capillary length $L_c=\sqrt{\gamma_0/E_0}$ (Eqs. (\ref{angle}), (\ref{probsig}), 
and (\ref{sigma_s3})) in units of  
the minimum  
wavelength $2\pi/q_{max}$ of the capillary waves for various values of the length $L_\kappa$ 
associated with the bending rigidity $\bar{\kappa}_0=(L_\kappa q_{max})^2$ of the 
phenomenological theory. $S_3$ remains 
finite in the limit $L_c\rightarrow\infty$ even for $\bar{\kappa}_0=0$. The variance 
$\sigma_g^2$ 
characterizes the distribution of the tilt angle (Eq. (\ref{angle}) 
and Figs. \ref{fig_verteilung} and \ref{sigma_angle}).  }
\label{figsigma}
\end{figure}

\subsection{Microscopic capillary wave theory} 

In the previous subsection we have discussed various variances as they are predicted 
by the phenomenological capillary wave theory which is characterized by a constant 
bending rigidity $\kappa_0$ and which leads to a monotonically increasing wave length 
depending surface tension $\gamma(q)=\gamma_0+\kappa_0q^2$

The actual microscopic capillary wave theory is defined by Eqs. (\ref{gaussenergy}), (\ref{sigmass}), 
and (\ref{finalsigma}) and enters into the variances $S_k(x;[\bar{\kappa}(q)])$ 
via $\kappa(q)=(\gamma(q)-\gamma_0q^2)/q^{2}\neq const$. Here $\gamma(q)$ has a 
nontrivial structure exhibiting a pronounced minimum. In the following we study the 
influence of this nontrivial structure of $\gamma(q)$ (or $\kappa(q)\neq const$) on the 
variances discussed in Subsec. \ref{phenomenological}  and compare this with the predictions 
of the phenomenological capillary wave theory.

\begin{figure}[htbp]
\begin{center}
\includegraphics[width=0.45\linewidth]{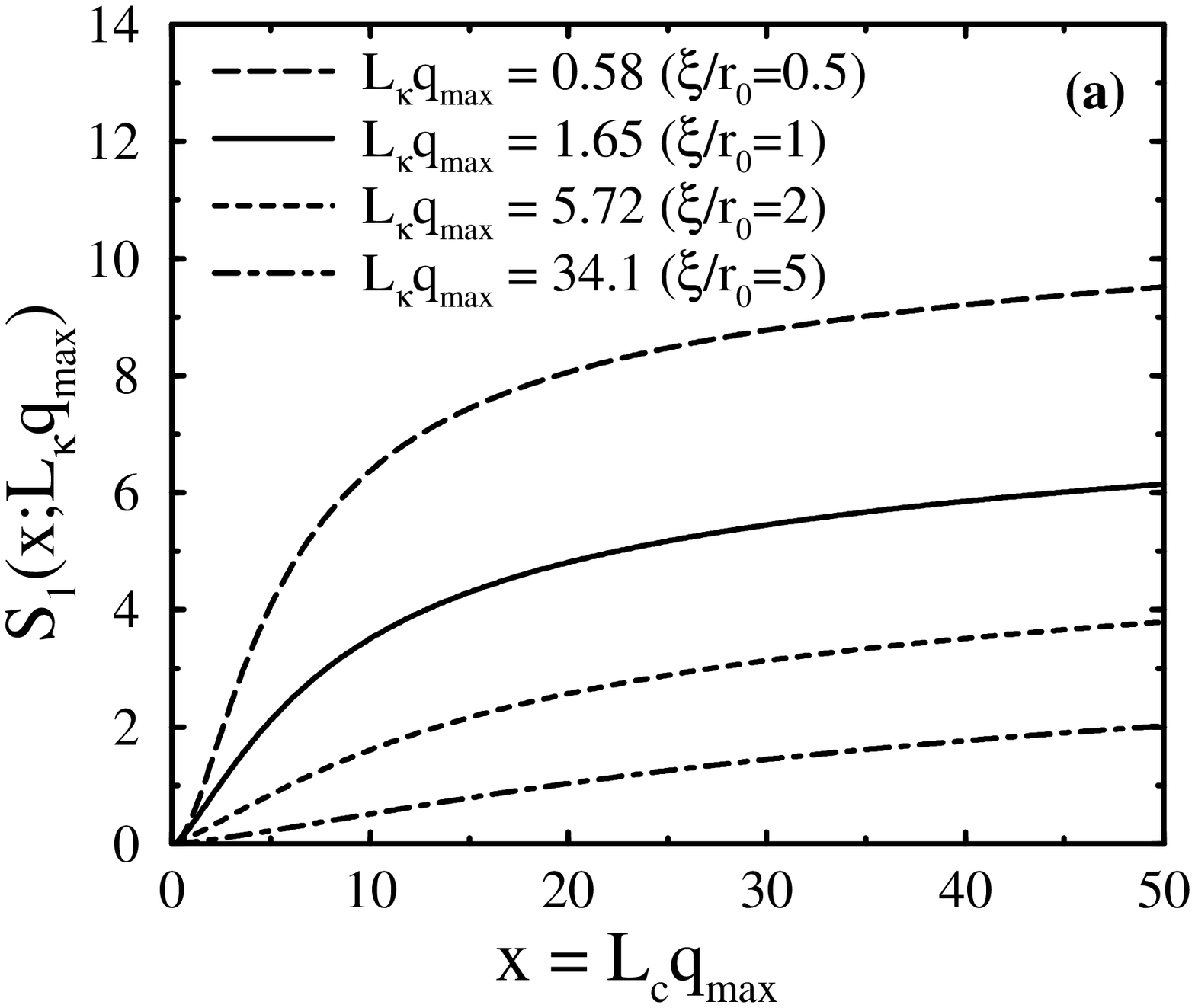}
\includegraphics[width=0.45\linewidth]{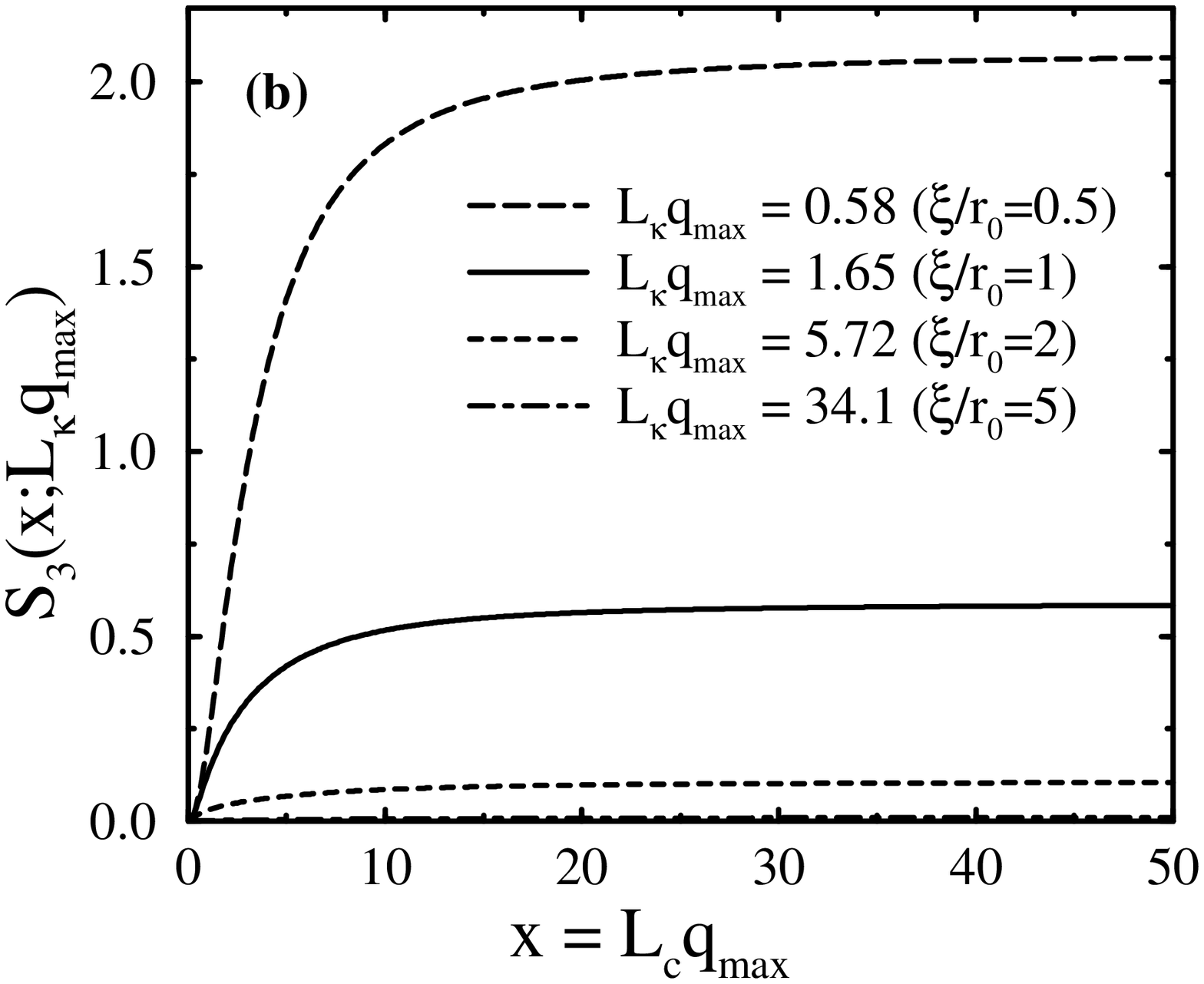}
\end{center}
\caption{
Normalized variances $S_1(x=L_cq_{max};L_\kappa q_{max}=\sqrt{\bar{\kappa}_0}$) (a) and $S_3(x;L_\kappa q_{max})$ (b) corresponding to the variances $\sigma^2$ and $\sigma_g^2$, respectively, using the microscopic expression for $\gamma(q)$ (Eq. (\ref{finalsigma})) with 
the choices $C_H=0.25$ and $r_0=2\pi/q_{max}$; 
$\bar{\kappa}_0\approx 1.82\left({\xi^4\over r_0^4}+{1\over 2} {\xi^2\over r_0^2}\right)$. These variances are larger than their phenomenological counterparts in Figs. \ref{figs1} and \ref{figsigma}. 
}
\label{figsigma_s1_micro_expression}
\end{figure}

In Fig. \ref{figsigma_s1_micro_expression} we show the normalized variances $S_1$ and $S_3$ corresponding to 
$\sigma_1^2=\sigma^2$ and $\sigma_3^2=\sigma_g^2$, repectively, as function 
of $x=L_cq_{max}$ (see Eqs. (\ref{bending})-(\ref{sn})). As in the phenomenological case 
(Figs. \ref{figs1} and \ref{figsigma}) $S_1$ diverges $\sim \ln x$ for $x\rightarrow \infty$ 
whereas $S_3$ reaches a finite limit $T_3=S_3(x=\infty)$ for $L_c\rightarrow \infty$. 
For Fig. \ref{figsigma_s1_micro_expression} the underlying $\gamma(q)$ is given 
by Eq. (\ref{finalsigma}) which is characterized 
by the bulk correlation $\xi$, the molecular diameter $r_0$, and the dimensionless 
parameter $C_H$; the forms of the functions $S_k$ are independent 
of $\gamma_0=\gamma(q=0)$ which enters only via $L_c$ (Eq. (\ref{caplength})). 
Here and in the following we choose $C_H=0.25$ which has turned out to 
describe properly available experimental data of $\gamma(q)$ \cite{nature,mora}. 
With this choice $S_k$ depends on $x=L_cq_{max}$, $r_0q_{max}$, $\xi q_{max}$, and 
$\xi/r_0$ (Eqs. (\ref{finalsigma}) and (\ref{sn})). In Fig. \ref{figsigma_s1_micro_expression} we choose $r_0=2\pi/q_{max}$ so that, 
with $\xi q_{max} = 2\pi \xi/r_0$, $S_k(x)$ depends parametrically only on $\xi/r_0$. 
By using Eq. (\ref{surfacetens}) this parametric dependence on $\xi/r_0$ can be expressed in 
terms of $\bar{\kappa}_0=(2\pi /r_0)^2 \kappa_0/\gamma_0 = (L_\kappa q_{max})^2$ 
(see Eq. (\ref{skg})) with $\bar{\kappa}_0 = {12\pi^2 \over 65} ((\xi/r_0)^4+ {1\over 2} (\xi/r_0)^2)$ 
which facilitates the direct comparision with Figs. \ref{figs1} and \ref{figsigma}. One finds similar functional 
forms but the variances as determined from the microscopic capillary wave theory are larger than their phenomenological counterparts. This increase of the variances indicates an increase of interfacial fluctuations due to the decrease of the actual surface tension $\gamma(q)$ for small wavelengths.

In Fig. \ref{figsigma_micro} we show the dependence of the limiting value $T_3=S_3(x=\infty)$ 
on various microscopic quantities. As stated above $T_3$ depends on $r_0q_{max}$, $\xi q_{max}$, 
and $\xi/r_0$ where only two of these parameters are independent so that 
Figs. \ref{figsigma_micro}(a) and \ref{figsigma_micro}(b) exhaust the available parameter space. 
Note, that Fig. \ref{figsigma_s1_micro_expression} explores the $x$ dependence within the 
parameter subspace $r_0q_{max}=2\pi$. Figure \ref{figsigma_micro}(c) shows the dependence 
on $L_\kappa q_{max}$ as introduced in Figs. \ref{figs1}, \ref{figsigma}, and 
\ref{figsigma_s1_micro_expression}. Using Eq. (\ref{surfacetens}) with $C_H=1/4$ one 
has $\kappa_0 = {3\over 65} \gamma_0 r_0^2 \left( {\xi^4\over r_0^4} 
+ {1\over 2} {\xi^2\over r_0^2} \right)$ so that with Eq. (\ref{caplength}) $L_\kappa q_{max} 
= \sqrt{3\over 65} r_0 q_{max} {\xi^2\over r_0^2} \sqrt{1+ {1\over 2} {r_0^2\over \xi^2}}$.  
Figure \ref{figsigma_micro} demonstrates that for $L_c\rightarrow \infty$ the variance for the tilt angle 
\be{variancetilt} 
\sigma_g^2 = {k_BT \over 2\pi \gamma_0} q_{max}^2 T_3\; , \quad L_cq_{max}=\infty \;, 
\ee 
as determined from the microscopic capillary wave theory deviates strongly from the predictions of the simple capillary wave theory ($\gamma(q)=\gamma_0$) and of the phenomenological capillary wave theory ($\gamma(q) =\gamma_0 + \kappa_0 q^2$ with $\kappa_0$ as given above). Moreover, Fig. \ref{figsigma_micro} shows that $\sigma_g^2$ depends strongly on the microscopic parameters $\xi$, $r_0$, and $q_{max}$. This means that the tilt angle distribution is a sensitive probe of the large-q behavior of $\gamma(q)$ and of molecular details.

\begin{figure}[htbp]
\begin{center}
\includegraphics[width=0.45\linewidth]{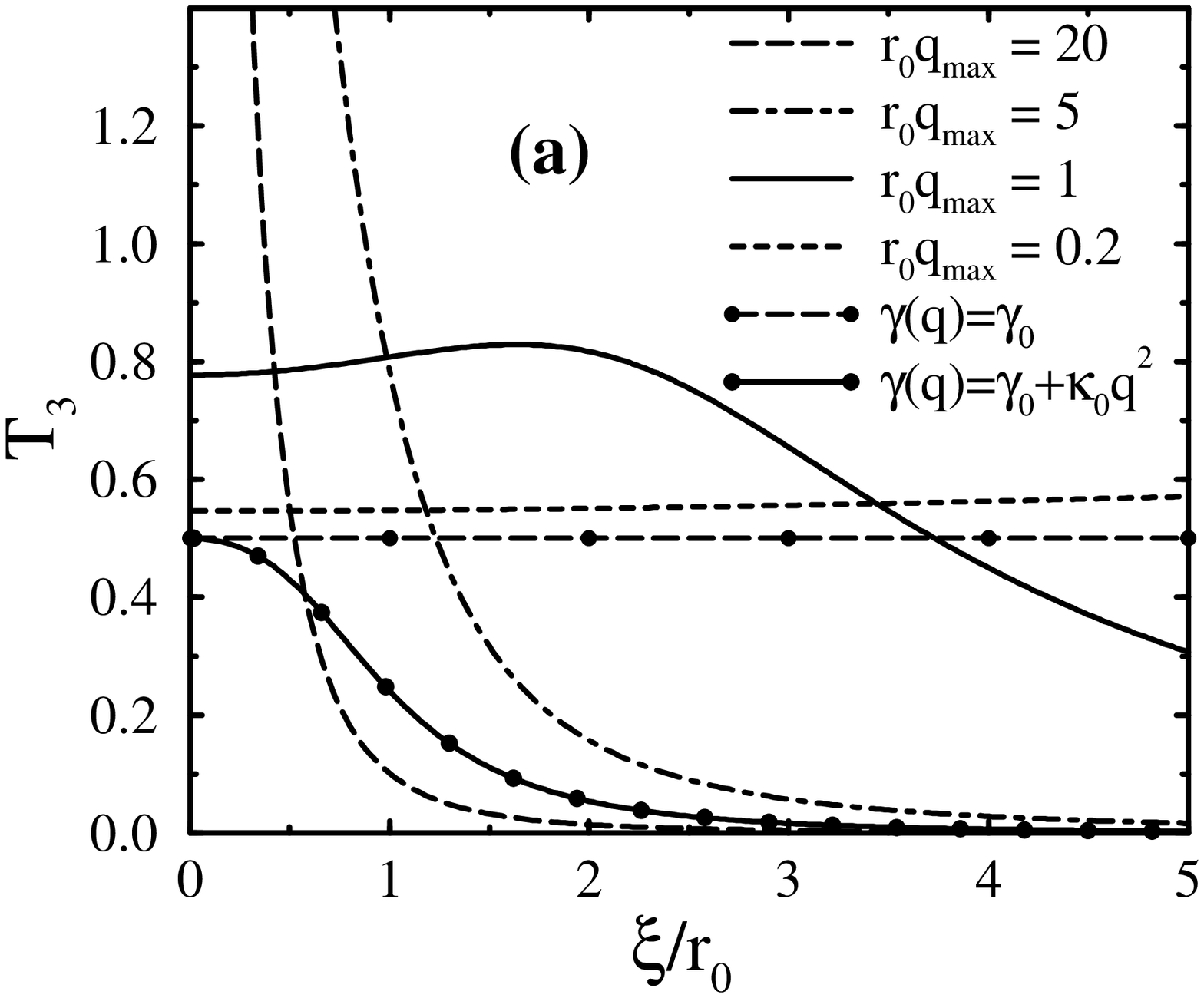}
\includegraphics[width=0.45\linewidth]{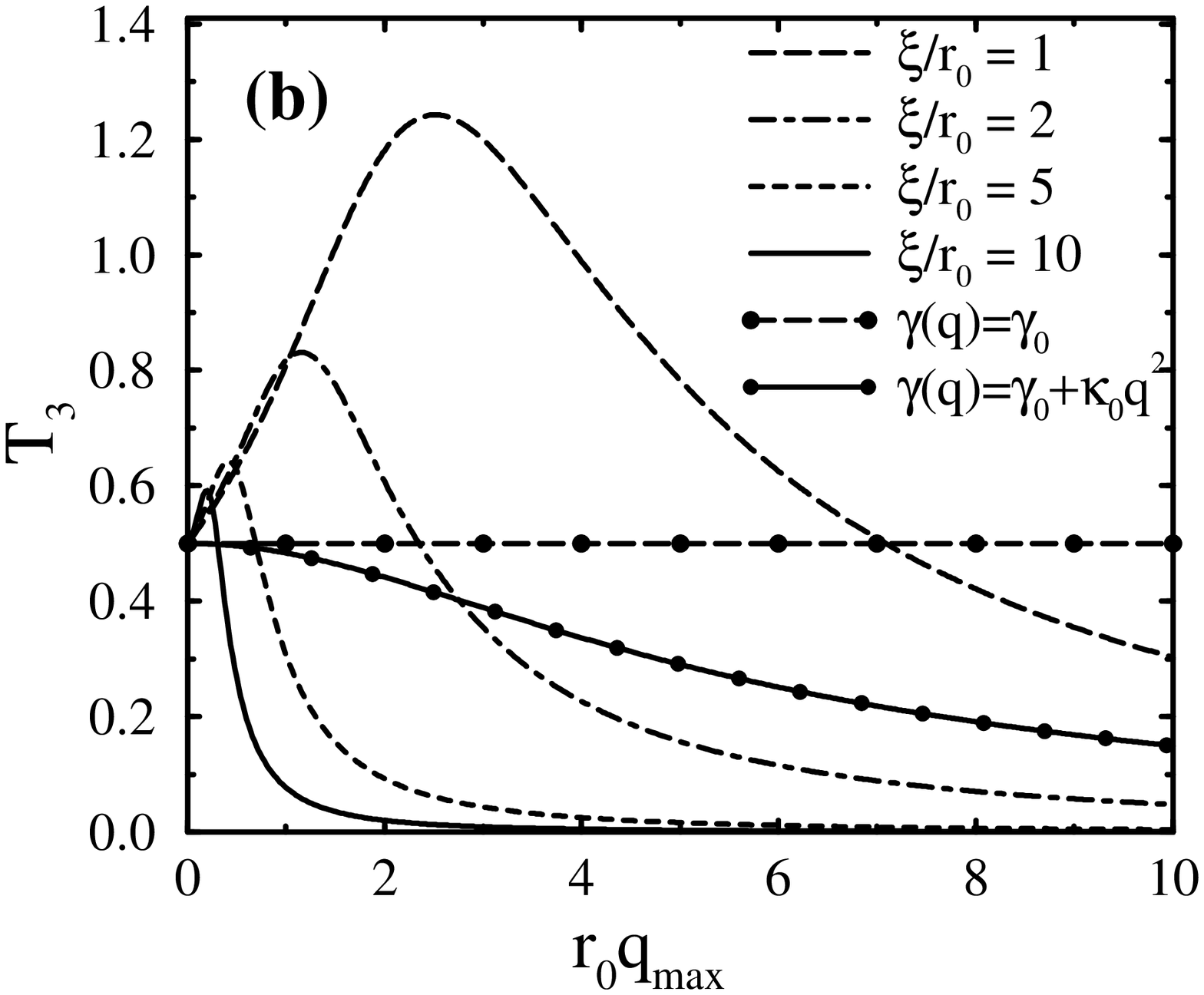}
\includegraphics[width=0.45\linewidth]{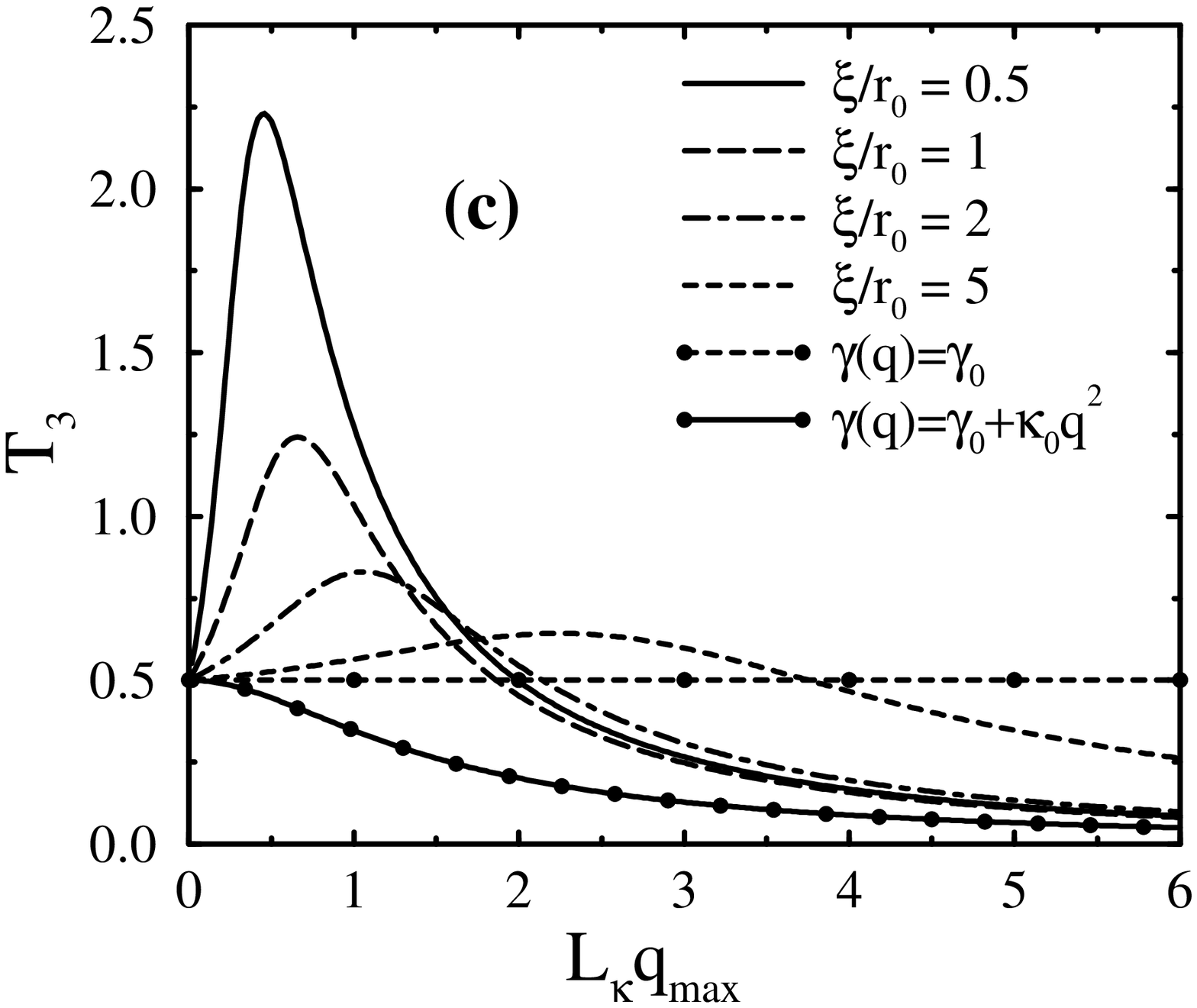}
\end{center}
\caption{
Dependence of the normalized variance $T_3=S(x=L_cq_{max}=\infty)$ proportional to 
the variance $\sigma_g^2$ of the tilt angle (Eq. (\ref{variancetilt})) on $q_{max}$ and on 
the bulk correlation length $\xi$ (a) and the molecular diameter $r_0$ (b) entering the 
microscopic capillary wave theory (Eq. (\ref{finalsigma})); $C_H={1\over 4}$. Comparision 
is made with the predictions of the simple capillary wave theory ($\gamma(q)=\gamma_0$) 
and the phenomenological capillary wave theory ($\gamma(q)=\gamma_0+\kappa_0q^2$ 
with $\kappa_0={3\over 65}\gamma_0r_0^2\left({\xi^4\over r_0^4}
+{1\over 2}{\xi^2\over r_0^2}\right)$ (Eq. (\ref{surfacetens})); for simplicity 
the phenomenological capillary wave theory is shown only for $r_0q_{max}=5$ in (a) and 
for $\xi/r_0=1$ in (b) since curves for other values of $r_0$ and $\xi$ can be obtained by 
rescaling the abscissa  accordingly. 
Figure \ref{figsigma_s1_micro_expression} 
corresponds to the parameter subspace $r_0q_{max}=2\pi$. 
(c) Plotting $T_3$ as function of $L_\kappa q_{max} = \sqrt{3\over 65}r_0q_{max}{\xi^2\over r_0^2}
\sqrt{1+{1\over 2}{r_0^2\over \xi^2}}$ instead of $\xi/r_0$ or $r_0q_{max}$ allows a comparision of the microscopic theory with the  phenomenological capillary wave theory for all bending 
rigidities $\kappa_0$ given by a single curve (circled solid line).  
}
\label{figsigma_micro}
\end{figure}

We note that for large correlation lengths  the dependence of $T_3=S_3(x=\infty)$ 
on $\xi/r_0$ differs qualitatively for the three theories under discussion, because 
of different functional forms of the bending rigidity function $\bar{\kappa}(y)$ 
introduced in Eq. (\ref{sn}): whereas $\bar{\kappa}(y)=0$ and $T_3(\xi)=1/2$ for 
the simple theory with $\gamma(q) =\gamma_0$, a constant bending rigidity 
$\bar{\kappa}(y)=\bar{\kappa}_0$ leads to 
$T_3(\xi)={1\over 2\bar{\kappa}_0} \ln(1+\bar{\kappa}_0)$ (see Eq. (\ref{skg3})) with 
$\bar{\kappa}_0 = {3\over 65} (r_0q_{max})^2\left({\xi^4\over r_0^4}+{1\over 2} {\xi^2\over r_0^2}\right)$ and therefore to the scaling  behavior $T_3(\xi>>r_0) \sim \xi^{-4}\ln(\xi/r_0)$ for 
the phenomenological capillary wave theory with $\gamma(q) =\gamma_0 + \kappa_0 q^2$. 
However, for the full microscopic theory given by Eq. (\ref{finalsigma}) one finds for 
large correlation lengths  
\be{bendingfunction} 
\bar{\kappa}(y) \; \;\; 
\longrightarrow_{\!\!\!\!_{\!\!\!\!\!\!\!\!\!\! \!\!  
\xi \rightarrow \infty }}  \; \;\;   0.74 \; C_H^2 {\xi^4\over r_0^4} (r_0q_{max})^2
\left(1-(1+r_0q_{max}y)e^{-r_0q_{max}y}\right)  \;, 
\ee
i.e., $\bar{\kappa}(y) \simeq \bar{c}_0 y^2$ for ${1\over \xi q_{max}} \lesssim  y \lesssim
 {1\over r_0 q_{max}}$ with $\bar{c}_0 = 0.37 C_H^2 (\xi q_{max})^4$. Thus 
$T_3=\int_0^1 {y\; dy \over 1+y^4\bar{c}_0}={1\over 2\sqrt{\bar{c}_0}} \arctan \sqrt{\bar{c}_0}$ 
exhibits  
the scaling behavior $T_3(\xi\rightarrow \infty ) \sim \xi^{-2}$ for the  microscopic theory. 
These differences are important for the temperature dependence $T\rightarrow T_c$ 
which will be discussed in the following section.

\begin{figure}[htbp]
\begin{center}
\includegraphics[width=0.45\linewidth]{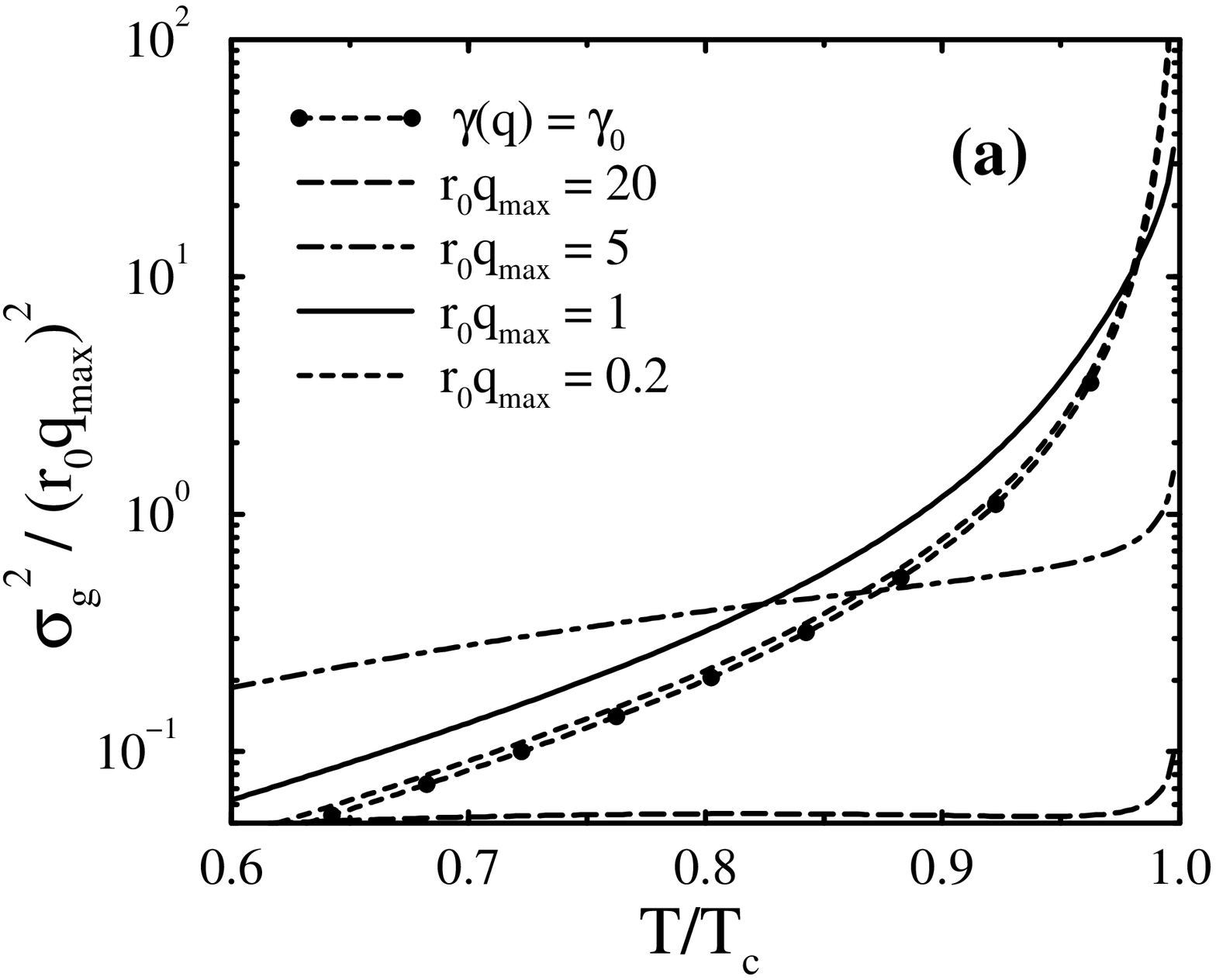}
\includegraphics[width=0.45\linewidth]{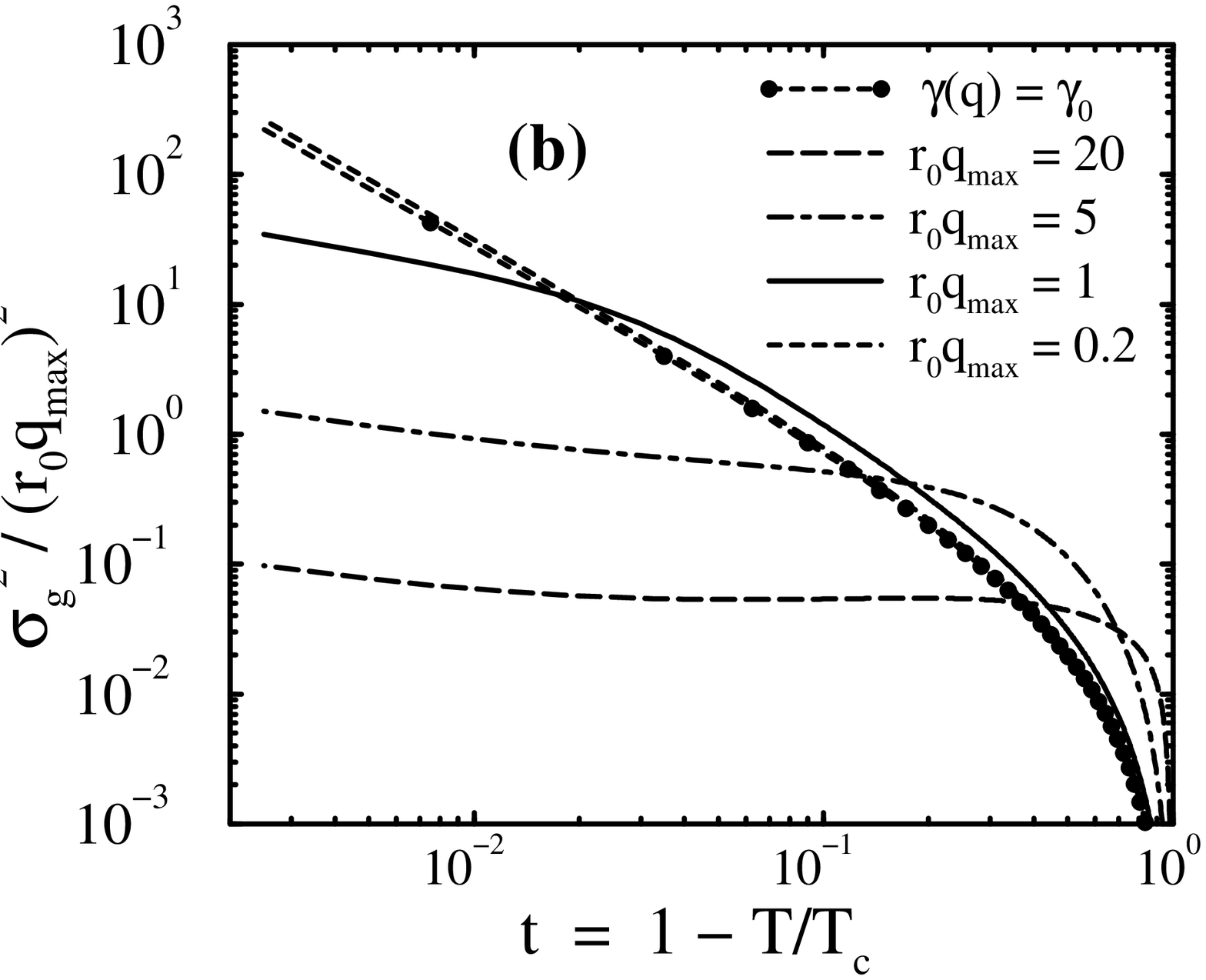}
\end{center}
\caption{ Linear (a) and logarithmic (b) 
temperature dependences   of the variance $\sigma_g^2$ characterizing the tilt 
angle distribution (Eq. (\ref{angle})) for the microscopic capillary wave theory as described
 in the main text (Eq. (\ref{tempsigma})); $L_cq_{max}=\infty$. $\sigma_g^2$ diverges upon 
approaching the critical point $t\rightarrow 0$. Within the simple capillary wave 
theory ($\gamma(q)=\gamma_0$) $\sigma_g^2$ diverges $\sim t^{-3/2}$ whereas 
within the microcscopic theory $\sigma_g^2\sim t^{-1/2}$. The crossover to this asymptotic 
behavior $\sim t^{-1/2}$ is shifted closer towards $T_c$ for increasing values of $r_0q_{max}$.
}
\label{temperature2}
\end{figure}

\subsection{Temperature dependence }

The temperature dependence of the local orientations of  
the liquid-vapor interface is of particular interest. 
The explicit results given above allow one to obtain a transparent
view of the analytic structure of $<\theta>(T)$ and $<(\delta\theta)^2>(T)$, as well as of its parametric
dependence on various features of the intrinsic density profile (such as the 
correlation length $\xi$ and the  
stiffness parameter $C_H$) and of the interaction potential $w(r)$ (such as the molecular size $r_0$ 
and the range of the potential). This transparency follows from the product
approximation for $\gamma(q)$ (see Eqs.  (\ref{sigmass}) and (\ref{finalsigma})). 
We have carried out a full
numerical analysis of the 
temperature dependence for the model considered above so that we are able to
assess the reliability of the product approximation compared to the full theory described 
in Ref. \cite{meckedietrich}.

For $T\rightarrow T_c$ the correlation length $\xi = \xi_0^-
t^{-\nu}$, $t = (T_c-T)/T_c$, 
diverges for $t\rightarrow 0$ where $\nu$ is a universal bulk
exponent with $\nu=0.5$ for the present mean field theory
(Eqs. (\ref{functional}) and (\ref{carnahan})). $\xi_0^-$ is a nonuniversal
amplitude with $\xi_0^- = r_0/2$ for the potential defined by
Eq. (\ref{potential}) .
For temperatures well below $T_c$ the correlation lengths $\xi^{(l)}$  and $\xi^{(g)}$  in the liquid
and  in the vapor phase, respectively,   differ  from
each other and from
the limiting common form $\xi = 
\xi_0^- t^{-\nu}$. 
Although it is straightforward to determine $\xi^{(l)}$ and
$\xi^{(g)}$ numerically, we have opted for the advantage of using the
 analytic expression given in Eq. (\ref{finalsigma0}) with $\xi_0^-=r_0/2$. 
The expression
in Eq. (\ref{finalsigma0}) has the virtue of fulfilling the  relations
$\xi^{(l)}>\xi(T)>\xi^{(g)}$ 
and will be used in the following whenever an
explicit expression for  the 
correlation length is needed.  Moreover, it has the appealing property that
at the triple point $T_{tr}\simeq (2/3)T_c$ the correlation length
$\xi(T_{tr}) \approx 0.58 r_0$ is of the order of the microscopic
cutoff length $r_0$, i.e.,  comparable to the diameter of the particles.   
Accordingly, one obtains for the temperature dependence of the variance 
(Eqs. (\ref{sigma_n}), (\ref{surfacetens}), and (\ref{finalsigma0})) 
\be{tempsigma} 
\sigma_g^2  = 2\lambda{T^2 \over T_c^2} 
\left(1- {T \over T_c}\right)^{-{3\over 2}} \; S_3(T) 
\ee
with 
\be{lambda} 
\lambda = {3.97 \over \pi^3} {k_BT_c \over w_0} (q_{max}r_0)^2 \;\;. 
\ee
Within the density functional for fluids presented in 
Sec. \protect\ref{capillary} one has  ${k_BT_c\over w_0} \simeq 0.22$ 
so that  $\lambda \simeq 0.028 (q_{max}r_0)^2$.

Whereas for the simple capillary wave theory 
($\gamma(q) =\gamma_0$) $S_3(T)={1\over 2}$ is constant, one finds for the micrsocopic theory 
with the bending rigidity $\bar{\kappa}(y)$ given by Eq. (\ref{bendingfunction}) the temperature 
dependence $S_3(T\rightarrow T_c) \sim t$, $t=1-{T\over T_c}$, upon  approaching 
the critical point at $T_c$ so that with  $\gamma_0(t\rightarrow 0) \sim t^{3/2}$  
within mean field theory $\sigma_g^2(t\rightarrow 0) \sim t^{-1/2}\ln t$ diverges for $T\rightarrow T_c$.

\begin{figure}[htbp]
\begin{center}
\includegraphics[width=0.45\linewidth]{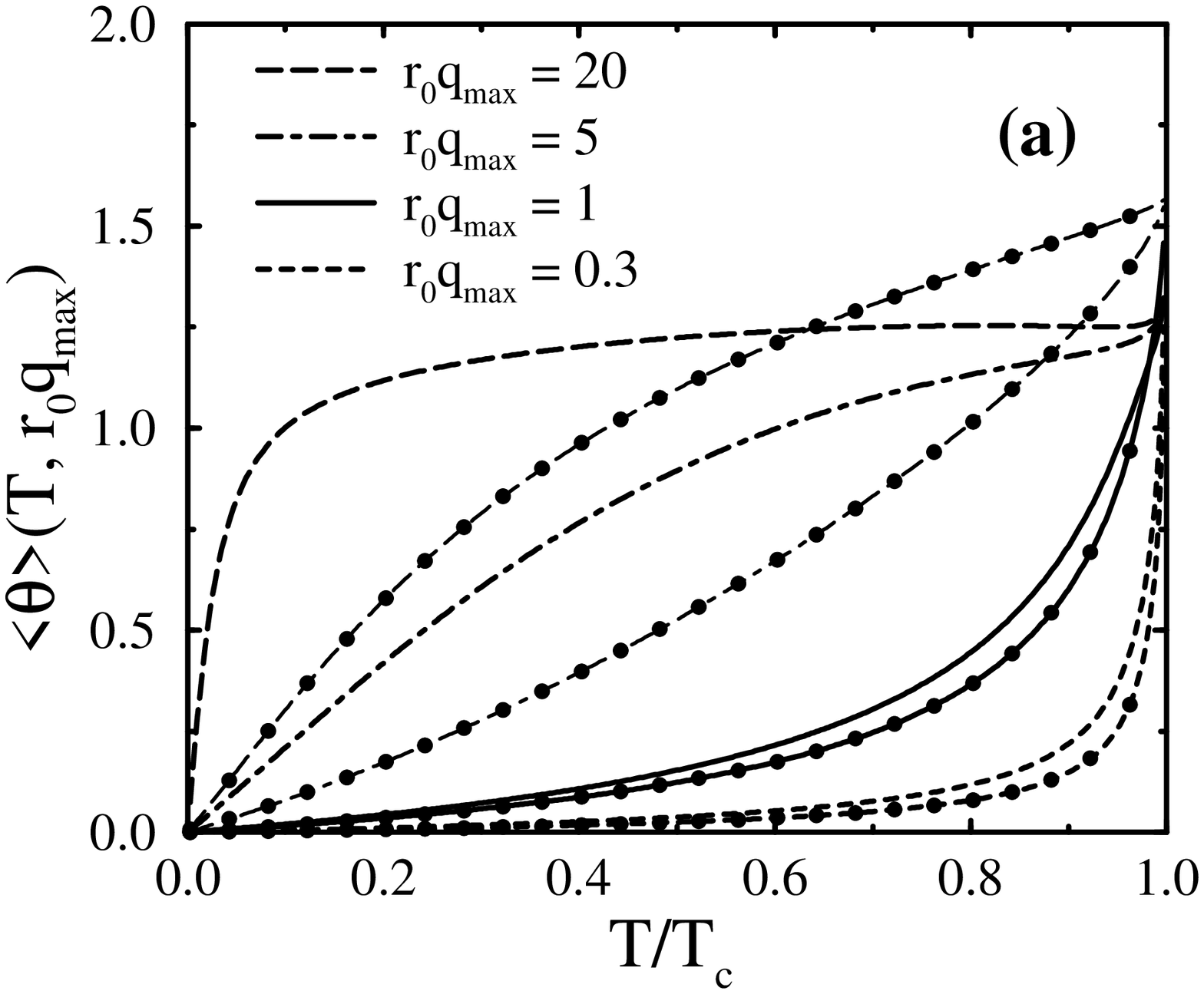}
\includegraphics[width=0.45\linewidth]{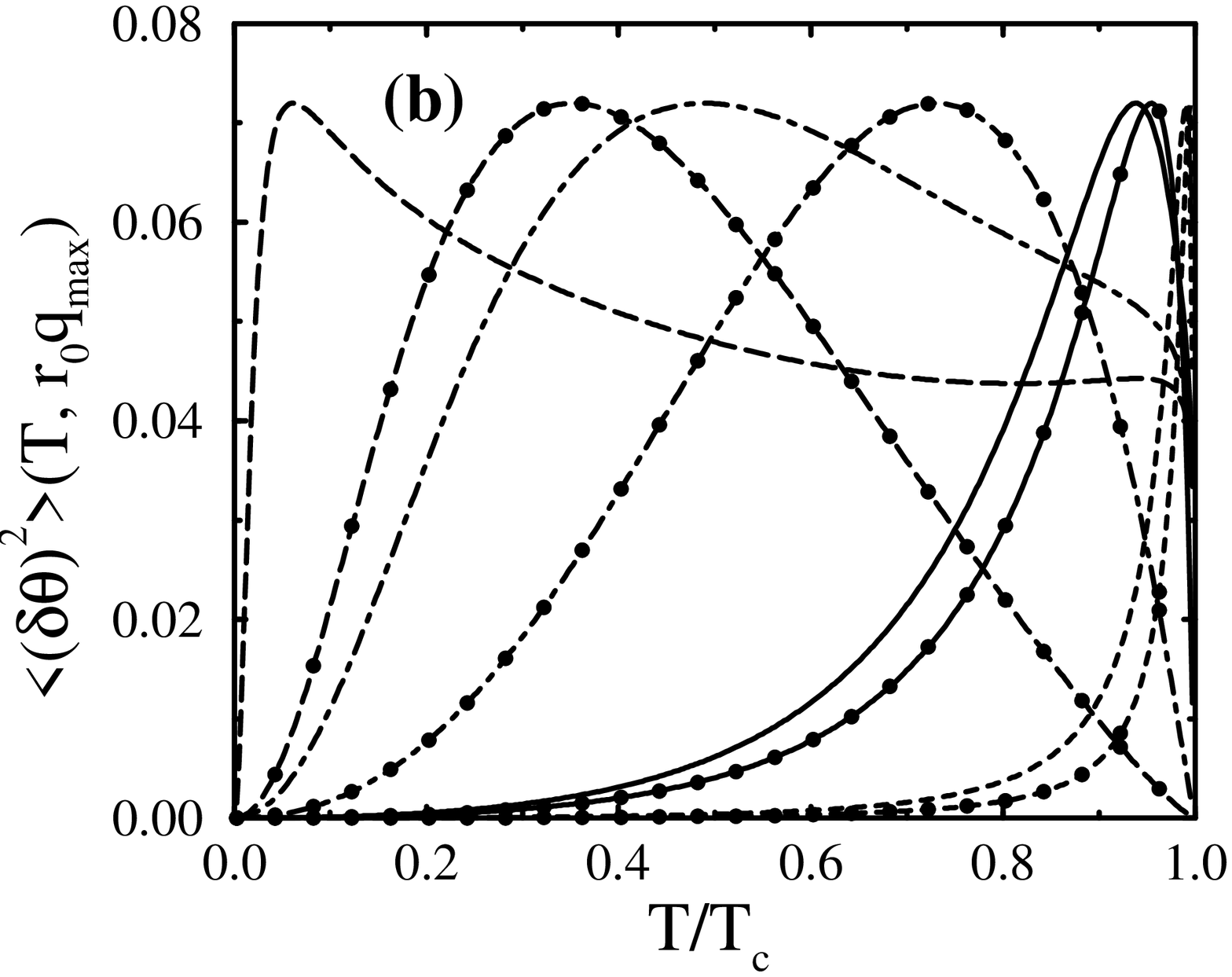}
\includegraphics[width=0.45\linewidth]{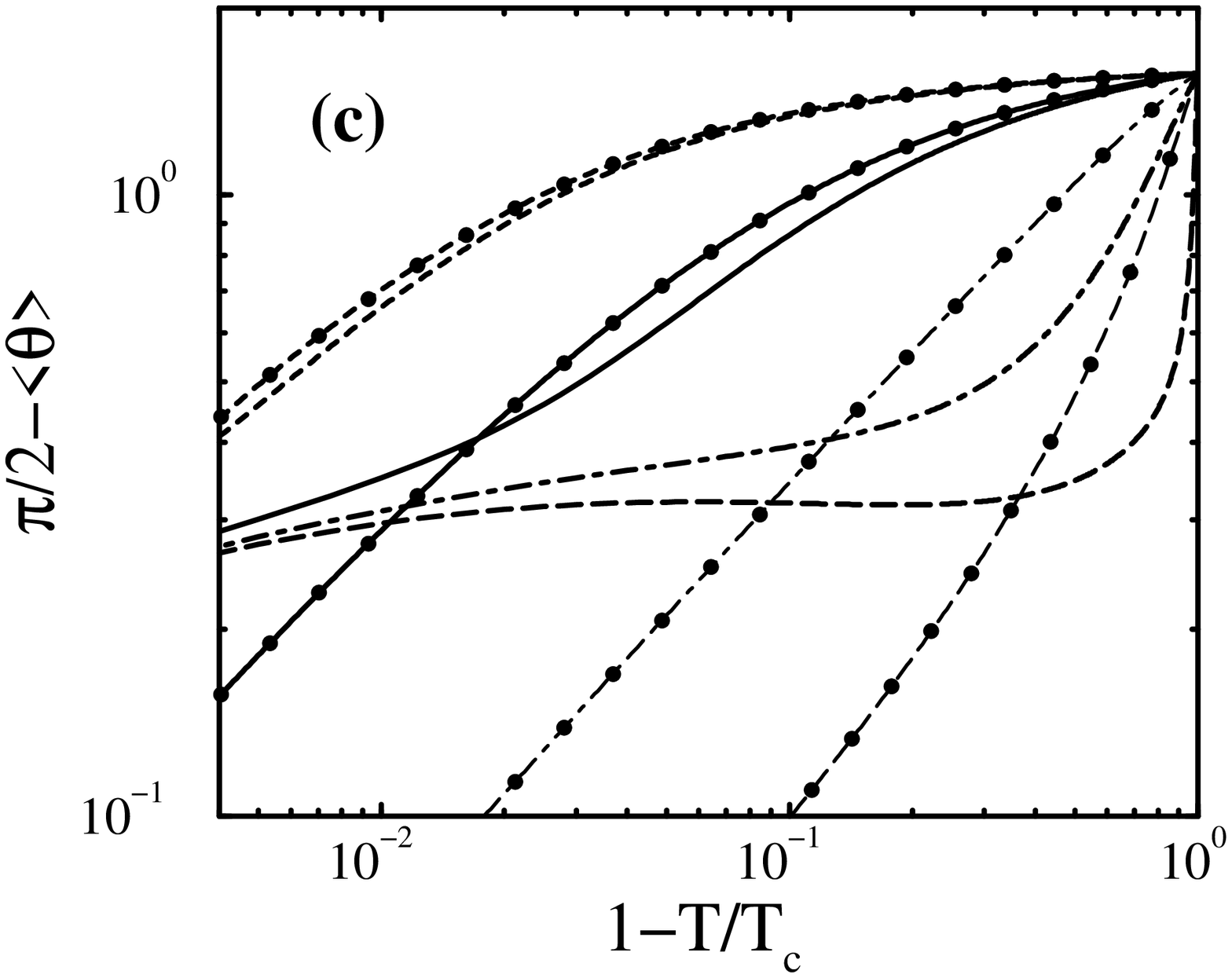}
\includegraphics[width=0.45\linewidth]{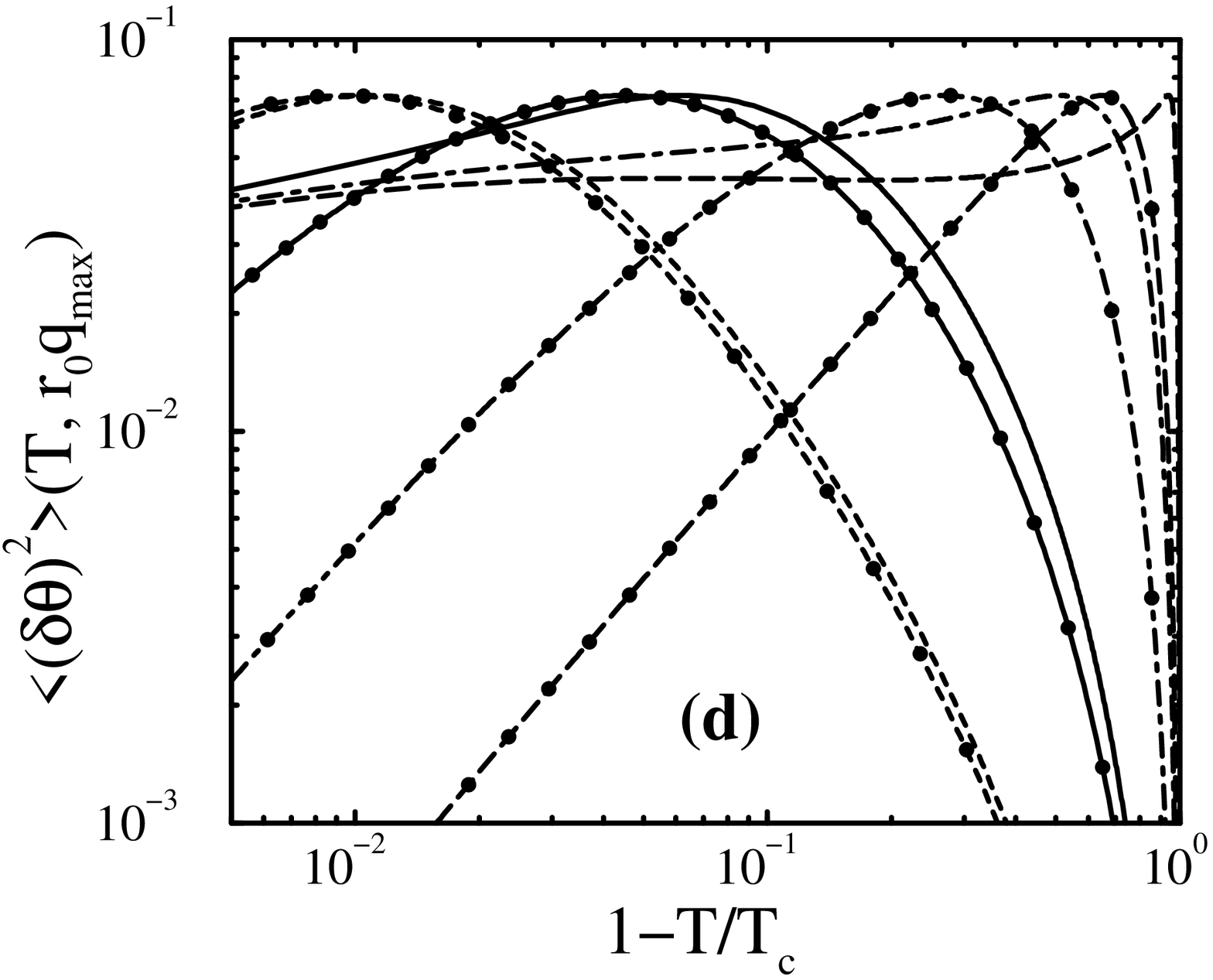}
\end{center}
\caption{Linear ((a), (b)) and logarithmic ((c), (d)) temperature dependences of the mean tilt 
angle $<\theta>(T)$ (Eq. (\protect\ref{meanangle}))   
and its mean squared deviation 
$<(\delta\theta)^2>(T)$ (Eq. (\protect\ref{varianceangle})) 
for the microscopic capillary wave theory (lines without filled circles) as described in the main text and 
in comparision with the simple capillary wave theory $\gamma(q)=\gamma_0$ 
(Eq. (\ref{tempangle})) shown with attached circles on the lines for the same values of the 
cut off length $r_0q_{max}$. 
The line code is the same for (a)-(d). 
 }
\label{temperature}
\end{figure}

In contrast one finds $S_3(T\rightarrow T_c) \sim t^2\ln t$ 
for  the phenomenological capillary wave theory with 
$\gamma(q) =\gamma_0 + \kappa_0 q^2$ and 
$\kappa_0={3\over 65}\gamma_0r_0^2\left({\xi^4\over r_0^4}
+{1\over 2}{\xi^2\over r_0^2}\right)$ (Eq. (\ref{surfacetens})), so that the 
variance $\sigma_g^2$ (Eq. \ref{variancetilt})) 
does not diverge but vanishes as $\sigma_g^2 \sim t^{1\over 2} \ln t$ for $T\rightarrow T_c$. 
However, for the microscopic theory one finds $S_3(T\rightarrow T_c) \sim t$ which 
implies $\sigma_g^2 \sim t^{-1/2}$.

For the simple capillary wave theory (constant surface energy $\gamma(q)=\gamma_0$) 
one has $S_3(T) = {1\over 2}$ (Eq. (\ref{skg3}) for $\bar{\kappa}_0=0$) so that  
one finds the temperature dependences (Eqs. (\ref{meanangle}) and (\ref{varianceangle})) 
\be{tempangle} 
\begin{array}{ll} 
<\theta> & \displaystyle 
\longrightarrow_{\!\!\!\!_{\!\!\!\!\!\!\!\!\!\! \!\!
T \rightarrow T_c }}  \; {\pi \over 2} - \sqrt{\pi \over \lambda} t^{3\over 4} 
+ {\pi \over 2} {1\over \lambda} t^{6\over 4} - \sqrt{\pi \over \lambda} t^{7\over 4} 
- \sqrt{4\pi \over 9\lambda^3} t^{9\over 4} 
+{\pi \over \lambda} t^{10\over 4} - \sqrt{\pi \over \lambda} t^{11\over 4} + {\cal O}(t^3)   \cr 
  &  \cr 
<(\delta \theta)^2> & \displaystyle 
\longrightarrow_{\!\!\!\!_{\!\!\!\!\!\!\!\!\!\! \!\!   
T \rightarrow T_c }}  \; - {3\over 2\lambda} t^{3\over 2} \ln t  
\; , \;\;  t= 1- {T\over T_c} \;,\;\; \gamma(q)=\gamma_0\;,  
\end{array} 
\ee
for  the mean value $<\theta>$ and the mean squared deviation $<(\delta\theta)^2>$ 
of the tilt angle $\theta$, respectively 
(see Fig. \ref{temperature}).

The values for the mean tilt angle $<\theta>$ and 
the mean squared deviations $<(\delta\theta)^2>$ depend strongly on the microscopic length 
scales $r_0 q_{max}$, due to the sensitivity on  the enhanced fluctuations 
of the interface on nanometer scales as predicted by the microsopic theory. 
In particular, the actual mean tilt angle is in general larger  for the microscopic theory than 
the one predicted by  the simple  phenomeneological  capillary wave 
theory with a constant surface tension $\gamma(q)=\gamma_0$.  
Only close to the critical point where the increase of the surface energy $\gamma(q)$ 
  for large $q$ 
becomes dominant (Eq. (\ref{finalsigma})), the phenomenological  capillary wave theory 
leads to larger values $<\theta>$ 
of the mean tilt angle, because it neglects  bending energies. Due to the decrease in surface 
energy  for wavevectors $q\sim 2\pi/\xi$  (see Eq. (\ref{finalsigma})) 
and the increase $\sim q^2$ for larger wavevectors the asymptotic behavior in the microscopic theory is given by $\pi/2 \; - <\theta>\; \sim t^{1/4}$ and $<(\delta\theta)^2>\; \sim t^{1/2}\ln t$ and not by the expressions given in Eq. (\ref{tempangle}) which hold 
for the  simple   capillary wave theory.  The change of the asymptotic 
slopes is  clearly visible in the double logarithmic plots in Figs. \ref{temperature}(c) and (d).

\section{Relations with experiments} 
\label{experiments}

Spectroscopic determinations of molecular orientations at fluid interfaces  provide 
valuable insights into the nature and structure of inhomogeneous fluid systems. 
These techniques render interfacial informations which are complementary to those 
which are obtained by X-ray and neutron scattering \cite{haase} which give no direct 
access to orientational ordering at interfaces. 
Recently Simpson and Rowlen have discussed relationships between the capillary wave 
spectrum and the local tilt angle of the surface \cite{simpson1,simpson2,simpson3}.

One finds in general two different types of approaches to measure local tilt angles of interfaces: 

(i) The interface is decorated with anisotropic particles which keep either a fixed angle with 
the fluctuating  interface or the thermal fluctuations of the molecule  depend in a systematic way on 
the interface orientation. Assuming that at low concentrations the decoration does not 
significantly modify  the interfacial fluctuations, one can use the dipole moments of 
these macromolecules as a probe 
to measure surface specific properties such as the distribution of tilt angles  
(see, for instance,  fluorescence and infrared spectroscopy, and Maxwell displacement 
current (MDC) measurements).

(ii) The fluctuating interface itself - albeit separating {\it isotropic} bulk phases of anisotropic fluid 
particles (e.g., carrying dipole moments) - can exhibit anisotropic features described by 
orientational profile (see, e.g., Ref. \cite{frodl}). 
Assuming that such an orientational  profile follows rigidly the fluctuations of the interface, 
one can probe 
selectively the interface, i.e., measure 
surface specific signals  
even without adding tracer macromolecules (see, for instance, second harmonic generation, 
AFM micrographs).

\subsection*{Fluorescence Spectroscopy} 

A standard way to measure molecular orientations at interfaces is using fluorescence spectrosopy. 
For such an excitation the polarization of the emitted light depends on the orientation $\theta_e$ of the emitting dipole 
and the orientation $\theta_a$ of the absorbing  dipole 
which both are  related 
to the orientation $\theta$ 
of the molecule at the interface. The relations $\theta_e(\theta)$ and $\theta_a(\theta)$ depend 
on the chemical structure of the fluorescence molecule and are in general fixed and time-independent. 
The so-called time-dependent anisotropy decay is usually defined as $r(t)={I_{zy}-(I_{yy}+I_{yx})/2\over I_{zy}+I_{yy}+I_{yx}}$ \cite{edmiston}, where the emission intensity $I_{ij}$ results from an 
excitation polarized along the $i$-axis and measured along the $j$-axis. Assuming that the absorption and emission dipoles are colinear and that the emission intensity is proportional to the projection of the dipole onto the detection plane,  the  measured decay of the anisotropy 
is given 
typically  by $r(t)=(r(0)-r(\infty))e^{-t/\tau}+r(\infty)$  (see Refs. \cite{edmiston,wirth,lakowicz}) and 
\be{fluorescence} 
r(0)  = {3\over 2} {<\cos^2\theta_a(\theta) \sin^2\theta_e(\theta)> \over <\sin^2\theta_e(\theta)>} - {1\over 2} \;\; \;\;
{\rm and}\;\; \;\; r(\infty) =  {3\over 2} <\cos^2\theta_a(\theta)> - {1\over 2}\;,  
\ee
where  $<\cdot>$ denotes an  average  over the orientations $\theta$ of the interface and thus 
probes the distribution function for the tilt angle $\theta$ (Eq. (\ref{angle})).

\subsection*{Infrared Spectroscopy and linear dichroism}

A classical technique to measure directly the orientation of macromolecules at interfaces is 
infrared reflection-absorption spectroscopy (see, e.g., Ref. \cite{duevel} and references therein).  
Molecular tilt angles of Langmuir-Blodgett films can be measured by  
the comparision of reflection  and transmission  intensities of  
infrared beams. The enhancement factor for the reflection-absorption to transmission-absorption 
intensities allows for a quantitative evaluation of molecular orientations \cite{umemura}. 
However, the dependence on the orientation is rather complex.

Anisotropic mean orientations of molecules at interfaces give rise to dichroism, i.e., a polarization 
dependent absorption due to the associated orientation of the transition dipole moments. 
The intensity $I$ of absorption is proportional to the mean 
squared cosine of the angle $\theta_a$ between  
the transition  moment axis and the electric field: 
\be{dichroism} 
I \; \sim \; <\cos^2 \theta_a>  \;\;. 
\ee
The angle $\theta_a(\theta)$ is determined by  the chemical structure of the absorbing molecule 
and by the orientation $\theta$ of the interface normal.  Because this relation is fixed, 
the orientational fluctuations of  $\theta_a$  are given by the probability distribution $P[\theta]$ of  the 
 local orientation $\theta$ of the supporting fluid interface  (Eq. \ref{angle}).  
For applications see Ref. \cite{simpson2} and references therein.

\subsection*{Generation of Maxwell displacement current (MDC)} 

It is  convenient  to investigate, for instance, the electrical properties of  monolayers 
of  macromolecules floating on  water surfaces (such as Langmuir-Blodgett films)  
by  the surface potential 
method with which the Maxwell displacement current is measured \cite{iwamoto91,iwamoto}. 
The induced charge $Q$ 
at the electrode  is directly proportional to the vertical component $m_z$ 
of the dipole moment of the molecule, i.e., for uniaxial polar molecules
\be{mdc} 
Q \; \sim \; <\cos\theta> \; = \; \int_0^{\pi/2} d\theta\; \sin\theta \cos\theta \; P[\theta]   \;\;. 
\ee

\subsection*{Second harmonic generation (SHG)} 

As a second-order nonlinear optical process, the frequency doubling of light, i.e., the second
 harmonic generation, is forbidden in isotropic media and thus provides a signal exclusively 
from spatial regions exhibiting deviations from isotropy, such as interfaces. Therefore, it can 
be used to measure orientational order at interfaces.  
Such measurements have been primarily 
performed for films of dye molecules of reasonably high symmetry, so that the number of
 nonzero components of the second-order nonlinear optical tensor $\beta^{(2)}$ of the isolated 
molecule is small. 
In these cases, with one component often dominating the nonlinear response, the  
nonlinear susceptibility 
tensor $\chi^{(2)}$ of the film can be 
used to extract information on molecular orientation.

For an uniaxial system of rod-like molecules  information about the 
 molecular tilt angle can be inferred 
from the relation \cite{heinz,hayden,goh,higgins,simpson4,simpson02} 
\be{shg} 
{2\chi^{(2)}_{xzx} 
\over \chi^{(2)}_{zzz}+2\chi^{(2)}_{zxx}}   \;=\; 
{<(\sin \theta)^3\cos \theta> \over <\cos\theta>} \;\;. 
\ee
The value  of each tensor component is proportional to the square root of the 
second harmonic intensity measured under the polarization conditions indicated by the indices. 
The first subscript denotes  the direction of the generated polarization driven by 
polarized electric fields indicated  by the second and the third index.

\subsection*{Direct Measurement of the tilt angle distribution}

Finally, the tilt angle distribution $P[\theta]$ (see Eq. (\ref{angle})) may be measured directly 
from an AFM micrograph provided that the tip diameter is smaller than the typical 
feature length of the interface. This experiment was done for solid interfaces such as silica surfaces 
(see Refs. \cite{simpson1,simpson2}) but may be achieved also for fluid interfaces when 
the capillary modes can be shock frozen  quickly enough. For simple liquids the typical relaxations 
are too fast  for any freezing technique to succeed in freezing  the capillary waves. But  
for interfaces involving polymers such as polystyrene-air interfaces at temperatures  close to the glass 
transition $T_g$, the tilt angle distribution may be measured by cooling fast 
below the critical temperature $T_g$ (see Ref. \cite{becker}).

\subsection*{Laser scanning confocal microscopy}

Adding polymers to a colloidal suspension under suitable conditions induces a fluid-fluid demixing transition into a colloid-rich phase and a colloid-poor phase which are separated by a fluid interface. This phase separation is driven by entropic forces. The mesoscopic size of the colloidal particles leads to ultralow values of the surface tension which scales up the thermal interface roughness into the $\mu m$ regime. The corresponding interface fluctuations can be monitored directly in real space by laser scanning confocal microscopy and can be interpreted in terms of the capillary wave spectrum \cite{aarts}. It seems to be possible to infer from such images also the local interface orientations and thus the full tilt angle distribution $P(\theta)$ as well as the lateral correlation function of tilt angles. This approach would be particular promising in that these systems allow one to tune the effective interaction between the colloidal particles within a wide range and thus generate designed forms of the wavelength dependent surface tension $\gamma(q)$ which governs the capillary waves. It might be possible to achieve a resolution down to the scale of the colloidal particles, which corresponds to short wavelengths on the atomic scale of simple fluids, and thus opens up the possibilities to study the influence of the large momentum cutoff.

\section{Summary and Conclusions}
\label{conclusion}

We have studied the local orientations of liquid-vapor interfaces with  a particular emphasis 
on the thermal fluctuations of the polar tilt angle $\theta$ between the local surface normal 
and the vertical direction given by the normal of the flat mean interface (Fig. \ref{interfacefig}). 
Our analysis 
is based on describing the interface by an effective interface Hamiltonian ${\cal H}$ which 
provides the statistical weight for capillary wavelike fluctuations. ${\cal H}$ is expressed in 
terms of a wavelength dependent surface tension $\gamma(q=2\pi/\lambda)$ and we have 
used three expressions for it: the simple capillary wave theory for which $\gamma(q)=\gamma_0$ is 
constant and equals the macroscopic surface tension $\gamma_0$ (Eq. (\ref{helfham}) with $\kappa_0=0$), 
a phenomenological expression for $\gamma(q)$ which takes into account bending rigidities 
(Eq. (\ref{sigma}) with $\kappa_0\neq 0$), and 
a microscopic expression which is derived from density functional 
theory (Eqs. (\ref{sigmass}) and (\ref{finalsigma})). Within the Gaussian approximation we have 
calculated the ensuing 
variances for the local interface positions and for its derivatives  (Eqs. (\ref{prob2}) and (\ref{prob8})) as well as 
the variance for the distribution of finite interfacial height differences  (Eqs. (\ref{joint}), (\ref{joint4}),  
and (\ref{test3}) as well as  Figs. \ref{figsigma_finite} and \ref{figsigmazero}).  
Based on these considerations we have derived the probability distribution function $P[\theta]$ 
for the tilt 
angle (Eq. (\ref{angle})), which exhibits a nonmonotonic dependence on the 
variance (Fig. \ref{fig_verteilung}) and 
allows us to determine the mean tilt angle $<\theta>$ and its mean squared deviation 
$<\theta^2>-<\theta>^2$ (Fig. \ref{sigma_angle}). 
In order to probe also   non-local  properties 
of the interface the  correlation function 
$<\theta(\vec{0})\theta(\vec{a})>$ of the tilt angles at different lateral positions 
is calculated (see Eq. (\ref{corr13})) which offers an interesting future 
tool  to analyze recent 
experimental data \cite{aarts}. 
The mean tilt angle and its moments depend on the 
physical parameters of the system solely via the variance of the height fluctuations 
$\sigma_g=\sigma_{k=3}$ (Eq. (\ref{prob8}) and Fig. \ref{sigma_angle}). Therefore, Sec. \ref{results} discusses the 
dependences of the variances (Eq. (\ref{sigma_n})) on the 
capillary length $L_c$  (Eq. (\ref{caplength})), the minimum wavelength $2\pi/q_{max}$, the stiffness length $L_\kappa$  (Eq. (\ref{caplength})),   and 
temperature (Figs. \ref{figs1}-\ref{temperature2}). These results demonstrate that these 
quantities discriminate 
clearly between the three types of wavelength dependent surface tensions mentioned above. 
This capacity of discrimination between different sophistications of effective interface 
Hamiltonians is transferred also to the temperature dependence of the tilt angle $<\theta>$ and 
its mean squared deviation $<\theta^2>-<\theta>^2$ (Fig. \ref{temperature}). It turns out that 
the tilt angle 
is a particularly sensitive probe of the wavelength dependent surface tension at short wavelengths 
which are difficult to reach with $X$-ray scattering. In Sec. \ref{experiments} we have discussed 
a variety of 
experimental techniques which allow one to measure the mean tilt angle by optical means.  
We conclude that the local orientations expressed in terms of the mean tilt angle and its moments 
offer an additional characterization of the thermal states of fluid interfaces on equal footing with 
the surface tension and density profiles and with a particular emphasis on the capillary wavelike 
fluctuations. This underscores the transferability of the concept of the wavelength dependent 
surface tension from describing interfacial structure factors to local orientations of fluid interfaces.

%\references


\begin{thebibliography}{99}
\bibitem{barker}
Starting from a Lennard-Jones interaction potential $W(r)= \epsilon
[(\sigma/r)^{12} - (\sigma/r)^6]$ the standard separation into   
$w_s(r)$ and $w(r)$ according to WCA \cite{evans2} leads to $w_{WCA}(r\geq
2^{1/6}\sigma) = W(r)$ and $w_{WCA}(r\leq
2^{1/6}\sigma) = -\epsilon$. The resulting nonanalyticity of  
$w_{WCA}(r)$ at $r=2^{1/6}\sigma$ prevents an analytic evaluation of
certain integrals involving $w(r)$. Our choice in Eq. (\ref{potential}) reduces to
$w_{WCA}(r)$ for $r\rightarrow \infty$ with $r_0=\sigma$ and  
$w_0=\epsilon$ and resembles closely $w_{WCA}(r)$ for small $r$ 
($w(r=0)=w_{WCA}(r=0)$); furthermore $w(r)$ as given in Eq. (\ref{potential})  
has the advantage that it 
is analytic and thus facilitates our following  analysis. This
advantage outweighs the fact that Eq. (\ref{potential}) deviates slightly from the
standard WCA approach.
\end{thebibliography}
\end{document}